\documentclass[12pt,preprint]{aastex}
\usepackage{amssymb,amsmath}
\newcounter{address}
\newcommand{\ie}{i.e.}
\newcommand{\etal}{et al.}
\newcommand{\eg}{e.g.}
\newcommand{\viz}{viz.}
\newcommand{\eqnname}{equation}

\newcommand{\sectionname}{Section}
\newcommand{\documentname}{article}

\newcommand{\flag}[1]{\texttt{\lowercase{#1}}}

\newcommand{\dd}{\mathrm{d}}
\renewcommand{\vec}[1]{\mathbf{#1}} 
\newcommand{\ten}[1]{\mathbf{#1}} 

\newcommand{\mm}{\vec{m}}
\newcommand{\rr}{\vec{r}}
\newcommand{\VV}{\ten{V}}
\newcommand{\TT}{\ten{T}}
\newcommand{\SSS}{\ten{S}}

\newcommand{\project}[1]{\emph{#1}}
\newcommand{\sdss}{\project{SDSS}}
\newcommand{\sdssiii}{\project{SDSS-III}}
\newcommand{\galex}{\project{GALEX}}
\newcommand{\ukidss}{\project{UKIDSS}}
\newcommand{\ukidsslas}{\ukidss\ \emph{LAS}}

\newcommand{\lsst}{\project{LSST}}

\newcommand{\slaq}{2SLAQ}
\newcommand{\xdqso}{\project{XDQSO}}
\newcommand{\xdqsoz}{\project{XDQSOz}}

\newcommand{\exd}{XD}

\newcommand{\boss}{\project{BOSS}}
\newcommand{\twoslaq}{\project{2SLAQ}}
\newcommand{\baf}{BAF}

\newcommand{\uv}{UV}
\newcommand{\ir}{IR}
\newcommand{\nir}{N\ir}

\newcommand{\ugriz}{\ensuremath{ugriz}}
\newcommand{\nuv}{NUV}
\newcommand{\fuv}{FUV}
\newcommand{\yjhk}{YJHK}

\newcommand{\iband}{$i$-band}

\newcommand{\fluxi}{\ensuremath{f_i}}
\newcommand{\setofrfj}{\ensuremath{\{f_j/\fluxi\}}}

\newcommand{\nbins}{47}
\newcommand{\binwidth}{0.2}
\newcommand{\binspacing}{0.1}
\newcommand{\ilow}{17.7}
\newcommand{\ihigh}{22.5}

\begin{document}

\title{Photometric redshifts and quasar
  probabilities from a single, data-driven generative model}
\author{Jo~Bovy\altaffilmark{\ref{NYU},\ref{email}},
  Adam~D.~Myers\altaffilmark{\ref{Wyoming},\ref{MPIA}},
  Joseph~F.~Hennawi\altaffilmark{\ref{MPIA}},
  David~W.~Hogg\altaffilmark{\ref{NYU},\ref{MPIA}},
  Richard~G.~McMahon\altaffilmark{\ref{Cambridge},\ref{KavliCambridge}},
  David~Schiminovich\altaffilmark{\ref{Columbia}},
  Erin~S.~Sheldon\altaffilmark{\ref{Brookhaven}},
  Jon~Brinkmann\altaffilmark{\ref{APO}},
  Donald~P.~Schneider\altaffilmark{\ref{PennState},\ref{PennGrav}}, 
  Benjamin~A.~Weaver\altaffilmark{\ref{NYU}}}
\altaffiltext{\theaddress}{\label{NYU}\stepcounter{address} Center for
  Cosmology and Particle Physics, Department of Physics, New York
  University, 4 Washington Place, New York, NY 10003, USA}
\altaffiltext{\theaddress}{\label{email}\stepcounter{address}
  Correspondence should be addressed to jo.bovy@nyu.edu~.}
\altaffiltext{\theaddress}{\label{Wyoming}\stepcounter{address}
  Department of Physics and Astronomy, University of Wyoming, 
  Laramie, WY 82071, USA}
\altaffiltext{\theaddress}{\label{MPIA}\stepcounter{address}
  Max-Planck-Institut f\"ur Astronomie, K\"onigstuhl 17, D-69117
  Heidelberg, Germany}
\altaffiltext{\theaddress}{\label{Cambridge}\stepcounter{address}
  Institute of Astronomy, University of Cambridge, Madingley Road,
  Cambridge, CB3 0HA, UK}
\altaffiltext{\theaddress}{\label{KavliCambridge}\stepcounter{address}
  Kavli Institute for Cosmology, University of Cambridge, Madingley Road,
  Cambridge, CB3 0HA, UK}
\altaffiltext{\theaddress}{\label{Columbia}\stepcounter{address}
  Department of Astronomy, Columbia University, New York, NY 10027,
  USA}
\altaffiltext{\theaddress}{\label{Brookhaven}\stepcounter{address}
  Brookhaven National Laboratory, Upton, NY 11973, USA}
\altaffiltext{\theaddress}{\label{APO}\stepcounter{address}
  Apache Point Observatory, P.O. Box 59, Sunspot, NM 88349}
\altaffiltext{\theaddress}{\label{PennState}\stepcounter{address}
  Department of Astronomy and Astrophysics, The Pennsylvania State
  University, 525 Davey Laboratory, University Park, PA 16802, USA}
\altaffiltext{\theaddress}{\label{PennGrav}\stepcounter{address}
 Institute for Gravitation and the Cosmos, The Pennsylvania State University,
  University Park, PA 16802}

\begin{abstract}
We describe a technique for simultaneously classifying and estimating
the redshift of quasars. It can separate quasars from stars in
arbitrary redshift ranges, estimate full posterior distribution
functions for the redshift, and naturally incorporate flux
uncertainties, missing data, and multi-wavelength photometry. We build
models of quasars in flux--redshift space by applying the
\emph{extreme deconvolution} technique to estimate the underlying
density. By integrating this density over redshift one can obtain
quasar flux--densities in different redshift ranges. This approach
allows for efficient, consistent, and fast classification and
photometric redshift estimation. This is achieved by combining the
speed obtained by choosing simple analytical forms as the basis of our
density model with the flexibility of non-parametric models through
the use of many simple components with many parameters. We show that
this technique is competitive with the best photometric quasar
classification techniques---which are limited to fixed, broad redshift
ranges and high signal-to-noise ratio data---and with the best
photometric redshift techniques when applied to broadband optical
data. We demonstrate that the inclusion of \uv\ and \nir\ data
significantly improves photometric quasar--star separation and
essentially resolves all of the redshift degeneracies for quasars
inherent to the \ugriz\ filter system, even when included data have a
low signal-to-noise ratio. For quasars spectroscopically confirmed by
the \sdss\ 84 and 97\,percent of the objects with \galex\ \uv\ and
\ukidss\ \nir\ data have photometric redshifts within 0.1 and 0.3,
respectively, of the spectroscopic redshift; this amounts to about a
factor of three improvement over \ugriz-only photometric redshifts.
Our code to calculate quasar probabilities and redshift probability
distributions is publicly available.
\end{abstract}


\section{Introduction}\label{sec:intro}

The last decade has seen the first instances of statistical studies
with quasars using purely photometric samples. Examples of these
include the measurement of the integrated Sachs--Wolfe effect
\citep{Giannantonio06a,Giannantonio08a} and cosmic magnification bias
\citep{Scranton05a}, and studies of the clustering of quasars on large
\citep{Myers06a,Myers07a} and small \citep{Hennawi06a,Myers07b}
scales. The importance of photometrically classified quasar samples
will only increase during the next decade as large new imaging surveys will
uncover large samples of quasars at fainter magnitudes, with 
minimal spectroscopy for the faintest objects. While efficient
photometric classification is one requirement to facilitate studies of
quasars without extensive spectroscopy, it has also been crucial 
to develop accurate methods for quasar
redshift estimation based on broadband photometry. Techniques for 
photometric redshift estimation have long been 
successful for galaxies \citep[\eg,][]{Baum62a,Connolly95a} and
became feasible for quasars with the advent of precise multi-filter
photometry \citep{Richards01b,Richards01a,Budavari01a,Wolf04a}.

Closely related to the quasar photometric-redshift
problem---traditionally seen as a \emph{regression} problem---is the
question as to how best to perform photometric \emph{classification}
of quasars. It has become clear that the best classifiers are
probabilistic in nature in that they calculate probabilities for
objects to be quasars based on accurately calibrated models for
stellar and quasar photometry
\citep[\eg,][]{Richards04a,Bovy11a}. These probabilities are often
calculated for quasars in certain broad redshift ranges, and they
therefore also act as low-resolution photometric redshifts for the
objects they classify as quasars. The object classification technique
of \citet{Suchkov05a} uses bins of width $\Delta z = 0.2$ and, thus,
achieves classification with a finer photometric-redshift
estimate. However, detailed photometric redshift estimates for
photometrically classified quasars utilize heterogeneous techniques,
such that the resulting redshift probability distributions are
inconsistent with the broad probabilities used for the initial quasar
classification. For instance, this is the case for the photometric
quasar catalogs of \citet{Richards04a,Richards09a}. For these
catalogs, a non-parametric kernel-density-estimation (KDE) technique
that ignores photometric uncertainties was used to classify quasars,
while a parametric model that convolves the quasar color locus with
the photometric uncertainties---a single Gaussian distribution in bins
of redshift $\Delta z \approx 0.075$---was applied to estimate
redshift \citep{Weinstein04a}.

For many purposes, one would like to target quasars in arbitrary
redshift ranges that differ from those predetermined and imposed by a
broad classification method. For example, the Baryon Oscillation
Spectroscopic Survey (\boss; \citealt{Eisenstein11a}) of \emph{The
  Sloan Digital Sky Survey III} (\sdssiii) aims to measure the baryon
acoustic feature in the Ly$\alpha$ forest of medium-redshift ($2.2
\lesssim z \lesssim 4.0$) quasars
\citep[\eg,][]{McDonald07a,Mcquinn11a}. The spectral range accessible
to the \boss\ spectrographs is $3600 < \lambda < 10000$
\AA\ (Barkhouser \etal, 2011, in preparation), thus \boss\ can only
study the Ly$\alpha$ forest as traced by redshift $z \gtrsim 2.2$
quasars. Therefore, \boss\ requires quasars to be targeted based on
their probability to be at redshift $\geq 2.2$, and the \boss\ quasar
classifiers were trained with this constraint
\citep[\eg,][]{Ross11a,Bovy11a}. However, other ground-based
instruments can observe at shorter wavelengths, \eg, the Multi-Object
Double Spectrograph for the Large Binocular Telescope, which can
observe the spectral range $3400 \AA\ < \lambda < 10000$
\AA\ \citep{Pogge10a}. This instrument could study the Ly$\alpha$
forest starting at redshift $z \gtrsim 2$. An Ly$\alpha$ forest
experiment designed for the Large Binocular Telescope might therefore
target quasars in the redshift range $2.0 \leq z < 2.2$ in addition to
those at higher redshift.

Another example of a project that requires accurate photometric
characterization of quasars is the search for binary quasars
\citep{Hennawi06a,Myers08a,Hennawi10a,Shen10a}, where the key metric
is the probability that two objects are both quasars {\em and}
proximate in redshift, \ie, the joint (or ``overlapping'') probability
that both components of a pair of objects are quasars of a particular
redshift. Similarly, the search for projected quasar pairs for
absorption line studies
\citep{Hennawi06b,Bowen06a,Hennawi07a,Prochaska09a} requires the joint
probability that both objects in a projected pair are quasars. Such
calculations require a full model of quasar probabilities and
redshifts. Ideally, therefore, photometric redshift estimation and
quasar classification ought to be performed together.

In the specific case of objects observed using the \ugriz\ filter
system \citep{Fukugita:1996qt}, quasar photometric redshifts are
plagued by a host of degeneracies at redshifts where various quasar
emission lines are mistaken for the Ly$\alpha$ line
\citep{Richards02a}; this results in ``catastrophic'' redshift
failures, although consideration of the full redshift posterior
distribution function (PDF) shows that most of these failures have a
significant integrated probability around the correct redshift (\eg,
\citealt{Ball08a}; see below). The addition of non-\ugriz\ data, \eg,
ultraviolet (\uv) and near-infrared (\nir) measurements, can both
alleviate these redshift degeneracies {\em and} improve quasar--star
separation. Quasar classification and characterization in the infrared
has been considered for simulated objects and for quasar samples with
a range of depths and areas
\citep[\eg,][]{Warren00a,Croom01a,Francis04a,Glikman06a,Maddox06a,Chiu07a,Richards09b,abrusco09a,Assef10a,Wu10a,Peth11a}. The
\nir\ is also the region to search for the highest redshift
quasars \citep[redshift $z \gtrsim 6$;][]{Mortlock11a}. These studies
show the great promise that \nir\ data hold for quasar
selection and redshift estimation. The \uv\ holds a similar
potential \citep[see,
  \eg,][]{Atlee07a,Trammell07a,Jiminez09a,Hutchings10a}.

The technique we introduce in this \documentname, which we denote
\xdqsoz, is the first that deals with the simultaneous classification
of quasars and assignation of quasar redshifts. This technique extends
the \xdqso\ quasar classification technique of \citet{Bovy11a} to
model the density of quasars in color--redshift space with a flexible
semi-parametric model consisting of a large set of Gaussian component
distributions. This model can be integrated analytically over any
redshift range to calculate probabilities from flux measurements for
individual objects. This, in turn, allows quasar probabilities to be
calculated over any redshift range. Thus, a probability distribution
in redshift space (a ``PDF'') is a natural component of the
model. Because we use the \emph{extreme deconvolution} (\exd) technique
\citep{Bovy09a} as our density estimation tool, the method can be
trained on and applied to low signal-to-noise ratio data, even with
missing values, \eg, to objects missing measurements in any arbitrary
collection of filters. This feature allows us to naturally include
\uv\ and \nir\ broadband fluxes, where sky coverages differ, as part
of our model space and to distinguish sources that are missing data in
a particular band from objects that are dropping out of that band. We
show that the addition of \uv\ and \nir\ broadband fluxes improves
quasar--star separation significantly and that it essentially breaks
all of the redshift degeneracies inherent to the \ugriz\ filter set.

This \documentname\ is organized as follows. In
\sectionname~\ref{sec:general}, we discuss general aspects of
photometric redshift estimation and classification in the context of
quasars. We briefly describe the data used to train and test the new
method in \sectionname~\ref{sec:data}. \sectionname~\ref{sec:xdqsoz}
contains a full description of the \xdqsoz\ quasar model and
\sectionname~\ref{sec:targeting} shows how this model is used to
calculate quasar probabilities over arbitrary redshift
ranges. \sectionname~\ref{sec:photoz} assesses the performance of the
photometric redshifts obtained using the \xdqsoz\ model. A discussion
of various extensions of the model is given in
\sectionname~\ref{sec:discussion} and we conclude in
\sectionname~\ref{sec:conclusion}. The \appendixname\ describes the
photometric classification and redshift estimation \xdqsoz\ code that
is made publicly available.

In what follows, AB magnitudes \citep{Oke83a} are used
throughout. Where dereddened fluxes and magnitudes are required we
have used the reddening maps of \citet{Schlegel98a}. All magnitudes
and fluxes should be considered as dereddened unless mentioned
otherwise.

\section{General considerations}\label{sec:general}

A technique for photometric redshift estimation of quasars should have
the following properties.
\begin{itemize}
\item It should provide full probability distributions for the
  redshift of the quasar based on its observed photometry, because
  this information has particular utility for quasars
  \citep[\eg,][]{Myers09a} as near-degeneracies in redshift estimation
  from broadband photometry are ubiquitous for quasars
  \citep[\eg,][]{Richards01a,Budavari01a}.
\item Upon the evaluation of the probability of the redshift the
  photometric uncertainties should be treated properly to allow
  photometric redshift estimation for faint objects.
\item If based on an empirical training set, the technique should be
  able to be trained on low signal-to-noise ratio data with
  potentially missing data. The training set and the evaluation set
  should also be allowed to have different noise properties, \eg,
  different distributions of signal-to-noise ratio. For example, while
  the optical fluxes are mostly well measured for a spectroscopic
  training sample, the addition of \uv\ and \nir\ data can help break
  redshift degeneracies (see below), but these measurements often have
  low signal-to-noise ratio, even for the training set.
\item The technique should allow an explicit redshift prior to be
  specified.
\end{itemize}

The key to photometric classification and redshift estimation for
quasars based on broadband fluxes is the joint probability of an
object's fluxes, its redshift, and the proposition that it is a quasar
$p(\mathrm{flux}, z, \mathrm{quasar})$. This joint probability can be
re-written in several ways that correspond to different ways of
approaching the problem
\begin{align}
p(\mathrm{fluxes}, z, \mathrm{quasar}) 
&= p(\mathrm{fluxes}| z, \mathrm{quasar})\,p(z|\mathrm{quasar})\,P(\mathrm{quasar})\label{eq:jointprob1}\\
&= p(\mathrm{fluxes}, z| \mathrm{quasar})\,P(\mathrm{quasar})\label{eq:jointprob2}\\
&= p(z|\mathrm{fluxes}, \mathrm{quasar})\,p(\mathrm{fluxes}|\mathrm{quasar})\,P(\mathrm{quasar})\,.\label{eq:jointprob3}
\end{align}
Photometric redshift estimation corresponds to the probability of an
object's redshift conditioned on its fluxes and assuming that it is a
quasar:
\begin{equation}
p(z|\mathrm{fluxes},\mathrm{quasar}) =
\frac{p(\mathrm{fluxes},z,\mathrm{quasar})}{p(\mathrm{fluxes},\mathrm{quasar})}\,.
\end{equation}
Quasar classification is the probability that an object is a quasar
based on its fluxes. To classify quasars in a certain redshift range
$\Delta z$, we integrate the joint probability that the object is a
quasar with redshift $z$ over redshift:
\begin{align}\label{eq:quasardeltazprob}
P(\mathrm{quasar\ in}\ \Delta z | \mathrm{fluxes}) &= 
\int_{\Delta z} \dd z\, p(\mathrm{quasar},z|\mathrm{fluxes})\\
&= \int_{\Delta z} \dd z\, \frac{p(\mathrm{quasar},z,\mathrm{fluxes})}{p(\mathrm{fluxes})}
\end{align}
The probability that an object is a quasar of any redshift is obtained
by setting the redshift range $\Delta z = [0,\infty]$. The
normalization factor $p(\mathrm{fluxes})$ in this \eqnname\ is given
by
\begin{equation}\label{eq:normalization}
p(\mathrm{fluxes}) = p(\mathrm{fluxes} , \mathrm{quasar}) + p(\mathrm{fluxes} , \mathrm{not\ a\ quasar})\,.
\end{equation}
The probability of an object not being a quasar can be obtained
empirically by modeling the fluxes of non-quasars
\citep[see][]{Richards04a,Bovy11a}.

The discussion above suggests that a unified approach to
classification and photometric redshift estimation is possible. Because
the method described in this \documentname\ is the first technique in
this class, we briefly discuss previous attempts at photometric
redshift estimation and how they fit in the framework outlined in this
section.

The $k$-nearest neighbors approach of \citet{Ball07a} is an
instance-based machine-learning technique that compares the colors of
test objects to the $k$ nearest objects in color-space in a training
set, and assigns a weighted combination of the redshifts of those
nearest neighbors to the test object. Its generalization to take
observational flux-uncertainties into account involves perturbing both
the test and the training data within their Gaussian noise ellipsoids
\citep{Ball08a}. In its noiseless implementation the method does not
return a full probability distribution for the redshift. When taking
the photometric uncertainties into account the technique essentially
returns samples from $p(z|\mathrm{flux},\mathrm{quasar})$ as in
\eqnname~(\ref{eq:jointprob3}), which can be binned to obtain the full
posterior distribution function. While the photometric uncertainties
of the test objects are handled correctly, the approach for dealing
with the uncertainties of the training data effectively convolves with
the uncertainties twice, as it adds scatter to the training data that
are already scattered from the intrinsic distribution due to
photometric noise. Because the technique directly uses the training
set, it also implicitly applies a redshift prior that approaches the
{\em observed} redshift distribution. This choice of prior does not
reflect the {\em intrinsic} redshift distribution, as the {\em
  observed} distribution is shaped by various selection effects
\citep{Richards06a}.

The approach taken by \citet{Hennawi10a} consists of fitting the
relative-flux--redshift distribution and its scatter to produce the
likelihood of the quasar redshift as in
\eqnname~(\ref{eq:jointprob1}). This fit is conducted without taking the
flux uncertainties into account, but upon evaluation of test objects 
the flux uncertainties are fully handled.

Closest to the approach taken in this \documentname\ is the technique
of \citet{Weinstein04a}. The distribution of colors in a set of narrow
bins in redshift is fit as a single multi-variate Gaussian
distribution. This approach is similar to quasar classification approaches 
where the color or relative-flux distributions
of quasars are fit in much broader redshift ranges using more general
density models (\citealt{Richards04a}, \citealt{Bovy11a}). \citet{Weinstein04a}
do not use the photometric uncertainties of the data they use for training. 
But, as in \citet{Hennawi10a}, photometric
uncertainties for test objects are fully taken into account.

All of the techniques described above could be extended to allow
quasar classification by specifying the necessary factors of
$P(\mathrm{quasar})$ or $p(\mathrm{fluxes},\mathrm{quasar})$ in
\eqnname s~(\ref{eq:jointprob1})--(\ref{eq:jointprob3}). The latter
could be taken from a quasar classification scheme such as NBC-KDE
\citep{Richards09a} or \xdqso\ \citep{Bovy11a}, although care should
be taken that the classification method uses the same redshift prior
as the photometric redshift technique for consistency.

The \xdqsoz\ technique introduced in this \documentname\ uses
\eqnname~(\ref{eq:jointprob2}) as the basis of both quasar
classification and photometric redshift estimation. Specifically, we
model the relative-flux--redshift distribution using a large number of
Gaussians by deconvolving this distribution for a training set using
the \exd\ technique
\citep{Bovy09a}. We use empirical relative fluxes that are
re-weighted using an explicit, magnitude-dependent redshift prior
(which can easily be divided out). As described in
\sectionname~\ref{sec:targeting}, conditioning on the fluxes to obtain
full photometric probability distributions for the redshift, and
marginalization over redshift to classify quasars, is simple and fast
in this approach. Because we deconvolve the relative-flux--redshift
distribution when training, we can straightforwardly incorporate
\uv\ and \nir\ data, both of which significantly improve the accuracy
and precision of the inferred redshifts.

\section{Training data}\label{sec:data}

\subsection{Optical data from the \emph{Sloan Digital Sky Survey}}

The Sloan Digital Sky Survey (\sdss; \citealt{York:2000gk}) has
obtained \emph{u,g,r,i} and \emph{z} CCD imaging of $\approx$ 10$^4$
deg$^2$ of the northern and southern Galactic sky
\citep{Gunn:1998vh,Stoughton:2002ae,Gunn06a}. \sdssiii\ \citep{Eisenstein11a}
has extended this area by approximately 2,500 deg$^2$ in the southern
Galactic cap \citep{Aihara11a}. All the data processing, including
astrometry \citep{Pier:2002iq}, source identification, deblending and
photometry \citep{Lupton:2001zb}, and calibration
\citep{Fukugita:1996qt, Hogg01a,Smith:2002pca,
  Ivezic:2004bf,Padmanabhan08a} are performed with automated
\sdss\ software. \sdss\ DR7 imaging observations were obtained over
the period 2000 March to 2007 July.

The \sdss\ training data used here are essentially the same as the
data used to train the \xdqso\ method; they are described in detail in
\citet{Bovy11a}.

We use a sample of 103,601 spectroscopically-confirmed redshift $z
\geq 0.3$ quasars from the SDSS DR7 quasar catalog
\citep{Richards02a,Schneider10a}. We use \emph{all} of these quasars
to essentially model the color--redshift relation for quasars (but see
below for the detailed description of our method). We combine the
color--redshift relation with an apparent-magnitude dependent redshift
prior obtained by integrating a model for the quasar luminosity
function over the apparent-magnitude range of interest
\citep{Hopkins07a}. This prior for a few bins in apparent magnitude is
shown in \figurename~\ref{fig:redshiftprior}; also shown is the
difference between the \citet{Hopkins07a} redshift prior and the prior
derived from the \citet{Richards06a} luminosity function. As the
sample of quasars from the \sdss\ DR7 quasar catalog spans a wide
range in luminosity that we apply to a narrow range in apparent
magnitude and that we extrapolate to largely unexplored faint flux
levels, we are ignoring correlations between quasar spectral
properties and luminosity \citep[\eg,][]{Baldwin77,Yip2004}. These
correlations mostly affect emission line shapes, such that they are
washed out in broadband colors, especially compared to the intrinsic
color-scatter.

\subsection{\uv\ data from the \emph{Galaxy Evolution Explorer}}\label{sec:galexdata}

In addition to \ugriz\ optical data, we use \uv\ data obtained by the
\emph{Galaxy Evolution Explorer} space mission
\citep[\galex;][]{Martin05a}. \galex\ has performed an all-sky imaging
survey in two \uv\ bands (\fuv: 1350 to 1750 \AA; \nuv: 1750 to 2750
\AA) down to $m_{\mathrm{AB}} \approx 20.5$ and a medium-deep imaging
survey that reaches $m_{\mathrm{AB}} \approx 23$
\citep[\eg,][]{Bianchi11}.  Some of the data used below to test the
technique described in this \documentname\ comes from the medium-deep
survey, while much of the data used to \emph{train} our technique
comes from the shallower all-sky survey (because our training sample
of quasars is drawn from the full $\approx 10,000$ deg$^2$
\sdss\ footprint), but this difference is largely offset by the fact
that our training set is brighter than the faint part of the test
set. \galex\ GR5 observations were obtained between April 2003 and
February 2009.

\begin{figure}
\begin{center}
\includegraphics[width=0.6\textwidth]{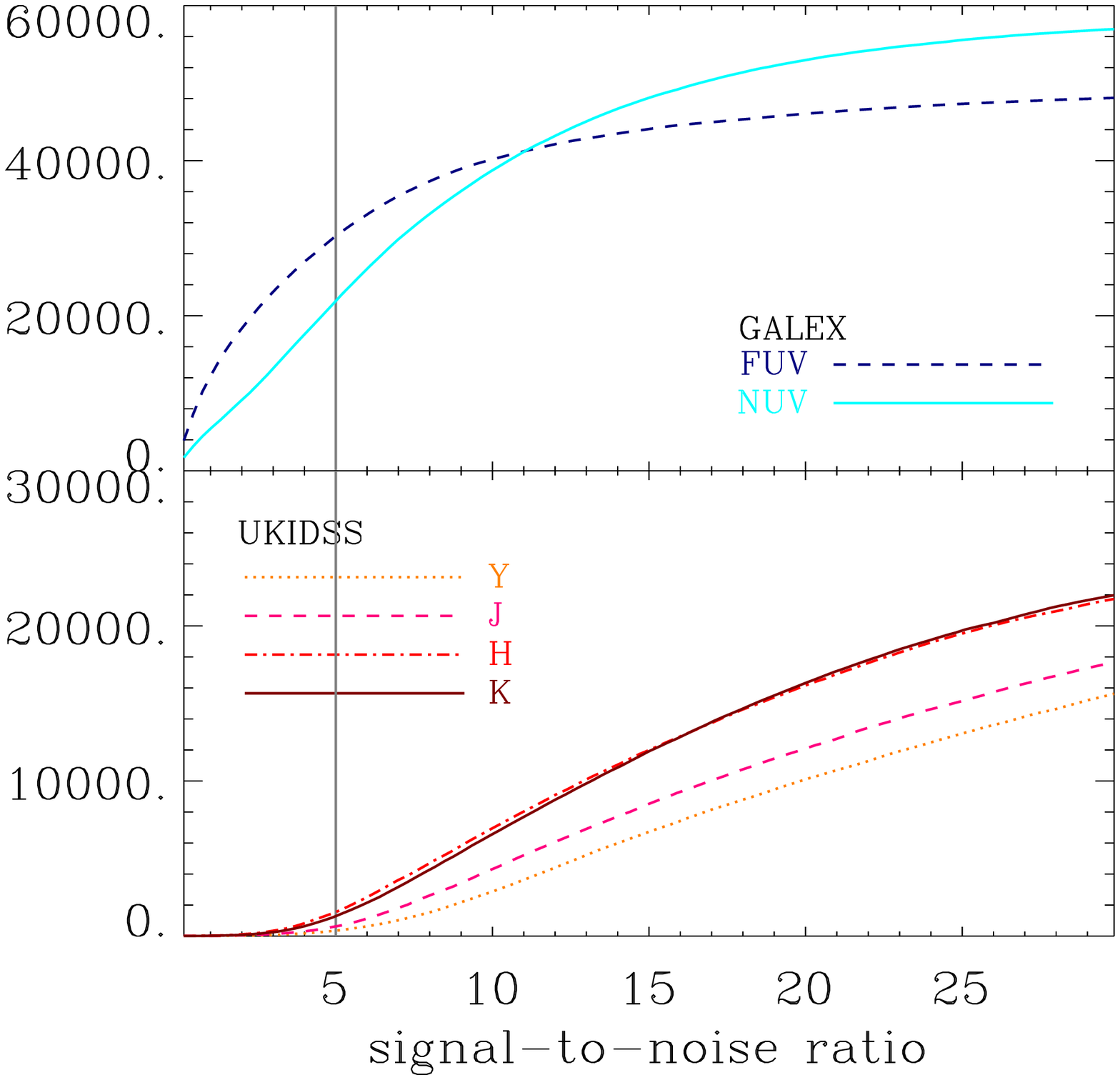}
\caption{Cumulative distribution of signal-to-noise ratio for those
  quasars in the \sdss\ DR7 quasar sample observed by
  \galex\ ($\approx$ 62,628 objects; \emph{top panel}) and
  \ukidsslas\ ($\approx$ 25,510 objects; \emph{bottom panel}). See
  \sectionname s~\ref{sec:galexdata} and \ref{sec:ukidssdata} for the
  number of objects in each individual bandpass. The five-sigma
  detection limit is indicated.}\label{fig:galexukidsssnr}
\end{center}
\end{figure}

Rather than using \galex\ catalog products \citep{Morrissey07a} we use
measurements of the \uv\ fluxes obtained by force-photometering
\galex\ images (from \galex\ Data Release 5) at the \sdss\ centroids
\citep{Aihara11a}, such that we obtain low signal-to-noise PSF fluxes
of objects not detected by \galex. As we show below, these low
signal-to-noise ratio observations are essential for better
classification of redshift $z \geq 2$ quasars. We expect these
measurements to be released as part of \sdss\ Data Release 9,
scheduled for 2012. The top panel of
\figurename~\ref{fig:galexukidsssnr} shows the distribution of
signal-to-noise ratio for \sdss\ quasars in the \galex\ footprint. A
total of 62,661 objects lie in the \galex\ \fuv\ footprint, 63,372 lie
in the \nuv\ footprint, and 62,628 are covered by both bandpasses.

\subsection{\nir\ data from the \emph{UKIRT Infrared Deep Sky Survey}}\label{sec:ukidssdata}

We also use \nir\ data to improve quasar classification and
photometric redshift estimation. The \emph{UKIRT Infrared Deep Sky
  Survey} (\ukidss) is defined in \citet{Lawrence07a} and consists of
five survey components with different wavebands, depths and
footprints.  For the study in this paper we use data from the
\ukidss\ \emph{Large Area Survey} (\emph{LAS}). Technical details
about the \ukidsslas\ observing strategy are described in
\citet{Dye06a}.

The \ukidsslas\ aims to cover 4,000 deg$^2$ of the SDSS footprint in
the Y, J, H and K wavebands. In this paper we use data from
\ukidsslas\ DR7 which includes observations obtained between May 2005
and July 2009 inclusive. The \ukidsslas\ DR7 overlaps the
\sdss\ imaging footprint over $\approx$ 2500 deg$^2$ and has median
point source 5-sigma AB magnitude limits in Y, J, H and K of 20.9,
20.6, 20.2, and 20.2, respectively.

The \ukidss\ data are acquired with the UKIRT Wide Field Camera
(WFCAM; \citealt{Casali07a}). The \ukidss\ photometric system is
described in \citet{Hewett06a}, and the calibration is described in
\citet{Hodgkin09a}. The pipeline processing and science archive are
described in M. J. Irwin et al. (2012, in preparation) and
\citet{Hambly08a}.

As in the case of the \galex\ data described in
\sectionname~\ref{sec:galexdata}, we use force-photometered
\nir\ fluxes at \sdss\ positions rather than \ukidss\ catalog
data. This ``list-driven'' information is derived from aperture
photometry on Data Release 7 of the \ukidsslas. We choose an aperture
radius of 1 arcsec\footnote{For a further description of the
  \ukidss\ data processing by the Cambridge Astronomy Survey Unit see
  \url{http://casu.ast.cam.ac.uk/surveys-projects/wfcam/technical/catalogue-generation}~.}. Of
the 103,601 quasars in the \sdss\ DR7 quasar sample, 29,726 lie within
the \ukidss\ DR7 K-band footprint. Approximately 22,000 of these
quasars are detected in the K band with also overlapping coverage in
all four observed wavebands (Y, J, H and K) in the \ukidsslas\ DR7
source catalog. The bottom panel of
\figurename~\ref{fig:galexukidsssnr} shows the distribution of
signal-to-noise ratio in the \nir\ for quasars in our training
sample. Unlike for \sdss\ imaging, measurements in all four filters of
the \ukidss\ survey are not obtained during the same observing run---H
and K observations are performed in the same observing block, while Y
and J are conducted separately. Thus the sky coverage in different
\ukidss\ bands varies, and many objects are missing data in one or
more of the four bands. The breakdown of training quasars with
observations in the \nir\ by bandpass is: Y: 26,876; J: 27,328; H:
28,911; K:29,726. A total of 25,510 objects have measurements in all
four bandpasses.

The differing epochs of the \ukidss, \galex, and \sdss\ can range up
to 9 years in the observed frame. For a quasar with a redshift of 2
this is 3 years in the rest frame. The observed optical variability in
radio quiet quasars over the rest frame 2 to 5 year timescale is
observed to be in the range 0.1 - 0.2 mag and is a function of
absolute magnitude \citep{Hook94a}. \citet{VandenBerk04a} find that
the variability amplitude decreases with rest-frame wavelength by a
factor of two between 1500\AA\ and 6000\AA\ with an amplitude of
$\approx$ 0.15 mag at 6000\AA\ (see also \citealt{Welsh11a}).

\citet{Kozlowski10b} have studied the mid-\emph{IR} variability using
multi-epoch Spitzer observations of a sample of $\approx$ 1000 active
galactic nuclei and find that the rest-frame J band variability
amplitude in the rest-frame timescale is $\approx$ 0.1 mag. In
summary the quasars in this study are expected to vary by $\approx$
0.3 and 0.1 magnitudes in the \uv\ and \nir, respectively, over the
elapsed period of the observations and this is less than the average
photometric errors in the individual wavebands and significantly less
than the range in colors.

\section{Flux--redshift density model}\label{sec:xdqsoz}

The photometric redshift technique \xdqsoz\ is an adaptation of the
\xdqso\ technique \citep{Bovy11a} to include redshift explicitly in
the model for the quasar population. \xdqsoz\ achieves this by
modeling the $p(\mathrm{flux},z|\mathrm{quasar})$ factor in
\eqnname~(\ref{eq:jointprob2}), where the \xdqso\ technique modeled
$p(\mathrm{flux}|\mathrm{quasar\ in}\ \Delta z)$ in three bins in
redshift (corresponding to low-, medium-, and high-redshift
quasars). As discussed in \sectionname~\ref{sec:general}, this
approach allows us to obtain full posterior distribution functions for
the redshift of a photometrically classified quasar based on its
broadband fluxes. Integrating this redshift probability distribution
over a range of redshifts and properly normalizing this result using
\eqnname~(\ref{eq:normalization}) gives a photometric quasar
probability in the chosen redshift range that is, as we show below,
competitive with the best available photometric quasar classification
techniques, \eg, \xdqso.

To estimate the density of quasars in flux--redshift space we use
\emph{extreme deconvolution}\footnote{Code available at
  \url{http://code.google.com/p/extreme-deconvolution/}~.}
\citep{Bovy09a}. As described in \sectionname~\ref{sec:data}, our
training set consists of the \sdss\ DR7 quasar sample, which consists
mostly of bright, \viz, dereddened $i < 19.1$ mag ($i < 20.2$ mag for
$z > 3$ sources), objects with small photometric uncertainties. The
\galex\ and \ukidss\ data described in \sectionname
s~\ref{sec:galexdata} and \ref{sec:ukidssdata} are much shallower than
the \sdss\ data and many objects are not detected at high significance
in these surveys, such that photometric uncertainties are not
insignificant (see \figurename~\ref{fig:galexukidsssnr}). Additional
complications are that these two supplemental surveys have not
observed the full \sdss\ footprint and that the \ukidsslas\ footprint
is different for the different \nir\ bands, such that we have
heterogeneous missing data and heteroscedastic uncertainties. \exd\ is
uniquely suited to deal with these complications in the proper
probabilistic manner. \exd\ assumes that the flux uncertainties are
known and that they are close to Gaussian, as is the case for PSF
fluxes for point-sources in
\sdss\ \citep{Ivezic03a,Scranton,Ivezic07a}. We assume that the
spectroscopic redshifts have vanishing uncertainties because their
typical value of $\sigma_z \approx 0.004$ \citep{Schneider10a} is
orders of magnitude smaller than typical uncertainties in broadband
photometric redshifts, which are set by the width of the quasar locus.

\exd\ models the underlying, deconvolved distribution as a sum of $K$
$d$-dimensional Gaussian distributions, where $K$ is a free parameter
that is set using an external objective (see
\sectionname~\ref{sec:modelconstruction}). \exd\ consists of a fast
and robust algorithm to estimate the best-fit parameters of the
Gaussian mixture.

\subsection{Construction of the quasar flux--redshift model}\label{sec:modelconstruction}

The full quasar-density model is constructed by fitting the
flux--redshift density of quasars in a number of bins in
the \iband\ magnitude. As we use the same set of quasars in each bin with
a different redshift prior---see the discussion in
\sectionname~\ref{sec:data}---we could instead have fit a single bin,
\eg, the brightest. The other bins could have been constructed by
dividing out the redshift prior of the first bin and multiplying in
the redshift priors for the fainter bins. However, as we will show
below, the advantage of a Gaussian representation of the
flux--redshift density is that it allows integrals of this density
over arbitrary redshift ranges to be calculated analytically. This
leads to fast quasar-probability estimation. If we were instead to
divide out the redshift prior and multiply in a different redshift
prior, the resulting function would no longer be Gaussian and the
numerical integration over redshift would be much more computationally
expensive\footnote{Alternatively, we could have modeled the density
  using a uniform prior over redshift and modeled the
  magnitude-dependent redshift prior as a polynomial or another
  mixture of Gaussians. Integrating a polynomial or mixture of
  Gaussians times a mixture of Gaussians could also be performed
  analytically and fast.}. Because our short-term objective is to run
this algorithm on essentially all of the $\approx 10^8$ \sdss\ point
sources and in the future on the $\approx 15$ PB of \lsst\ catalog
data \citep{Abell09a}, this computational advantage is
important. After fitting the first bin, all other fits are initialized
using the previous bin's optimal solution; these extra fits all
converge very quickly as the quasar flux--redshift density does not
vary strongly with apparent magnitude. The redshift prior is shown for
a few bins in apparent magnitude in
\figurename~\ref{fig:redshiftprior}. If a different redshift prior is
desired, one can divide out this prior and multiply in a new prior
(these priors are included in the code release described in
the \appendixname). For example, if one would prefer to use
the \citet{Richards06a} model for the quasar luminosity function, one
would multiply the posterior distribution function for the redshift
obtained using the fiducial \citet{Hopkins07a} prior with the factor
shown in the bottom panel of \figurename~\ref{fig:redshiftprior}. This
panel shows the ratio of the \citet{Richards06a} redshift prior to the
\citet{Hopkins07a} prior in a number of apparent-magnitude bin. It is
clear that there is only a significant difference at relatively large
redshift and at faint magnitudes, where constraints on the luminosity
function are sparse.

As for the \xdqso\ technique, we divide the quasar-density model into
a factor describing, essentially, the color--redshift density of
quasars---but we again use relative fluxes rather than colors---and
another factor describing the apparent-magnitude distribution of
quasars. We adopted this approach because the flux density of quasars
has a dominant power-law shape corresponding to the number counts as a
function of apparent magnitude, while the color distribution is much
flatter.  We write
\begin{equation}
p(\mathrm{fluxes},z|\mathrm{quasar}) = p(\mathrm{fluxes\ relative\ to}\
  i,z|\mathrm{quasar})\,p(i\mbox{-}\mathrm{band\ flux}|\mathrm{quasar})\,.
\end{equation}
The apparent-magnitude factor does not depend on redshift. So, this
factor is the same as used in the \xdqso\ method. The factor is
calculated by the sum of the apparent-magnitude priors in
\figurename~1 of \citet{Bovy11a} weighted by the quasar densities in
\tablename~1 of the same \documentname. As quasar redshifts are always
positive, we model the logarithm of the redshift. Since our training
sample consists of point-like objects at redshift $z \geq 0.3$, about
all at $z < 5.5$, our model should only be trusted to return
reasonable densities within this range.

\begin{figure}
\includegraphics{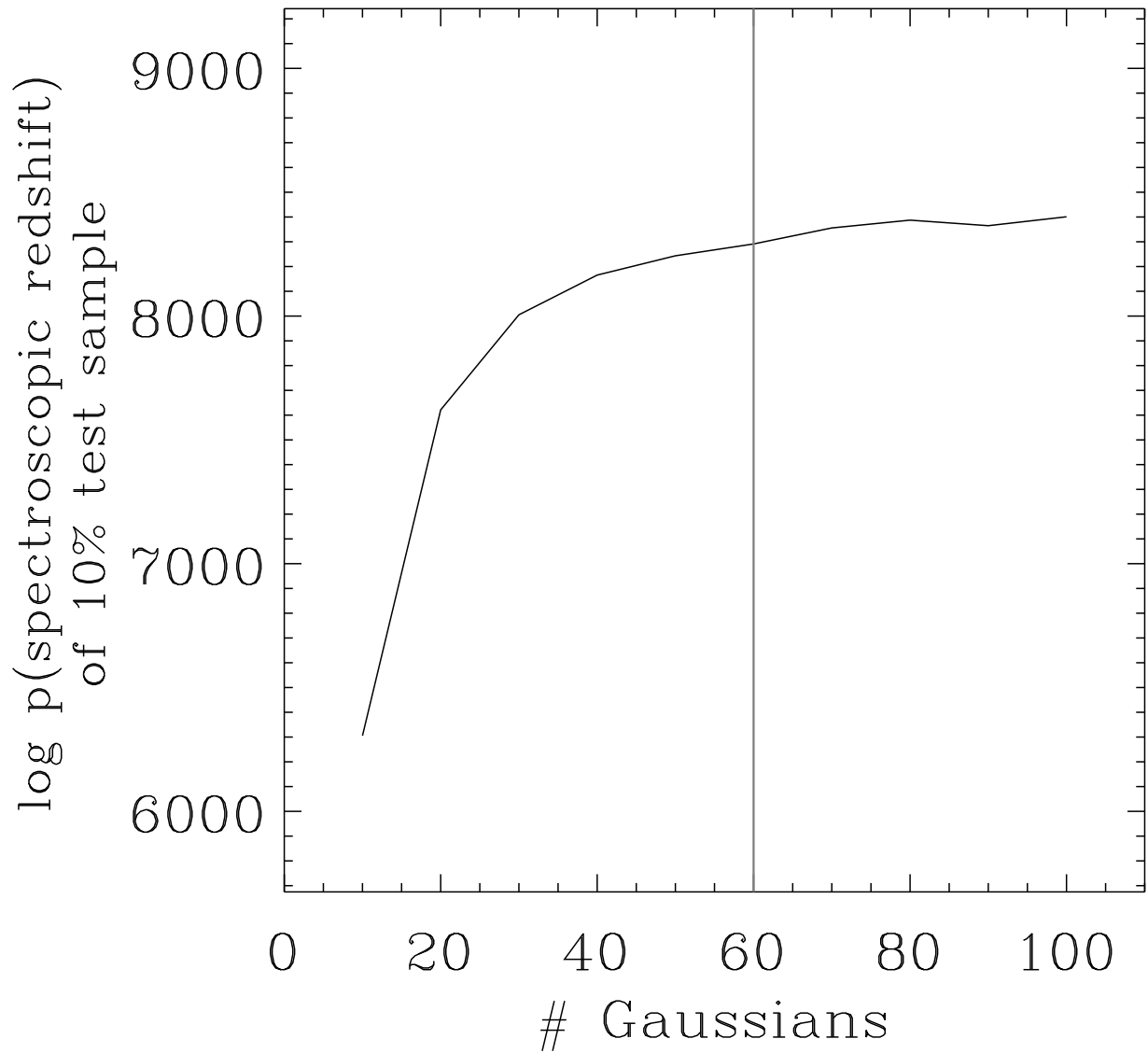}
\caption{Total probability of the spectroscopic redshifts of objects
  in the 10\,percent test sample given their \ugriz\ fluxes using
  models with different numbers of Gaussians trained on the remaining
  90\,percent of objects in the \sdss\ DR7 quasar
  catalog.}\label{fig:testloglike}
\end{figure}

In each bin we model the $d$-dimensional relative-flux--redshift
density, where $d$ is the number of independent colors plus one (for
redshift), of quasars using 60 Gaussians that are allowed to have
arbitrary means, variance matrices, and amplitudes---the amplitudes
are constrained to sum to one. We use the full set of 103,601, $z \ge
0.3$ quasars to train the final model, but in order to test whether we
are under- or overfitting the data we performed a cross-validation
test. To cross-validate we extract a random subset containing
10\,percent of the full sample to use as an independent test data
set. By training the model on the remaining 90\,percent of the sample
we can select the number of Gaussians that optimally predicts---\ie,
predicts with the highest probability---the redshifts of objects in
the test sample. The results from this procedure are shown in
\figurename~\ref{fig:testloglike}. Our ability to better predict the
test redshifts saturates around $K \approx 50$; we chose 60 Gaussians
to represent the relative-flux density of quasars. Compared to the
\xdqso\ method, which used 20 Gaussians each in three redshift bins,
this revised approach uses the same number of Gaussians while
representing an extra dimension (redshift). One might be concerned
that because the Gaussians will preferentially be found in
high-density, \viz, low-redshift, regions, the density of medium- and
high-redshift quasars is not adequately described in the
\xdqsoz\ model. We will see below that this is not the case and that
\xdqsoz\ performs similarly to \xdqso\ in selecting medium- and
high-redshift quasars.

The full model consists of \nbins\ bins of width \binwidth\ mag
between $i = \ilow$ and $i = \ihigh$, spaced \binspacing\ mag apart
(adjacent bins overlap). As described above, the \exd\ fits for all
but the brightest bin are initialized using the best-fit parameters
for the previous bin. Each bin uses the full set of 103,601 redshift
$z \ge 0.3$ quasars.

In each of 47 bins we fit 60 $n$-dimensional Gaussians, yielding a
total of $47 \times (60 \times [1+d+d(d+1)/2] - 1)$ parameters. The
\ugriz-only model has 59,173 parameters, the model that also uses the
two \uv\ bands has 101,473 parameters, the model that adds the four
\nir\ bands to the optical fluxes has 155,053 parameters, and the full
\uv-\ugriz-\nir\ 11-dimensional model has 219,913 parameters. To
obtain the total number of parameters for photometrically classifying
quasars using \xdqsoz, we need to add the number of parameters
describing the stellar relative-flux density in 47 bins---from the
\xdqso\ method---to this number, amounting to 14,053, 26,273, 42,253,
and 61,993 parameters for the \ugriz, \ugriz+\uv, \ugriz+\nir, and
\ugriz+\uv+\nir\ models, respectively. Models including \uv\ or
\nir\ data are trained using any available data, \ie, any object with
a measured flux in any of the bandpasses is included in the training
set.

\subsection{Comparison of the model and observations}

In this section, we assess the performance of the \exd\ technique
for modeling the relative-flux--redshift distribution of quasars, and
provide examples which demonstrate that the \exd\ technique produces
excellent fits to the data. We demonstrate that the \exd\ method does
an excellent job of empirically calibrating the color--redshift
relation; the ability of the \exd\ technique to model the
relative-flux density of quasars in the current \xdqsoz\ context is
excellent as well, but it is very similar to the performance in the
\xdqso\ context and we refer the reader to \citet{Bovy11a} for a
discussion of this performance.

\begin{figure}
\includegraphics[width=0.24\textwidth,clip=]{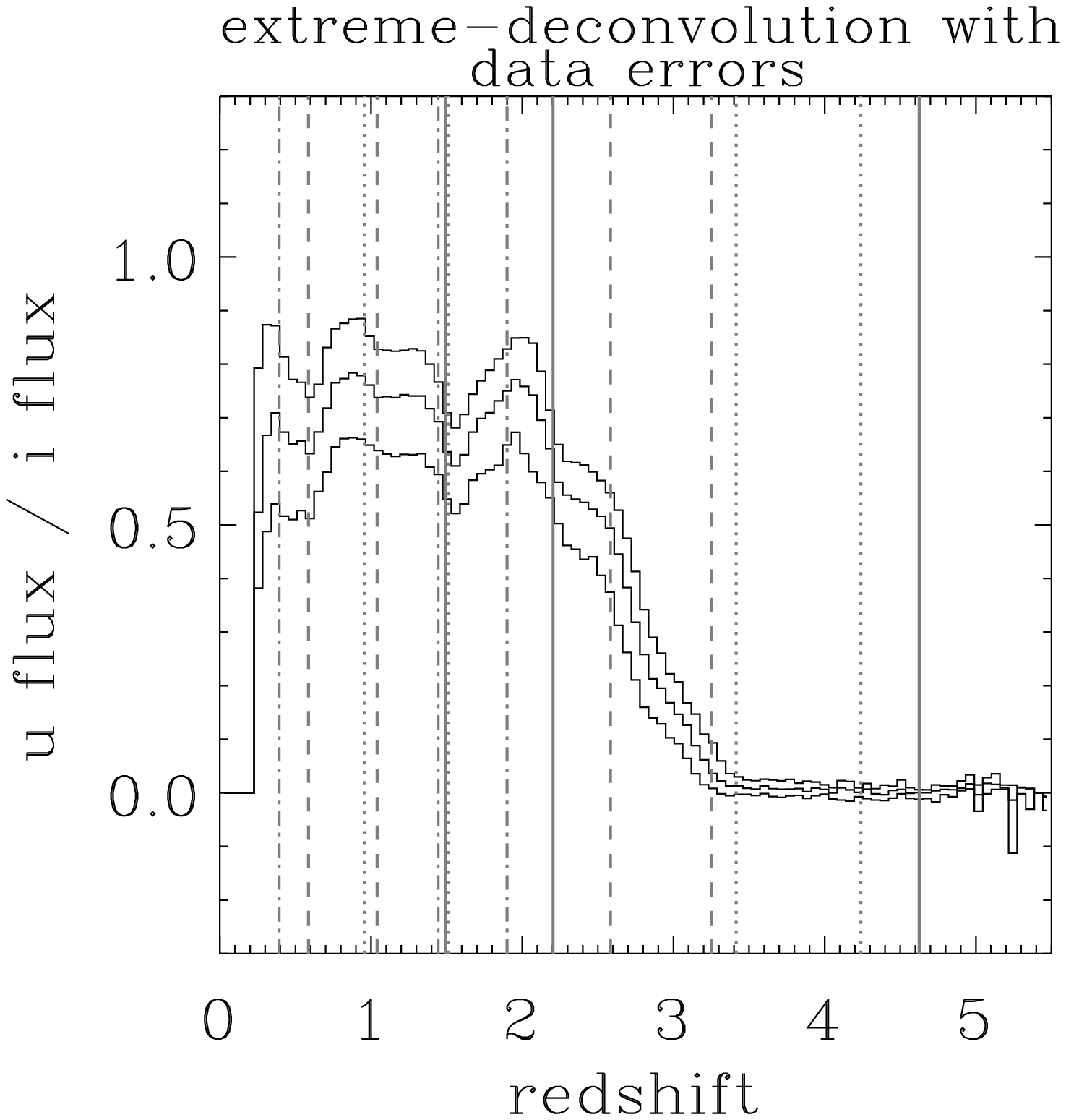}
\includegraphics[width=0.24\textwidth,clip=]{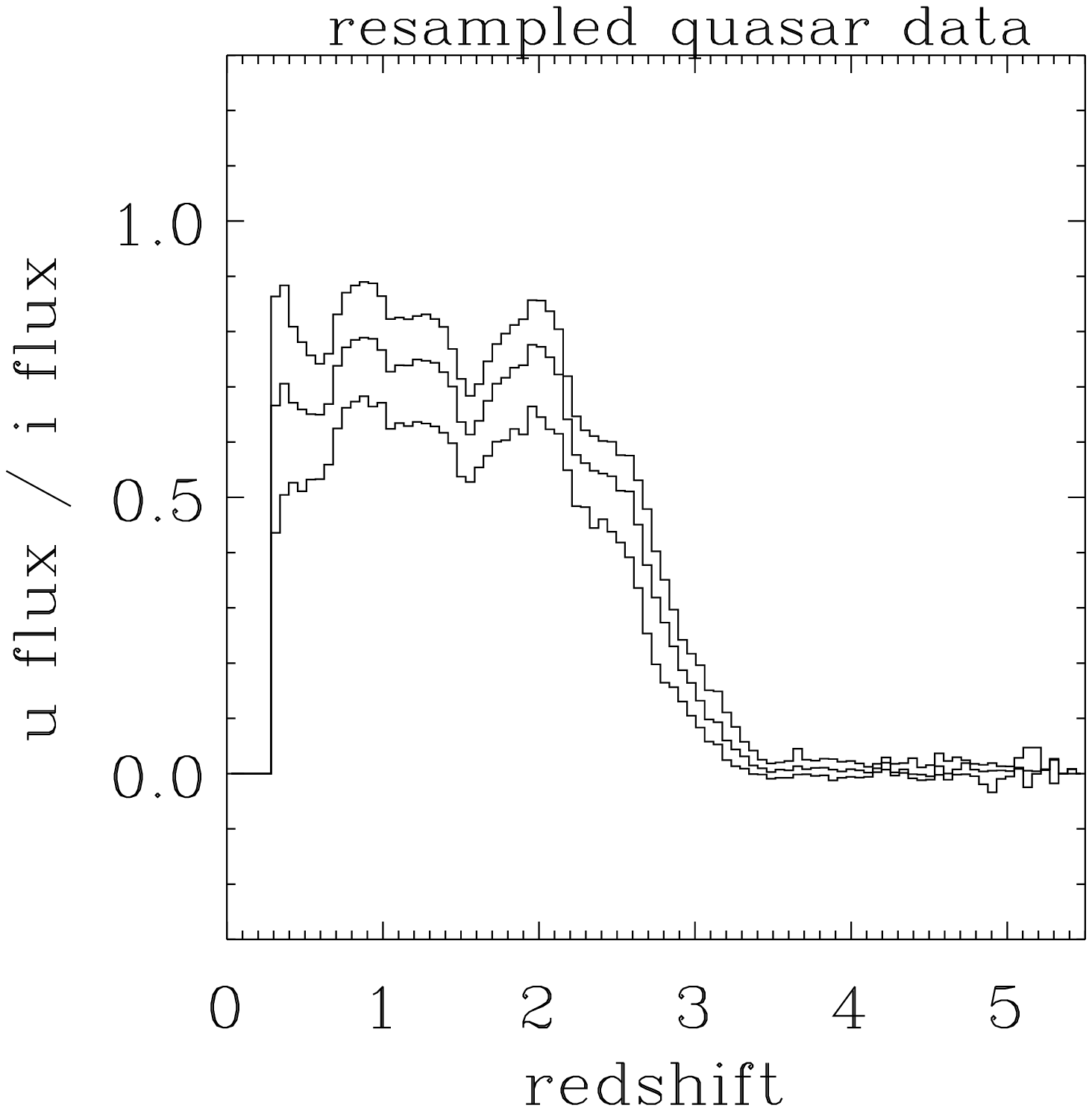}
\includegraphics[width=0.24\textwidth,clip=]{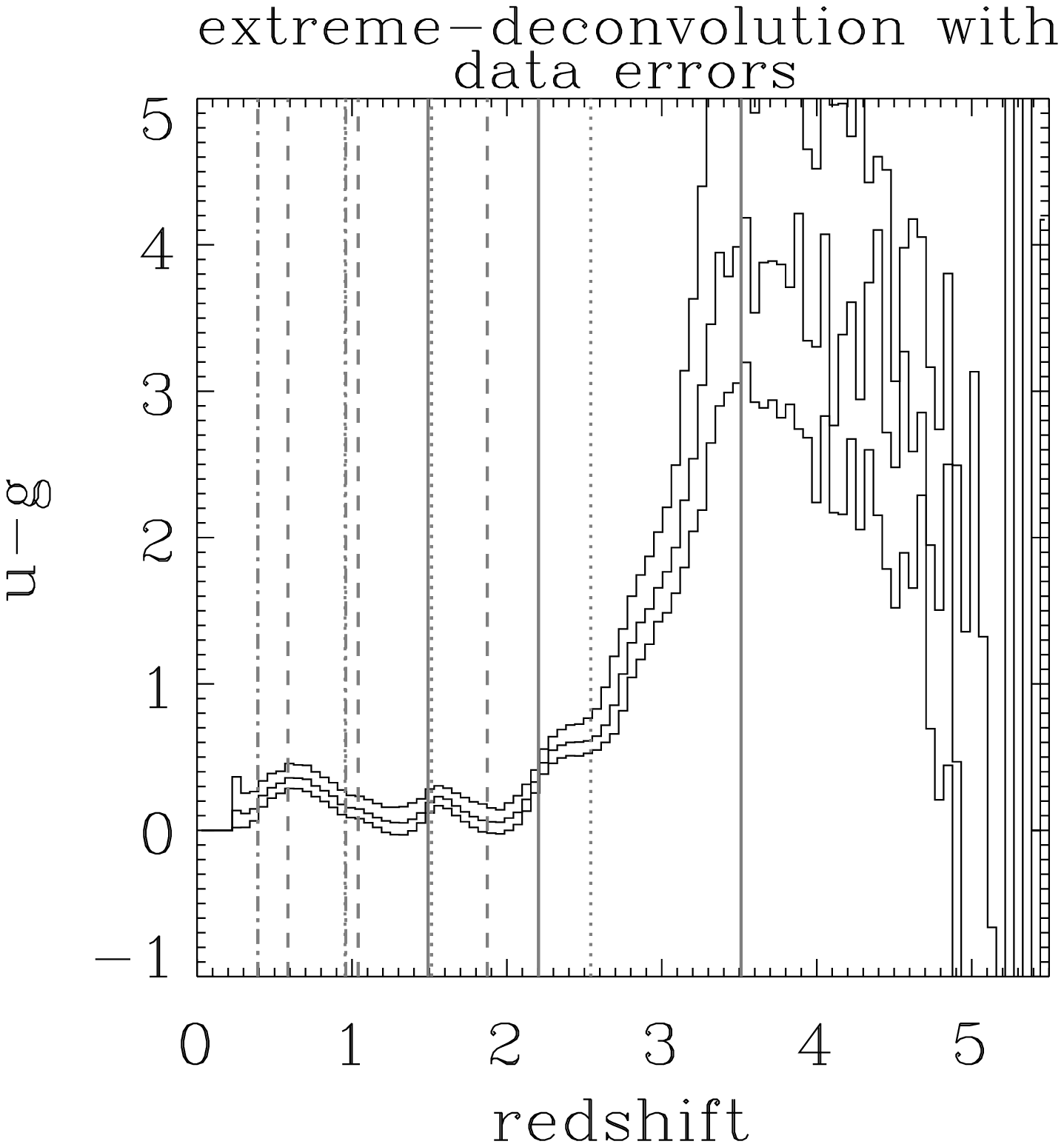}
\includegraphics[width=0.24\textwidth,clip=]{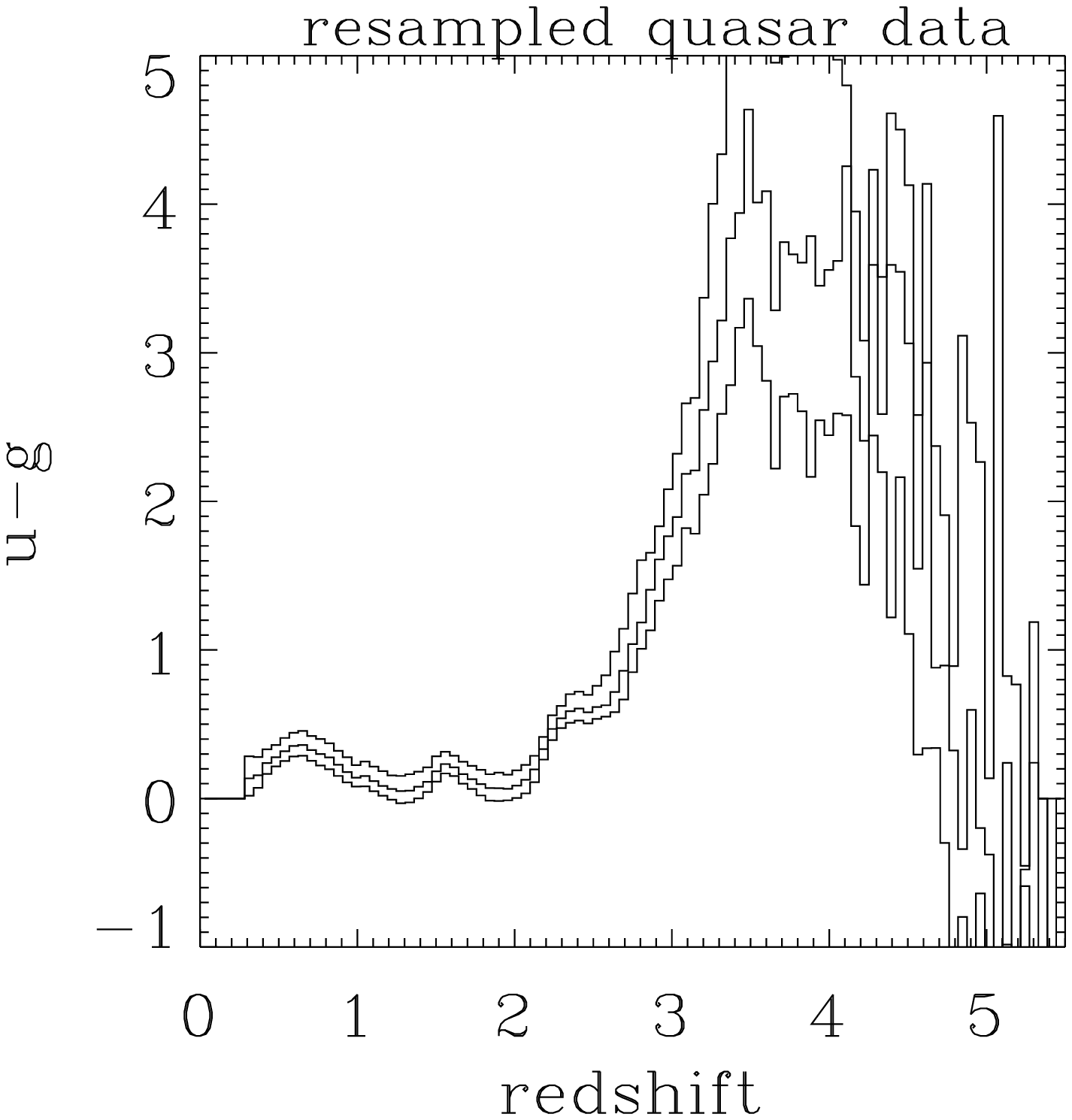}\\
\includegraphics[width=0.24\textwidth,clip=]{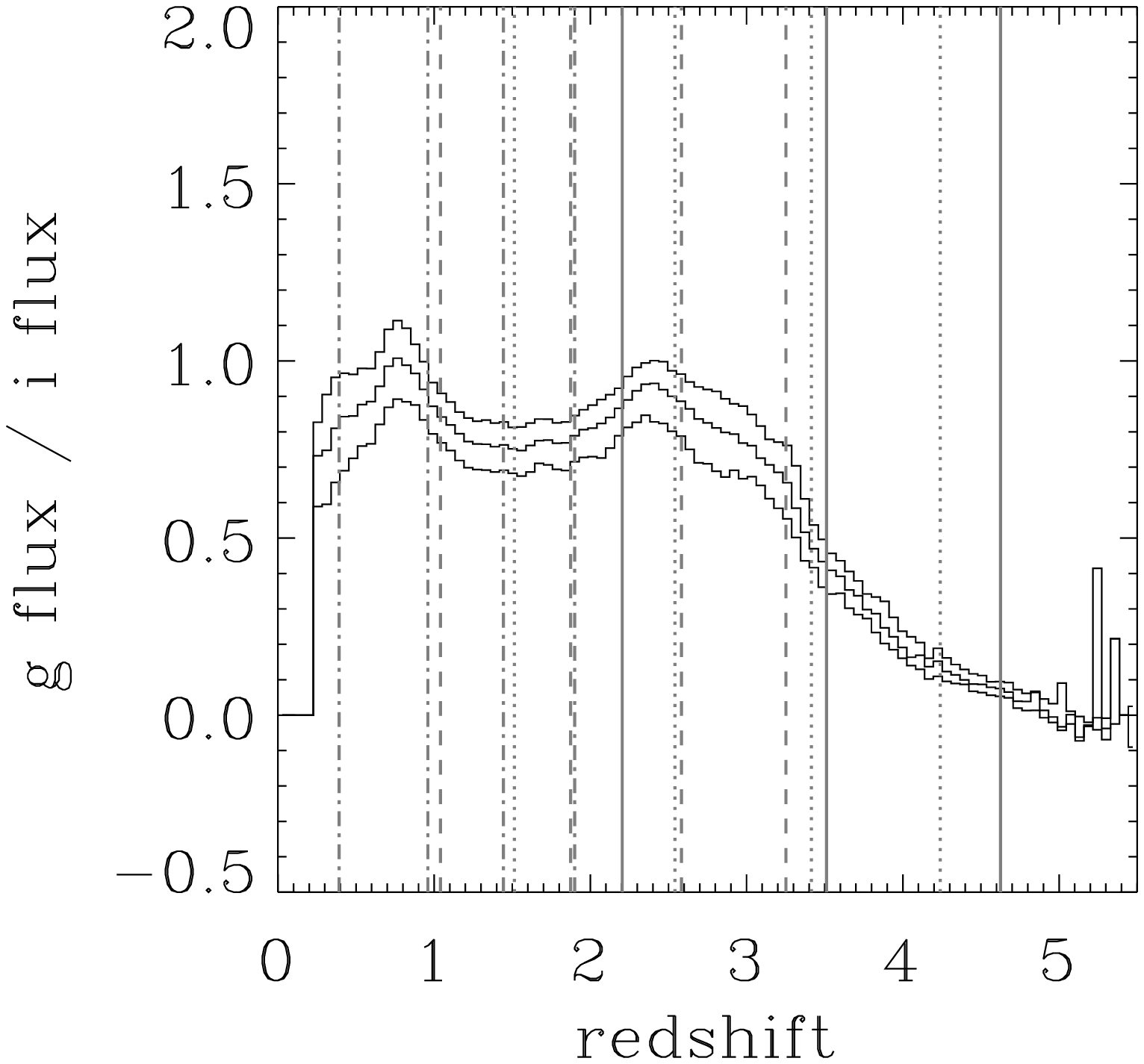}
\includegraphics[width=0.24\textwidth,clip=]{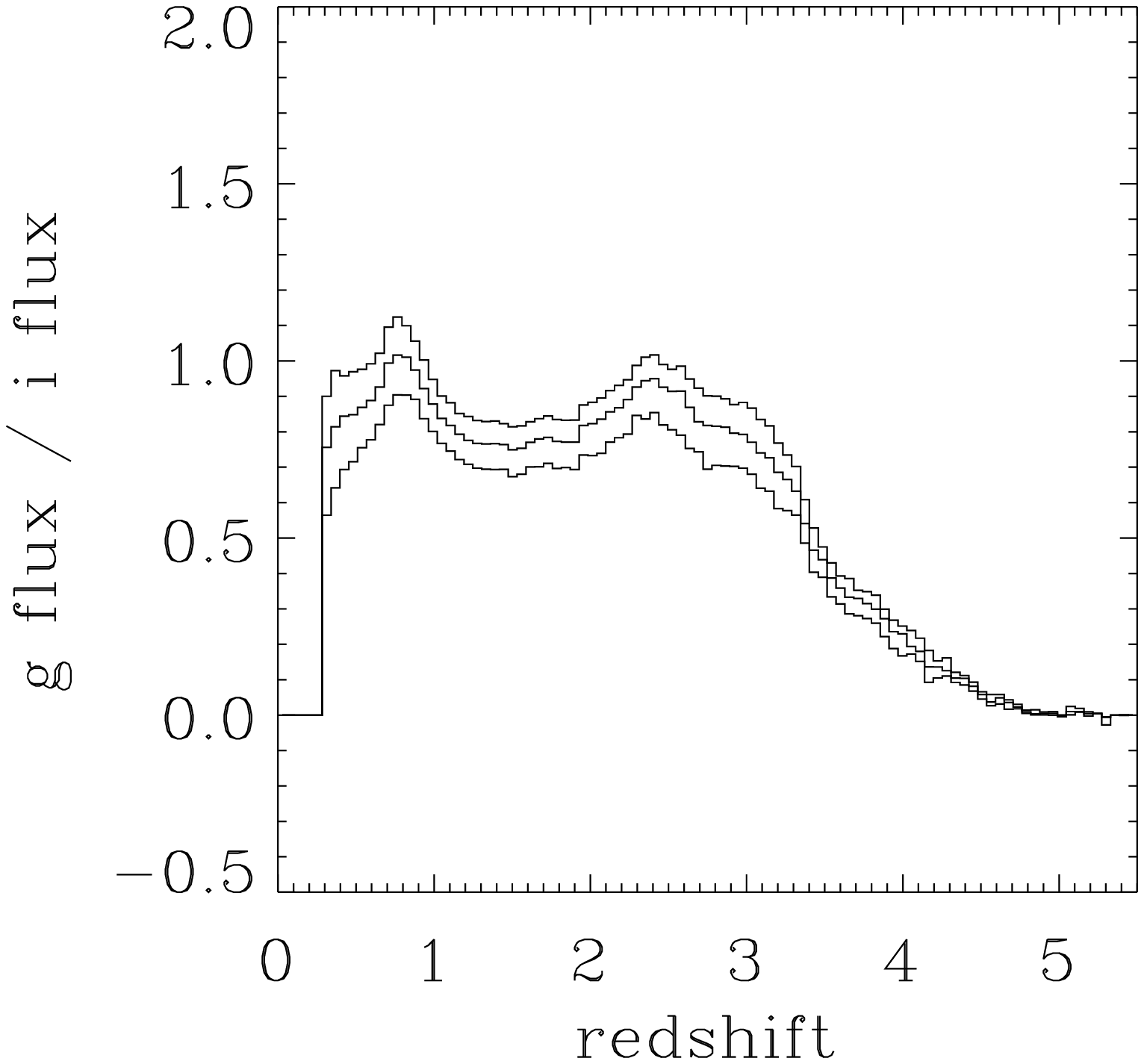}
\includegraphics[width=0.24\textwidth,clip=]{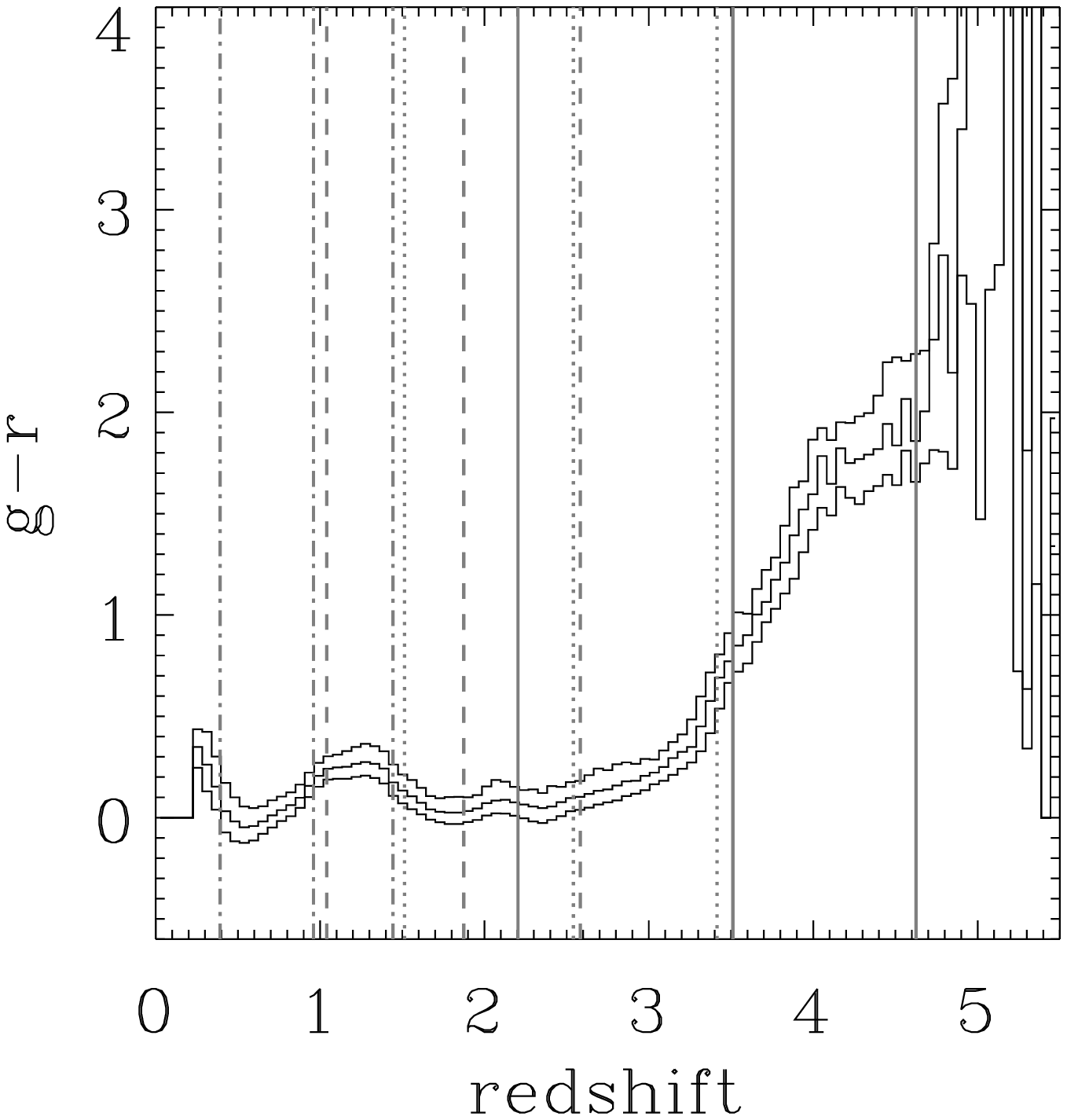}
\includegraphics[width=0.24\textwidth,clip=]{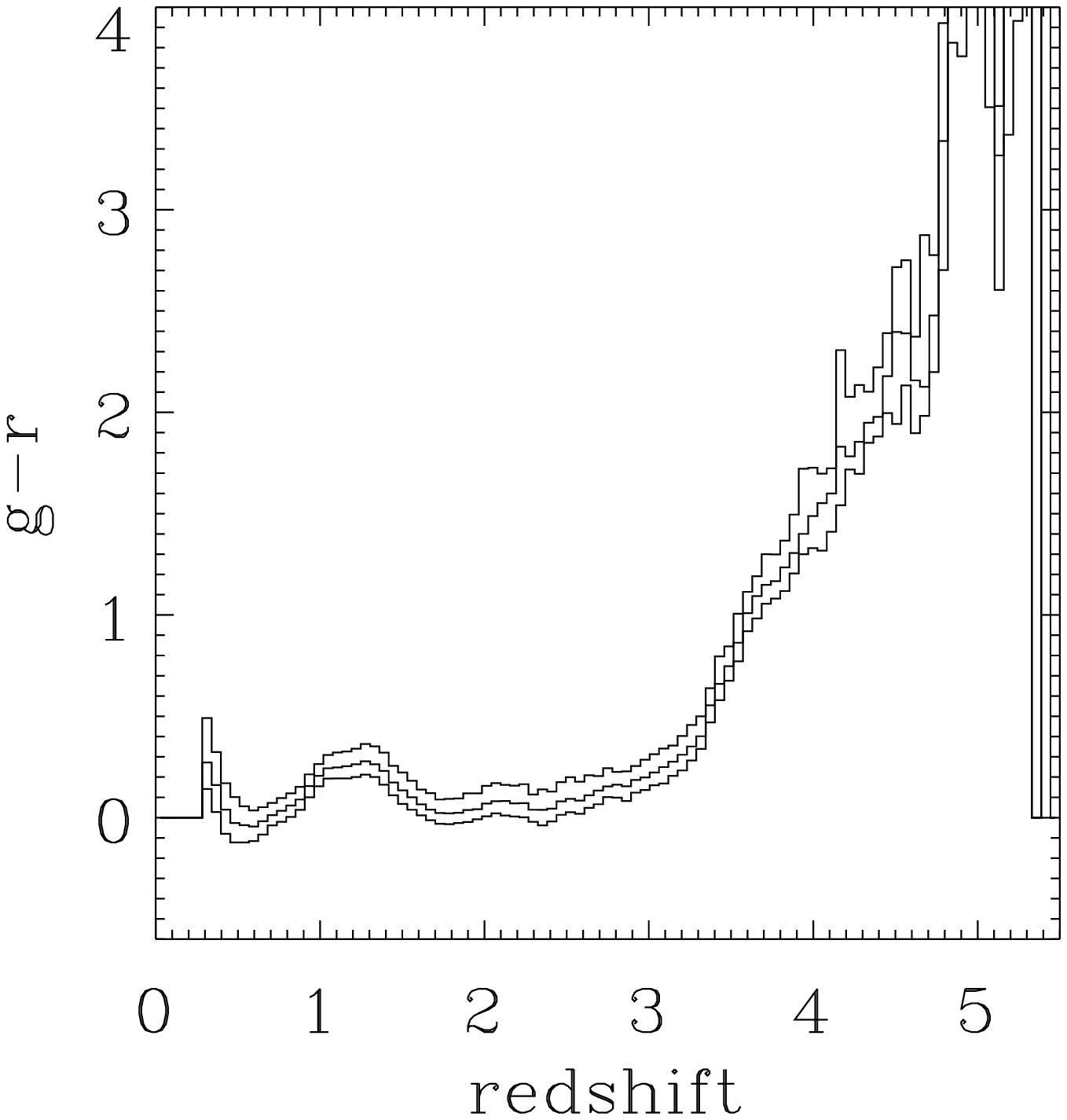}\\
\includegraphics[width=0.24\textwidth,clip=]{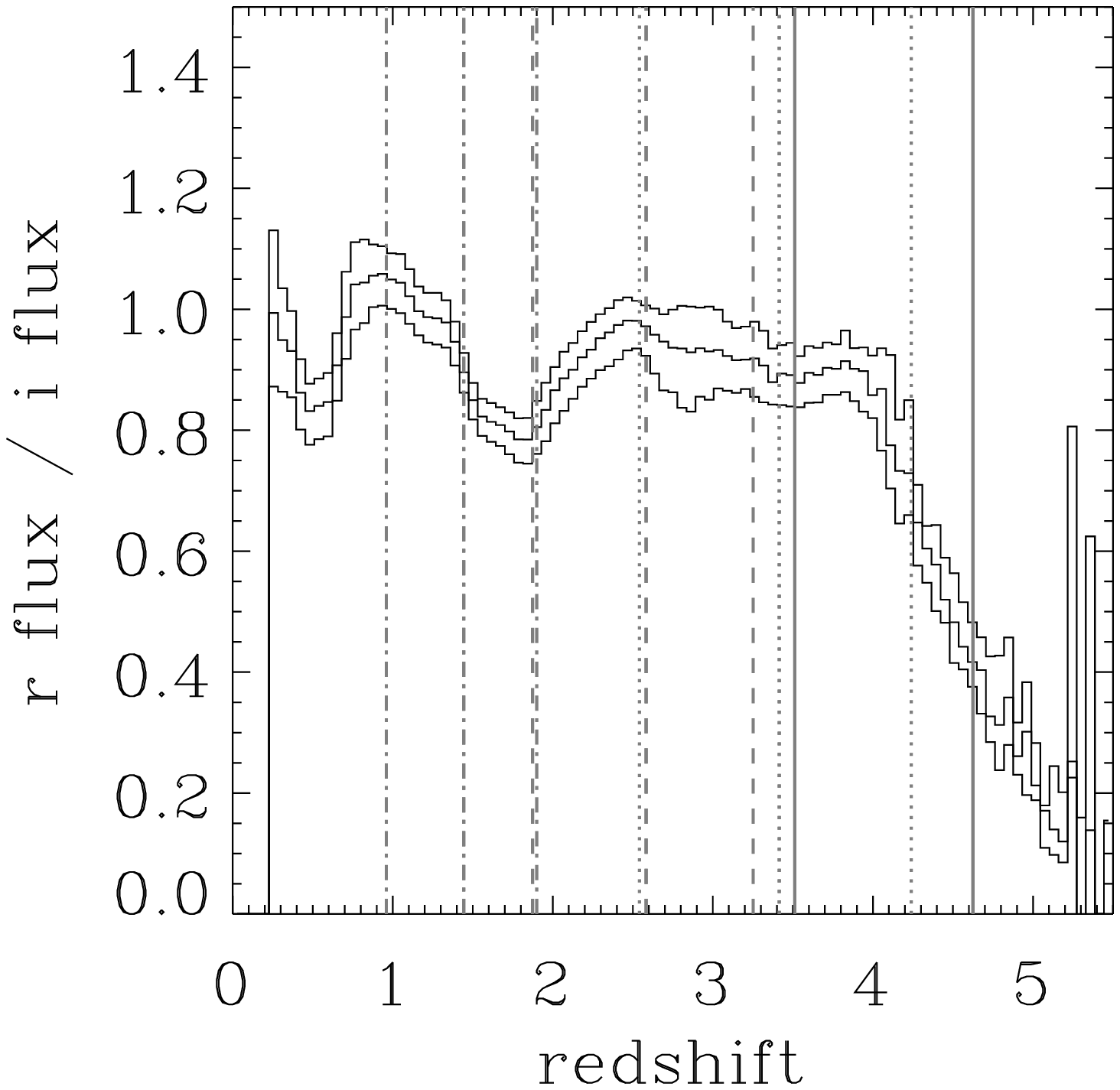}
\includegraphics[width=0.24\textwidth,clip=]{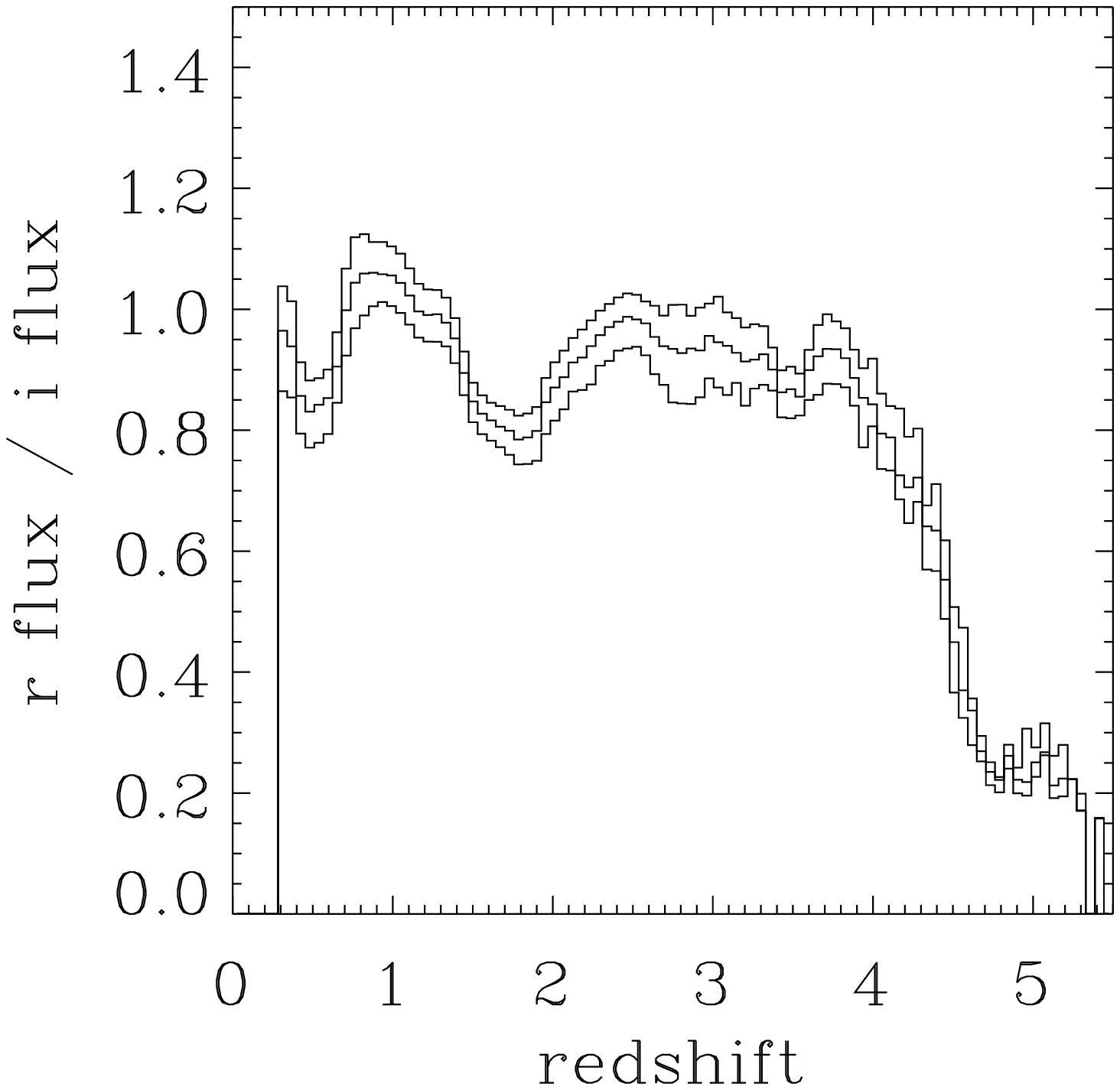}
\includegraphics[width=0.24\textwidth,clip=]{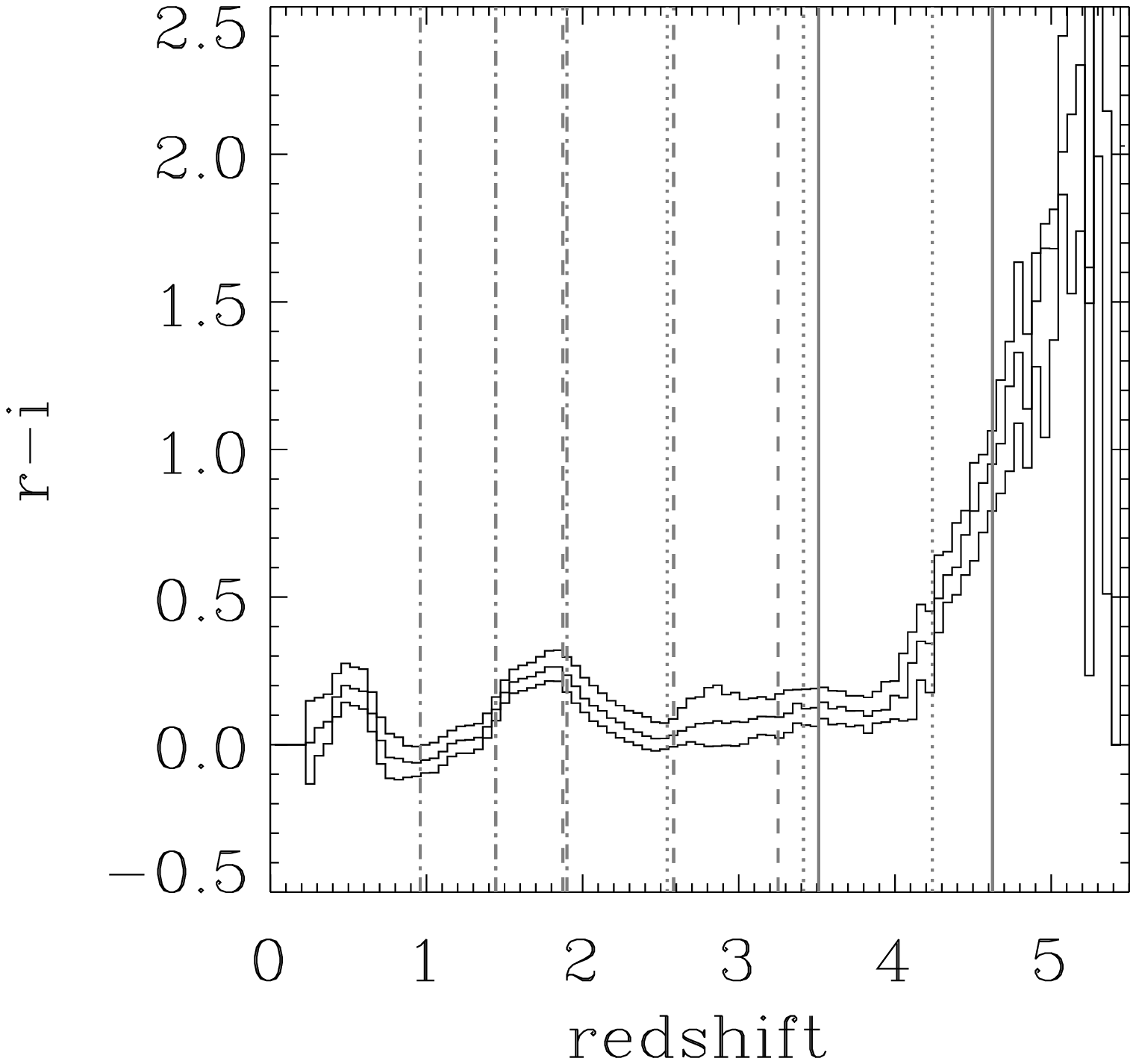}
\includegraphics[width=0.24\textwidth,clip=]{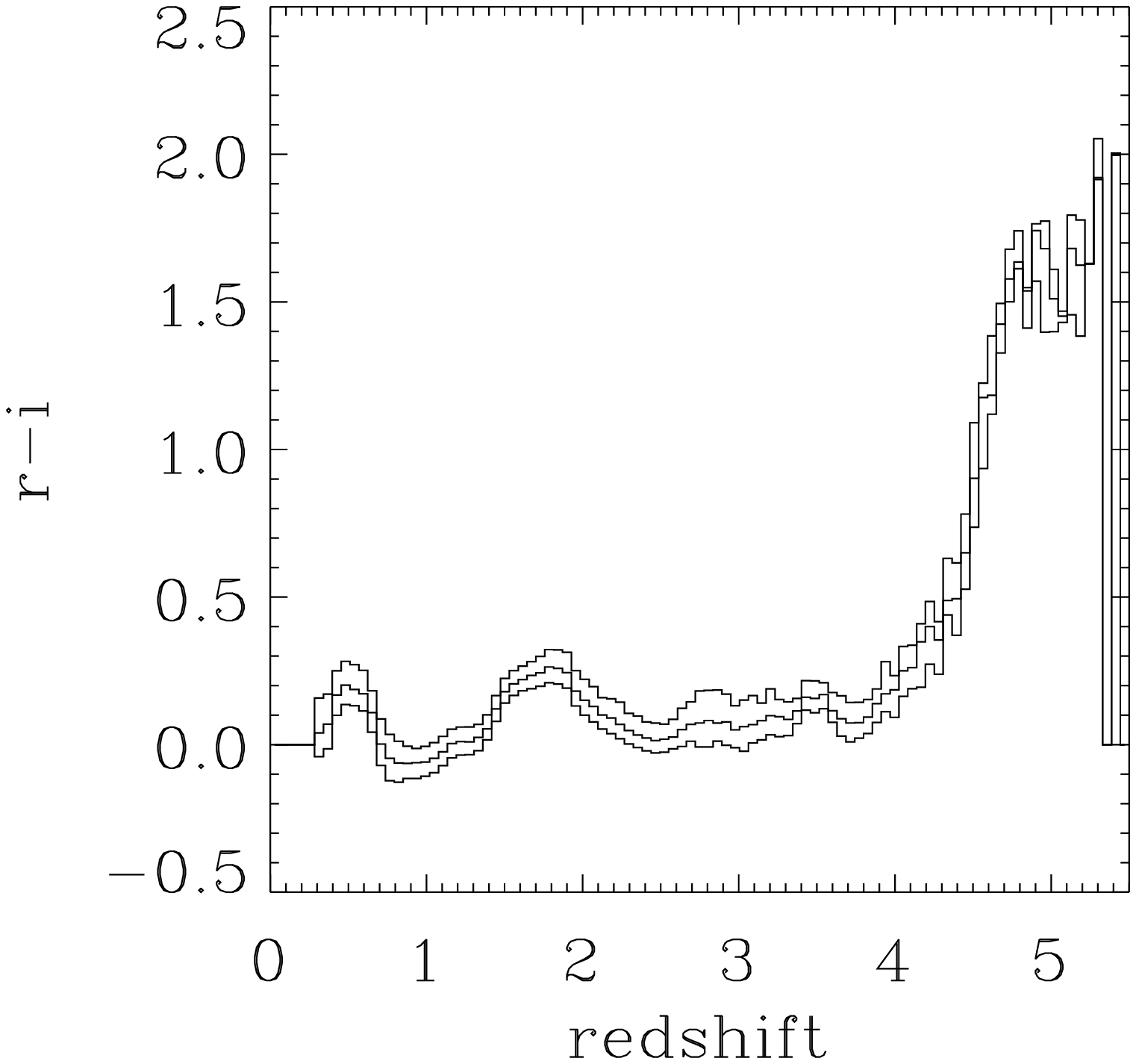}\\
\includegraphics[width=0.24\textwidth,clip=]{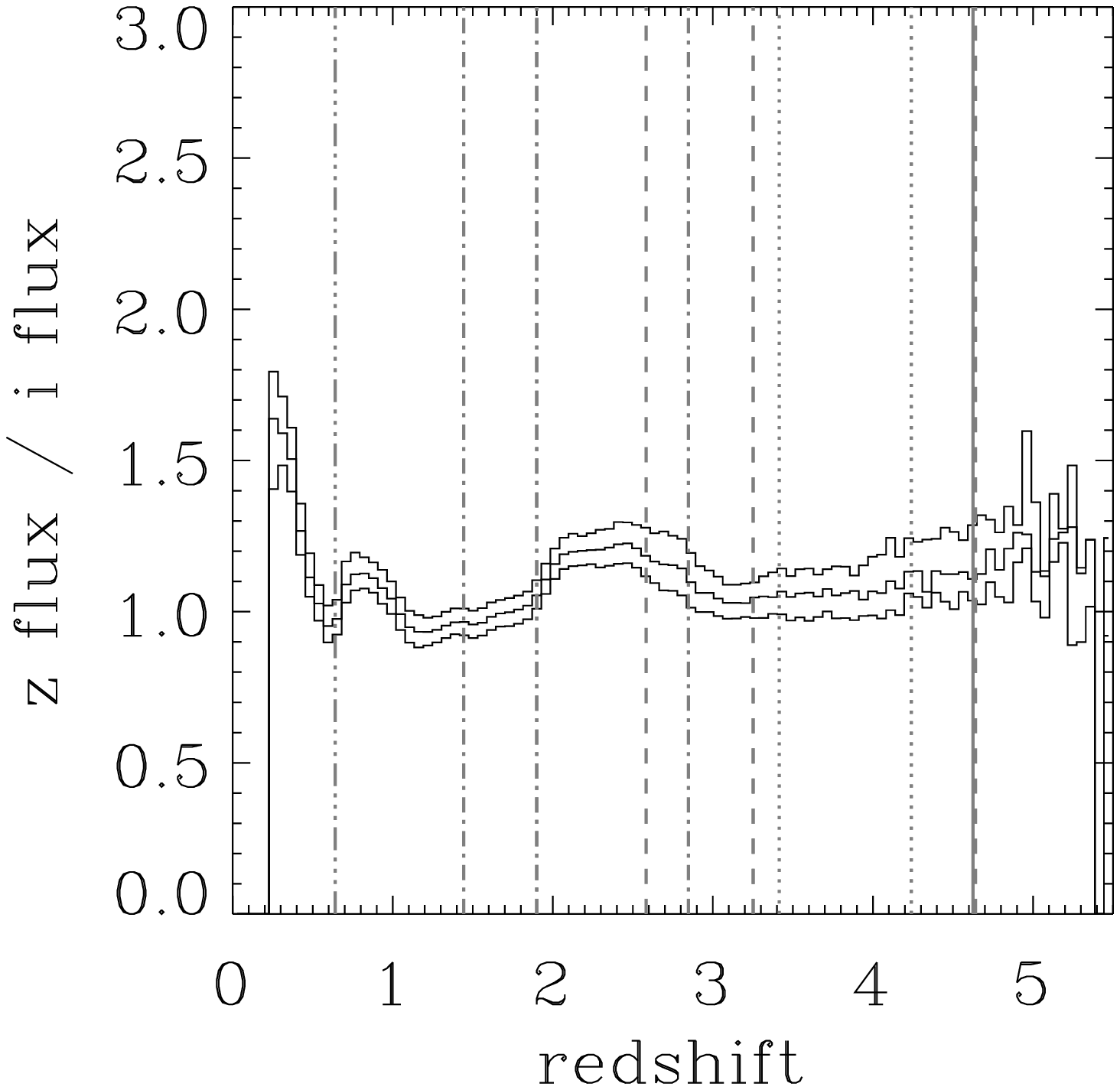}
\includegraphics[width=0.24\textwidth,clip=]{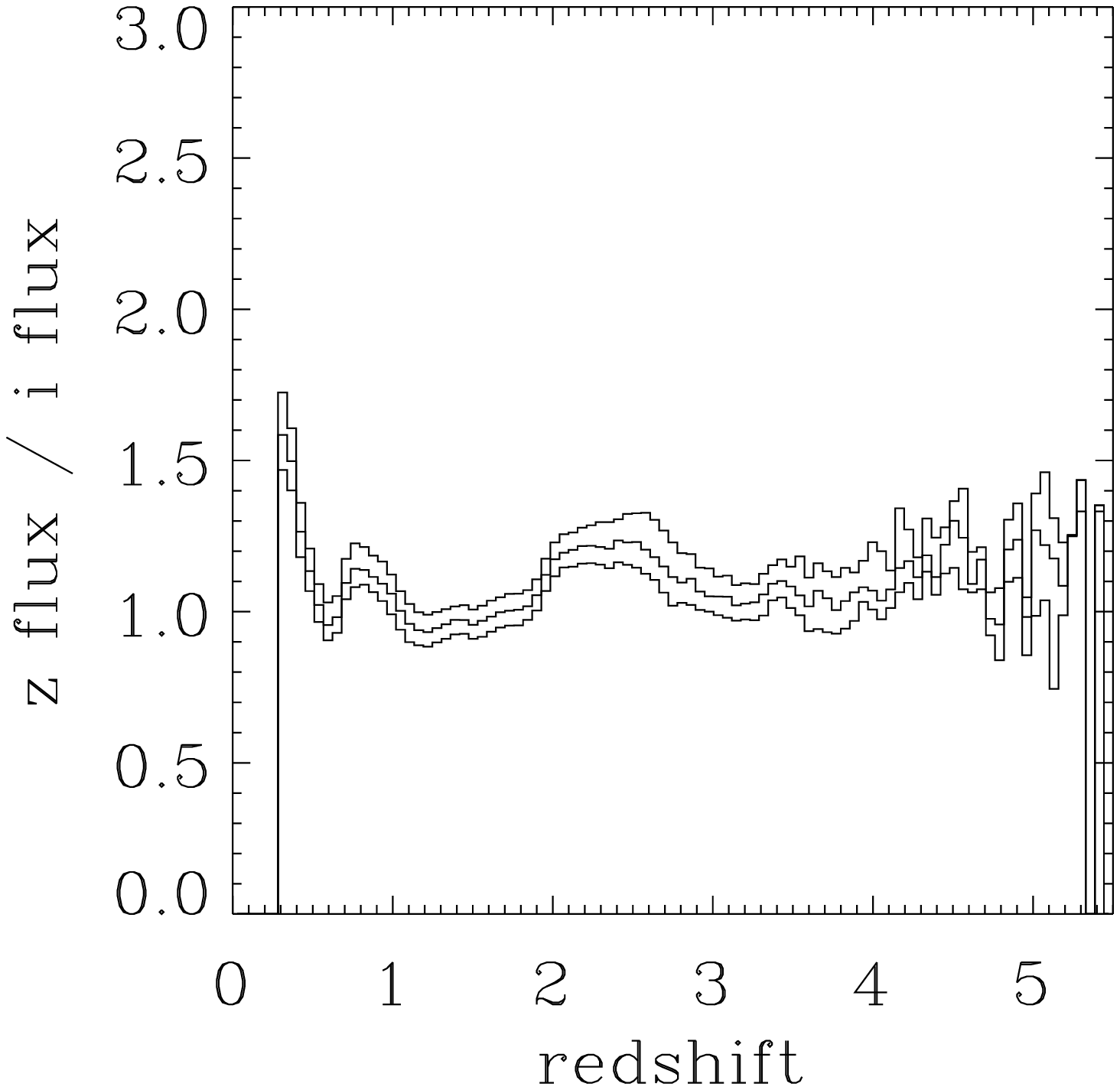}
\includegraphics[width=0.24\textwidth,clip=]{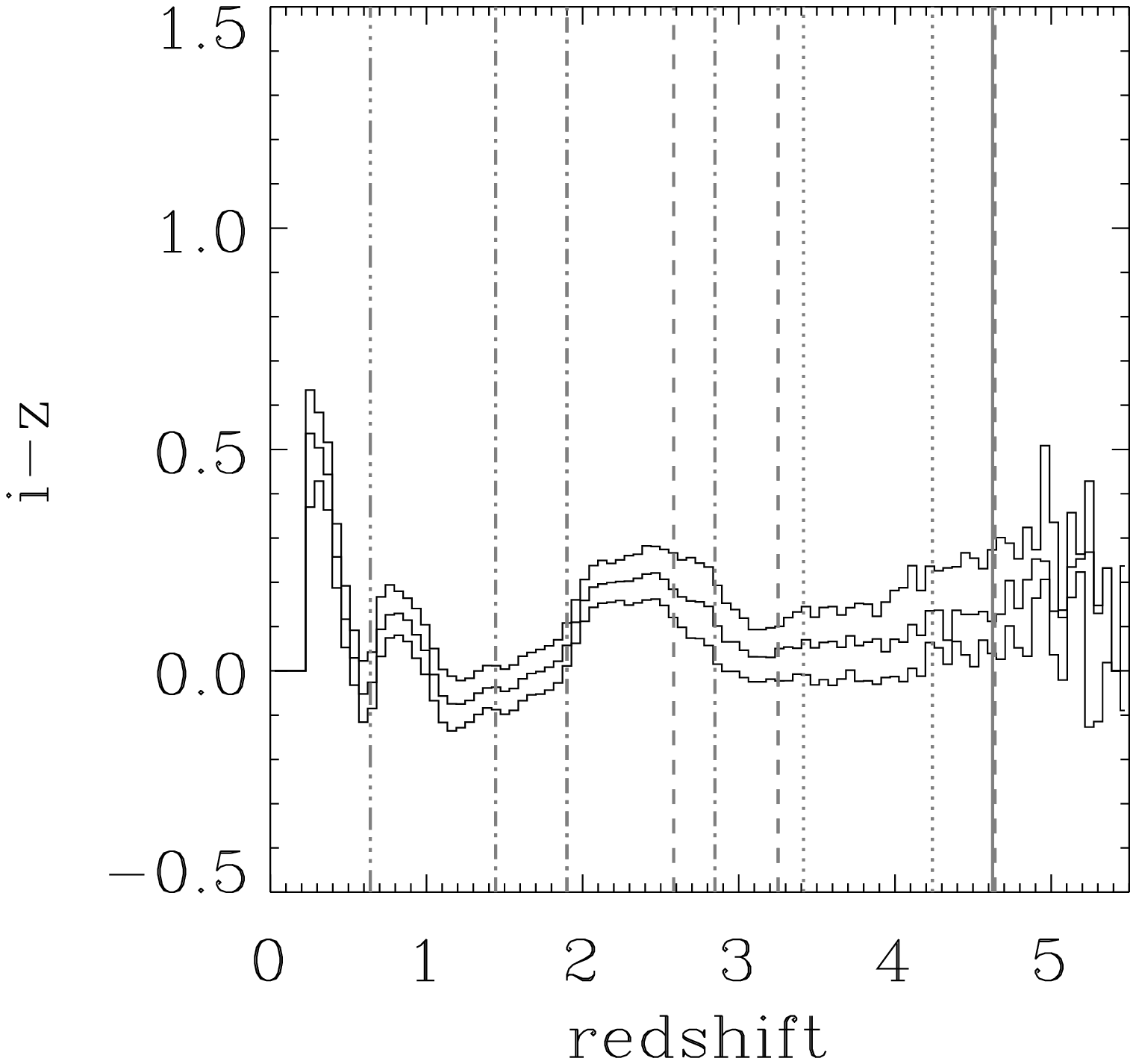}
\includegraphics[width=0.24\textwidth,clip=]{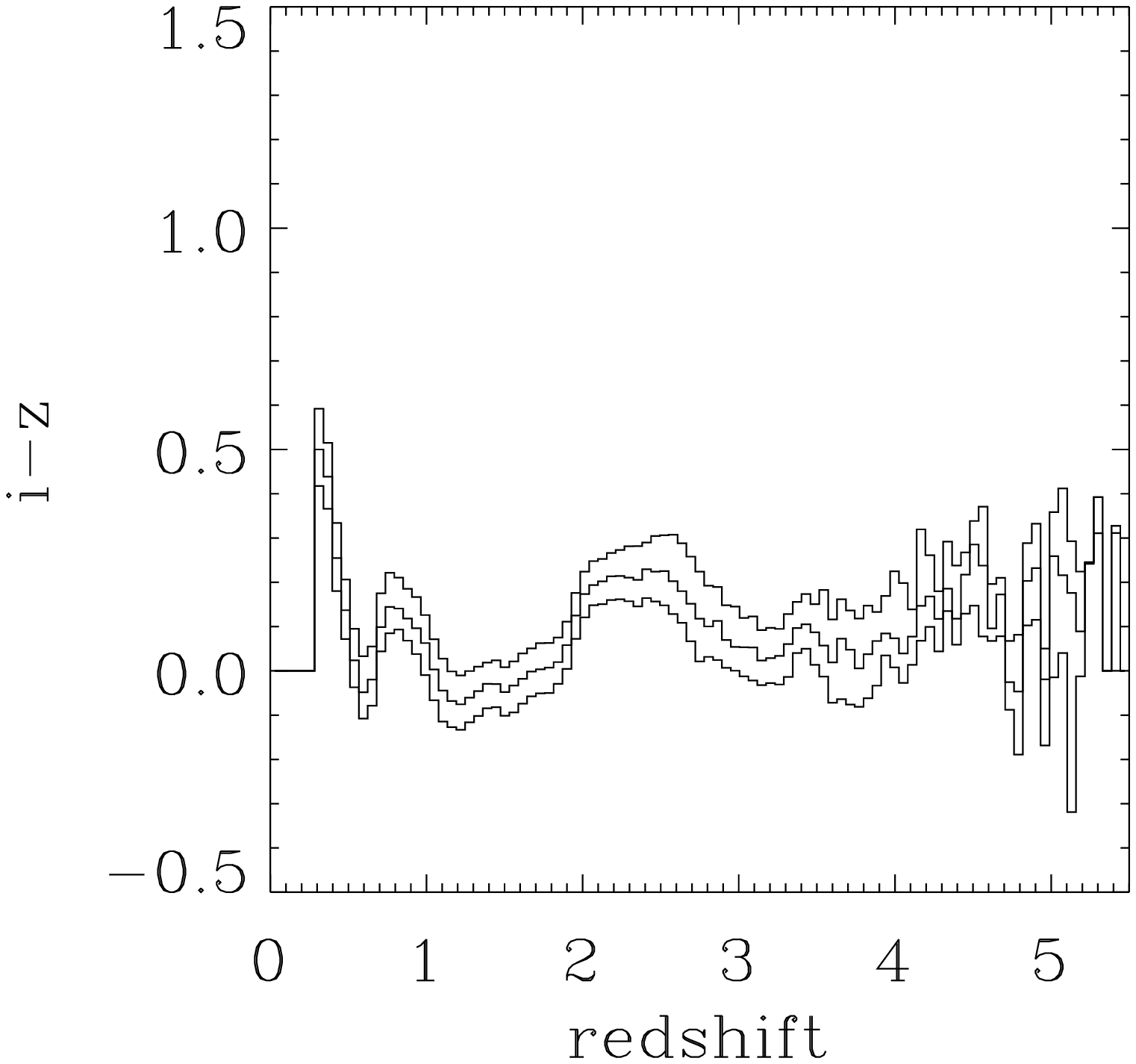}
\caption{Flux--redshift and color--redshift diagrams for the 18.6 $\leq
  i < 18.8$ bin in apparent magnitude for the 103,577 objects in the
  quasar catalog. The first column shows a conditional plot of a
  sampling from the \emph{extreme deconvolution} fit with the errors
  from the quasar data added; the second column presents the quasar
  data resampled according to the quasar luminosity function as
  described in \sectionname~\ref{sec:data} and in more detail in
  \citet{Bovy11a}. All fluxes are relative to the \iband\ flux of the
  object. The third and fourth columns show the same information as
  the first and second columns, but for colors. Linear conditional
  densities are shown as well as the 25, 50, and 75
  quantile-lines. The vertical lines denote where prominent emission
  lines pass in and out of the relevant filters (Ly$\alpha$: full;
  CIV: dotted; CIII: dashed; MgII: dash-dotted; H$\alpha$:
  dash-dot-dot-dotted). Although only the conditional relation between
  redshift and flux/color is shown here, we fit the full density in
  the flux--redshift space.}\label{fig:allzexfit}
\end{figure}

\figurename~\ref{fig:allzexfit} shows relative-flux--redshift and
color--redshift diagrams of quasars for a single \iband\ magnitude
bin. The conditional distribution of relative-flux as a function of
redshift is shown here (although the model contains a full model of
the density on this manifold); this emphasizes what is new in
\xdqsoz\ as compared to \xdqso. We see that the \exd\ technique is
superb at capturing the complexity of the quasar color locus, even at
higher redshifts where the data are sparse and noisy. The locations
where prominent emission lines cross the relevant \sdss\ filters are
indicated, and it is clear that this drives much of the structure in
the color--redshift relation.

\begin{figure}
\includegraphics[width=0.4\textwidth,clip=]{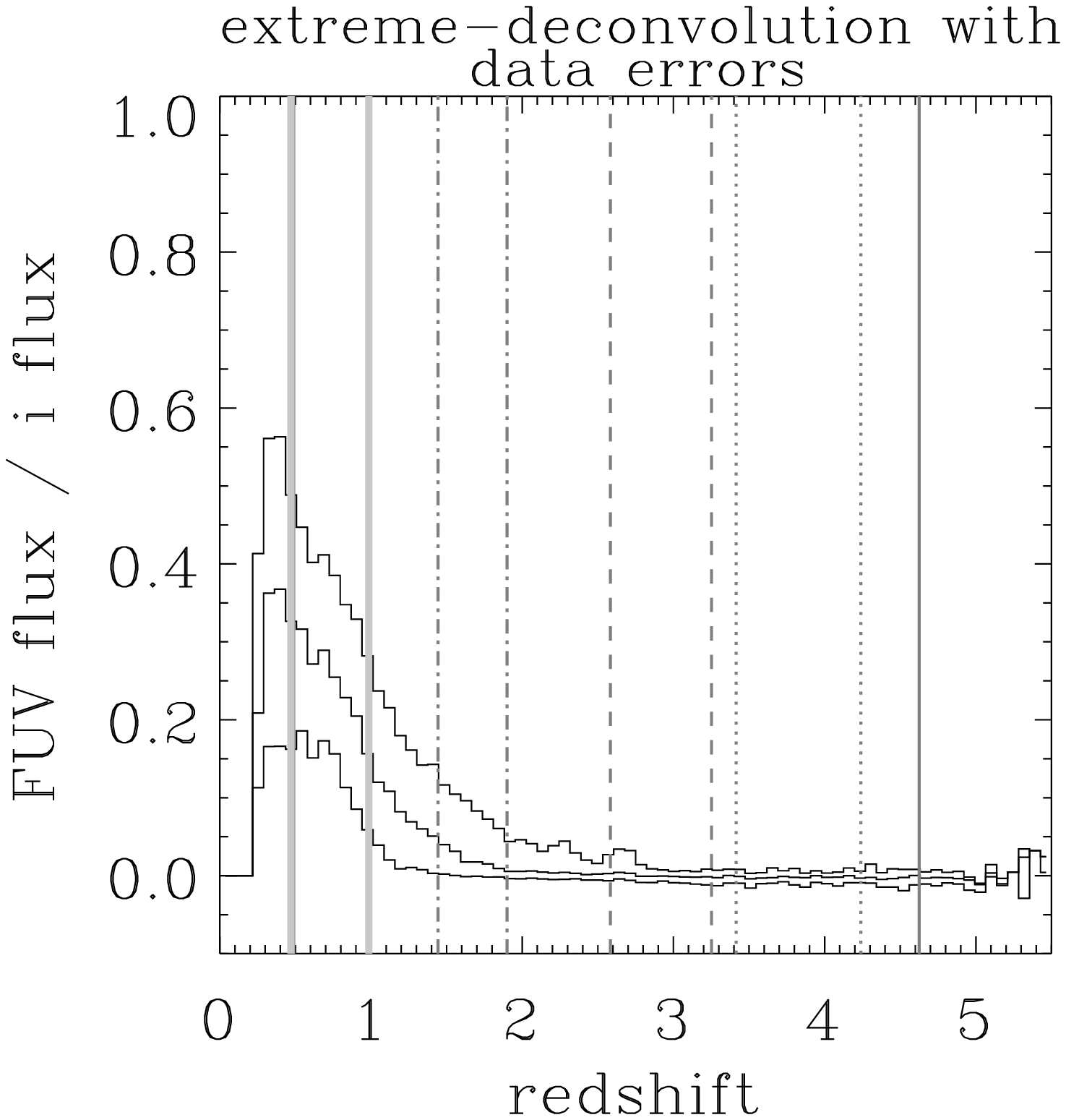}
\includegraphics[width=0.4\textwidth,clip=]{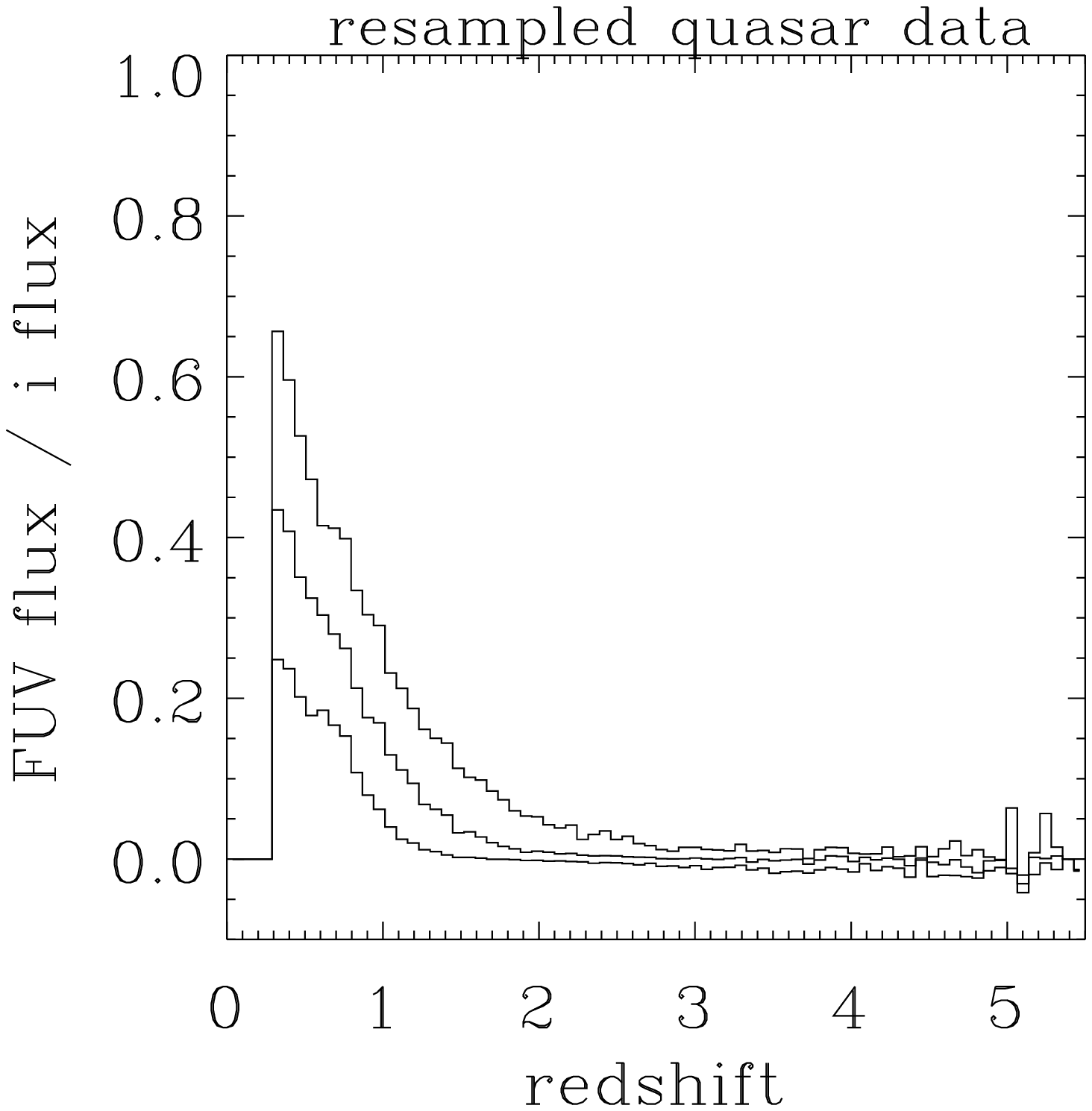}\\
\includegraphics[width=0.4\textwidth,clip=]{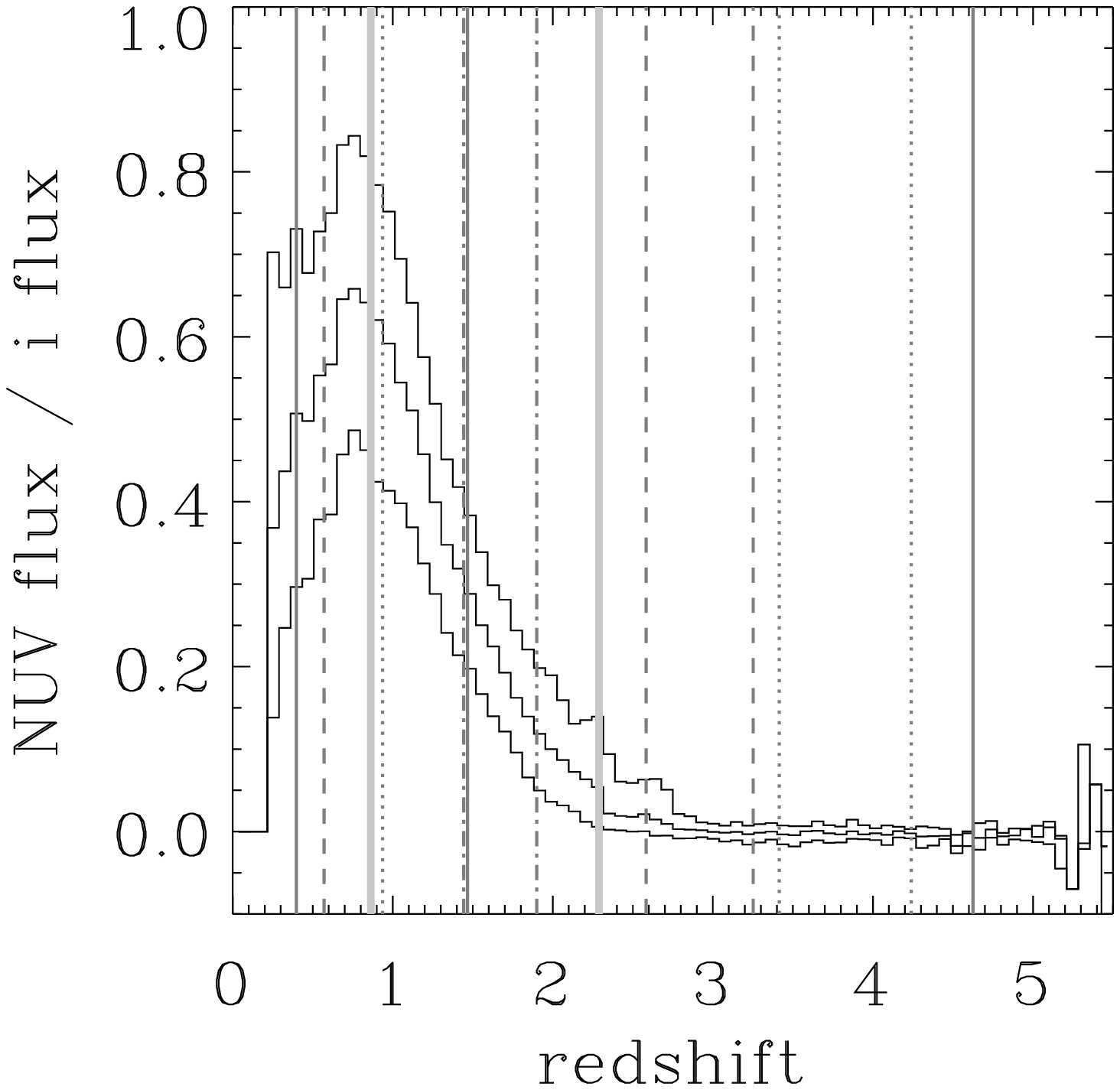}
\includegraphics[width=0.4\textwidth,clip=]{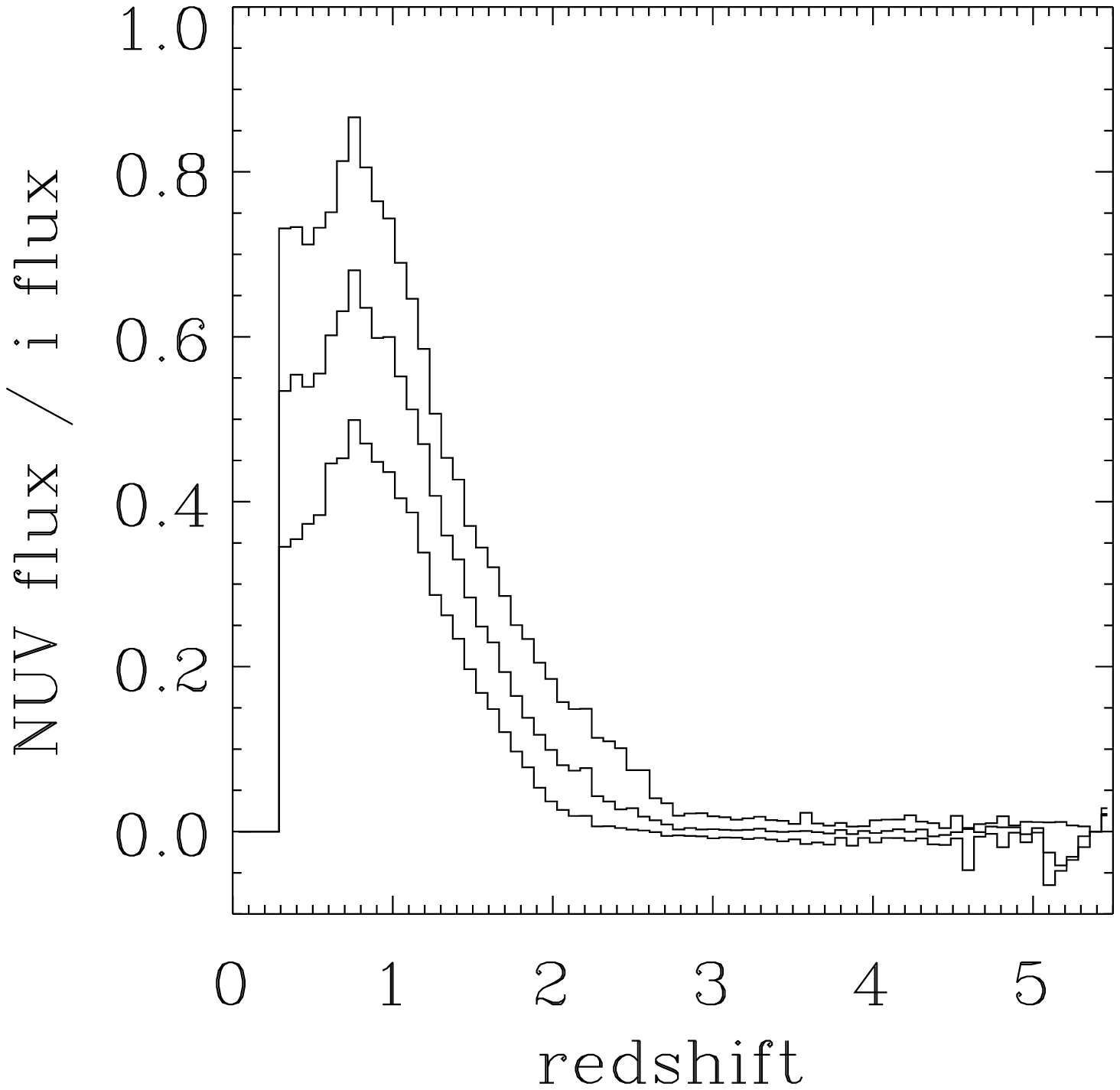}
\caption{Flux--redshift diagrams for the $18.6 \leq i < 18.8$ bin in
  apparent magnitude for the 62,628 \sdss\ quasars with
  \galex\ observations in both \galex\ bandpasses. The left column
  shows a conditional plot of a sampling from the \emph{extreme
    deconvolution} fit with the errors from the quasar data added; the
  right column displays the quasar data resampled according to the
  quasar luminosity function as described in
  \sectionname~\ref{sec:data}. All fluxes are relative to the
  \iband\ flux of the object. Densities, curves, and vertical lines
  are as in \figurename~\ref{fig:allzexfit}. The thick light-gray
  bands show where the Lyman limit ($\lambda912$ \AA) crosses the
  \uv\ filters.}\label{fig:allzexfitgalex}
\end{figure}

\figurename~\ref{fig:allzexfitgalex} shows similar
relative-flux--redshift diagrams for the \uv\ fluxes in the model
containing both optical and \uv\ data. The agreement between the
empirical model and the data is excellent. These diagrams clearly
demonstrate that the \uv\ flux of $z \gtrsim 1$ and $z \gtrsim 2.3$
quasars, for \fuv\ and \nuv\ respectively, is suppressed because of
absorption below the Lyman limit ($\lambda912$ \AA) by intervening
systems \citep{Moller90a,Picard93a,Worseck11a}. \uv\ observations are
an excellent tool to distinguish $z \approx 0.8$ quasars from $z
\approx 2.3$ quasars, which have degenerate \ugriz\ colors and plague
medium-redshift quasar selection \citep[\eg,][]{Ross11a}, even at low
\uv\ signal-to-noise ratio.

\begin{figure}
\includegraphics[width=0.24\textwidth,clip=]{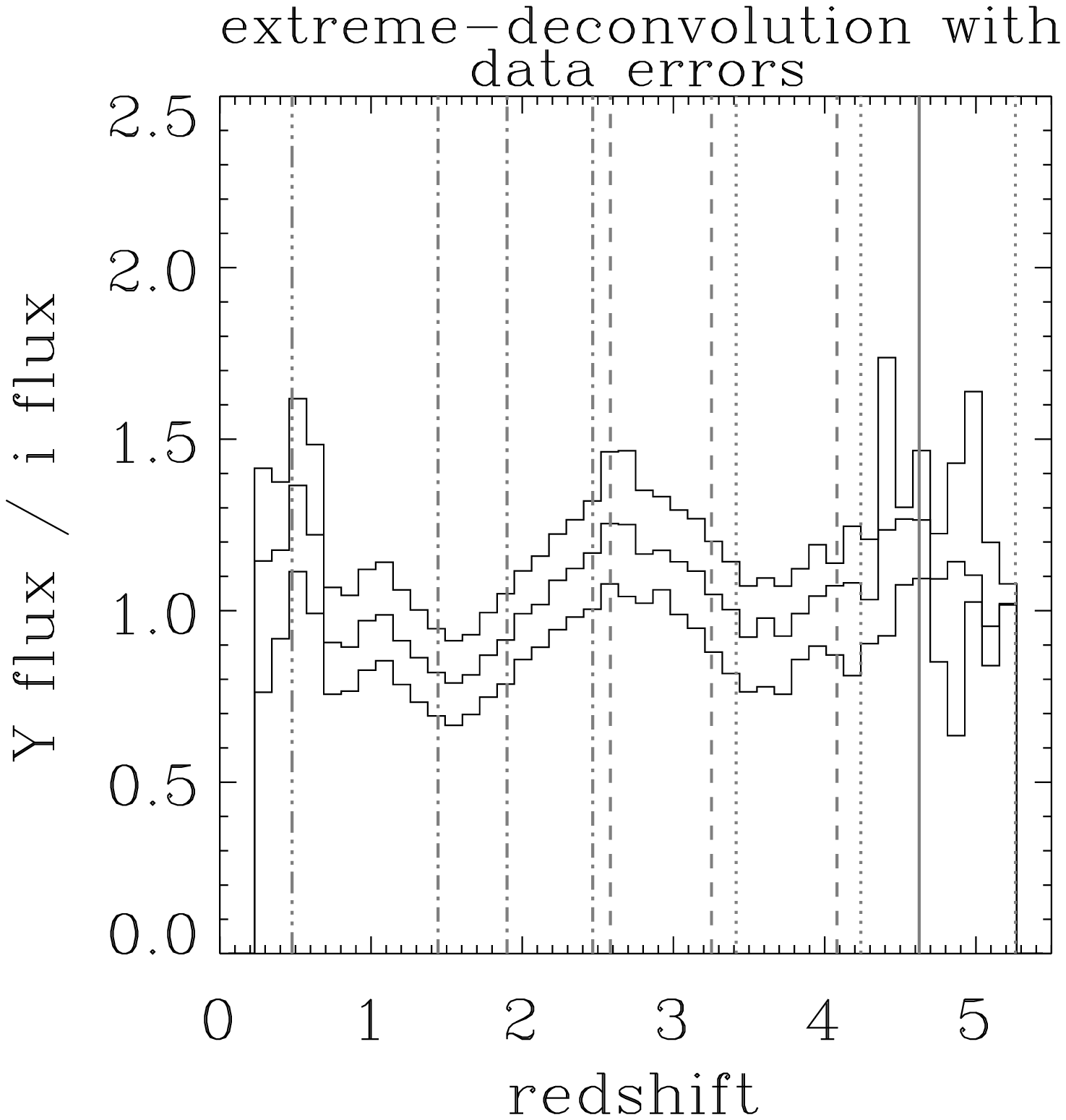}
\includegraphics[width=0.24\textwidth,clip=]{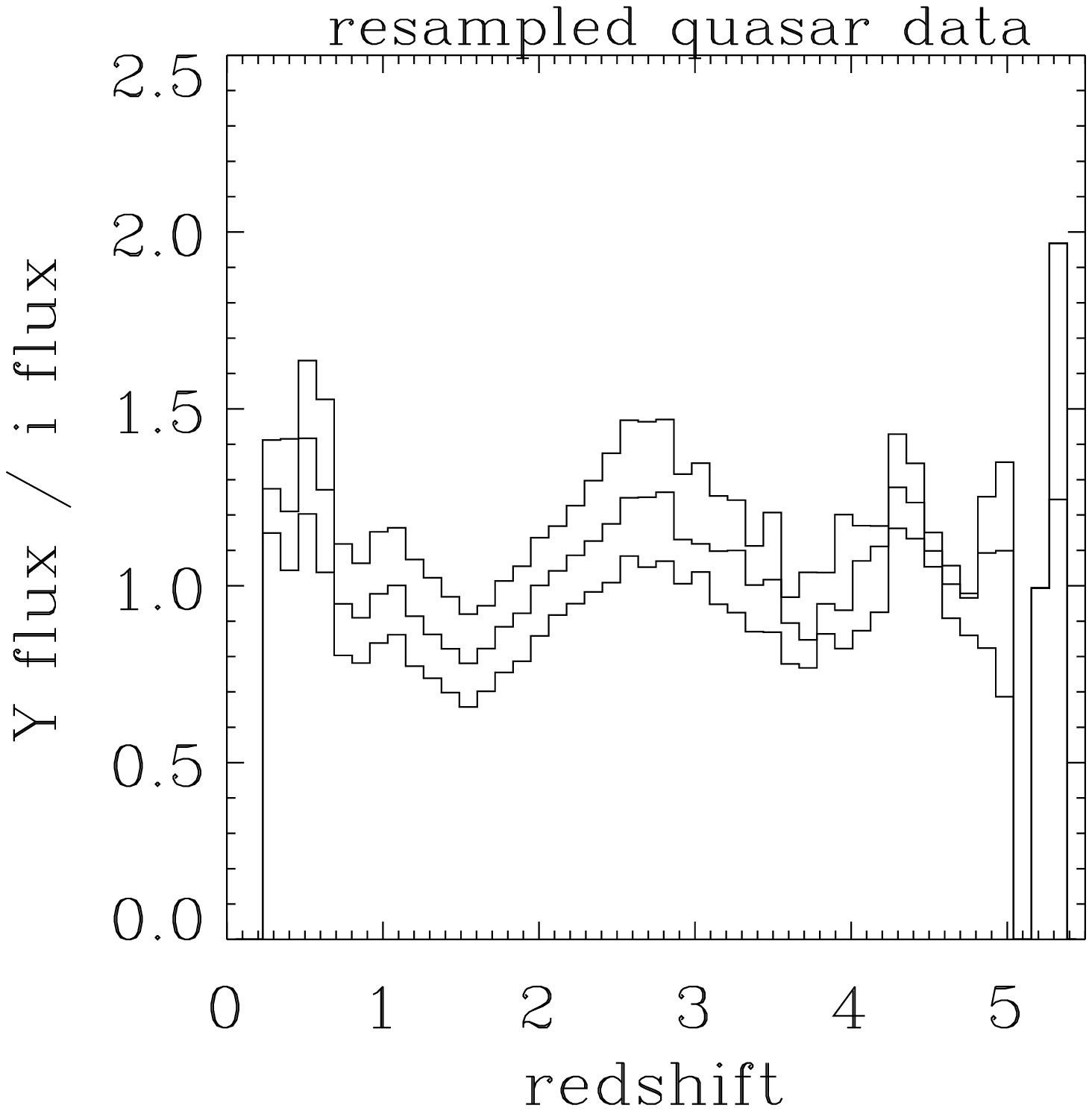}\\
\includegraphics[width=0.24\textwidth,clip=]{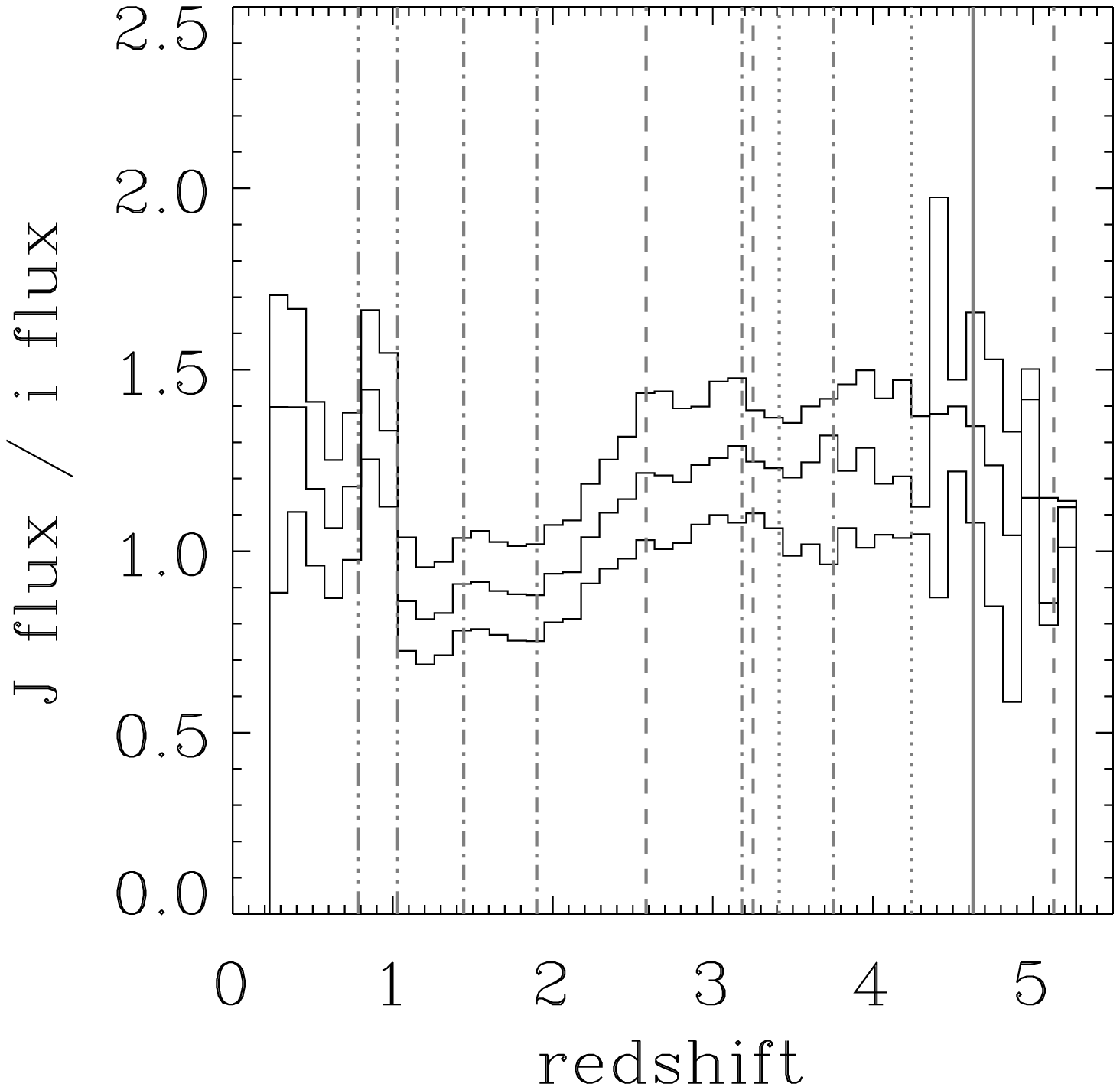}
\includegraphics[width=0.24\textwidth,clip=]{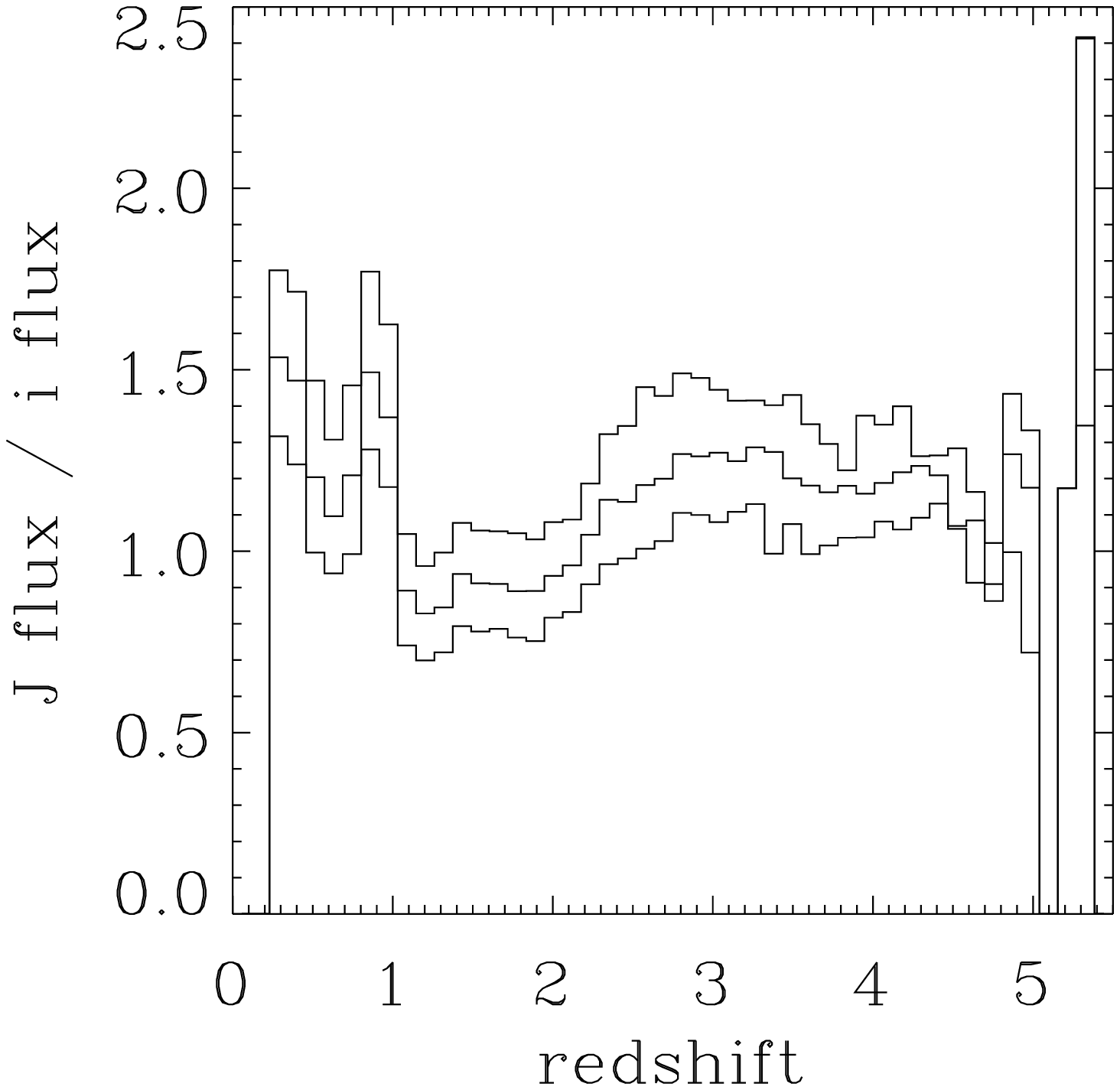}\\
\includegraphics[width=0.24\textwidth,clip=]{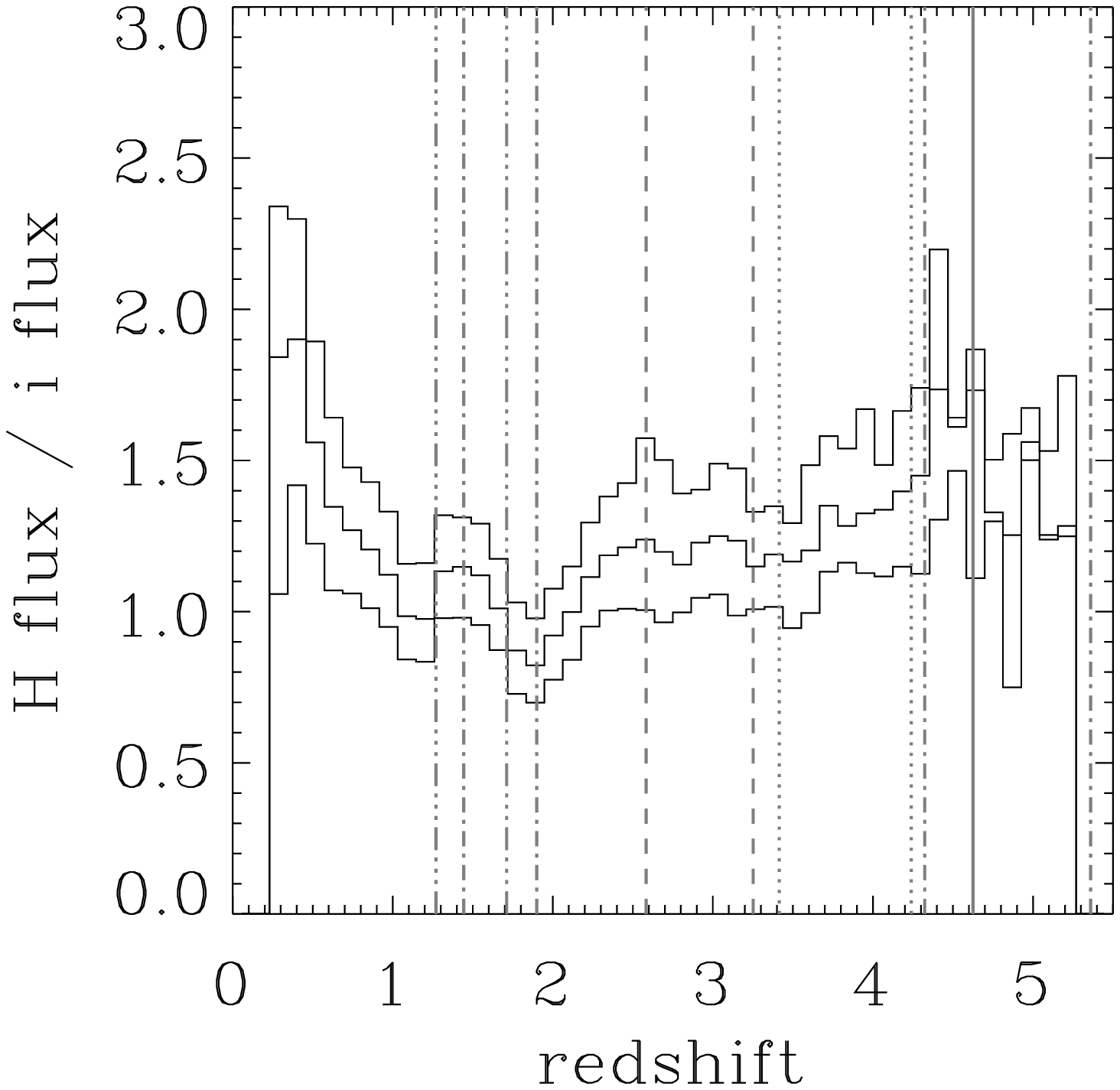}
\includegraphics[width=0.24\textwidth,clip=]{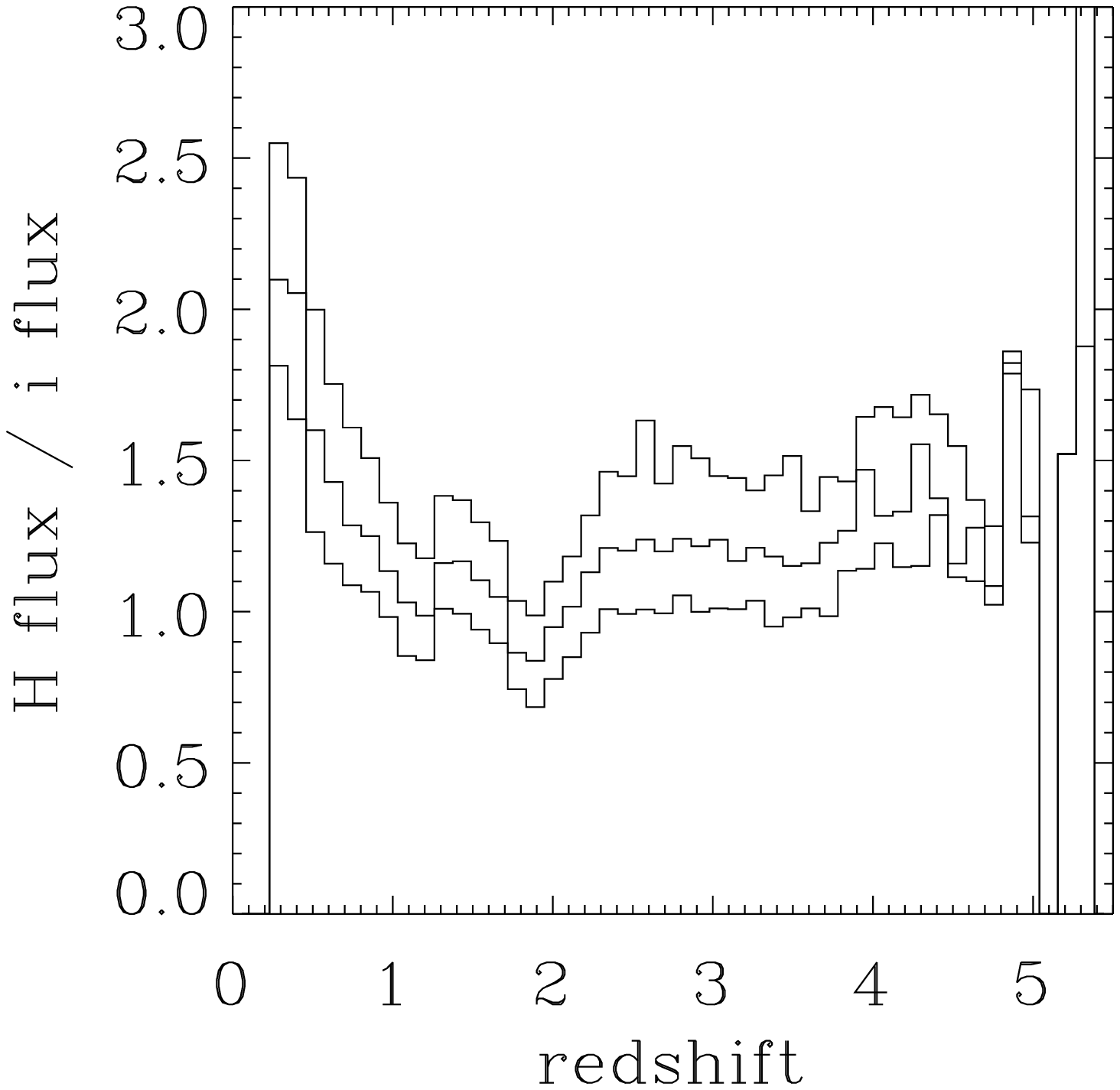}\\
\includegraphics[width=0.24\textwidth,clip=]{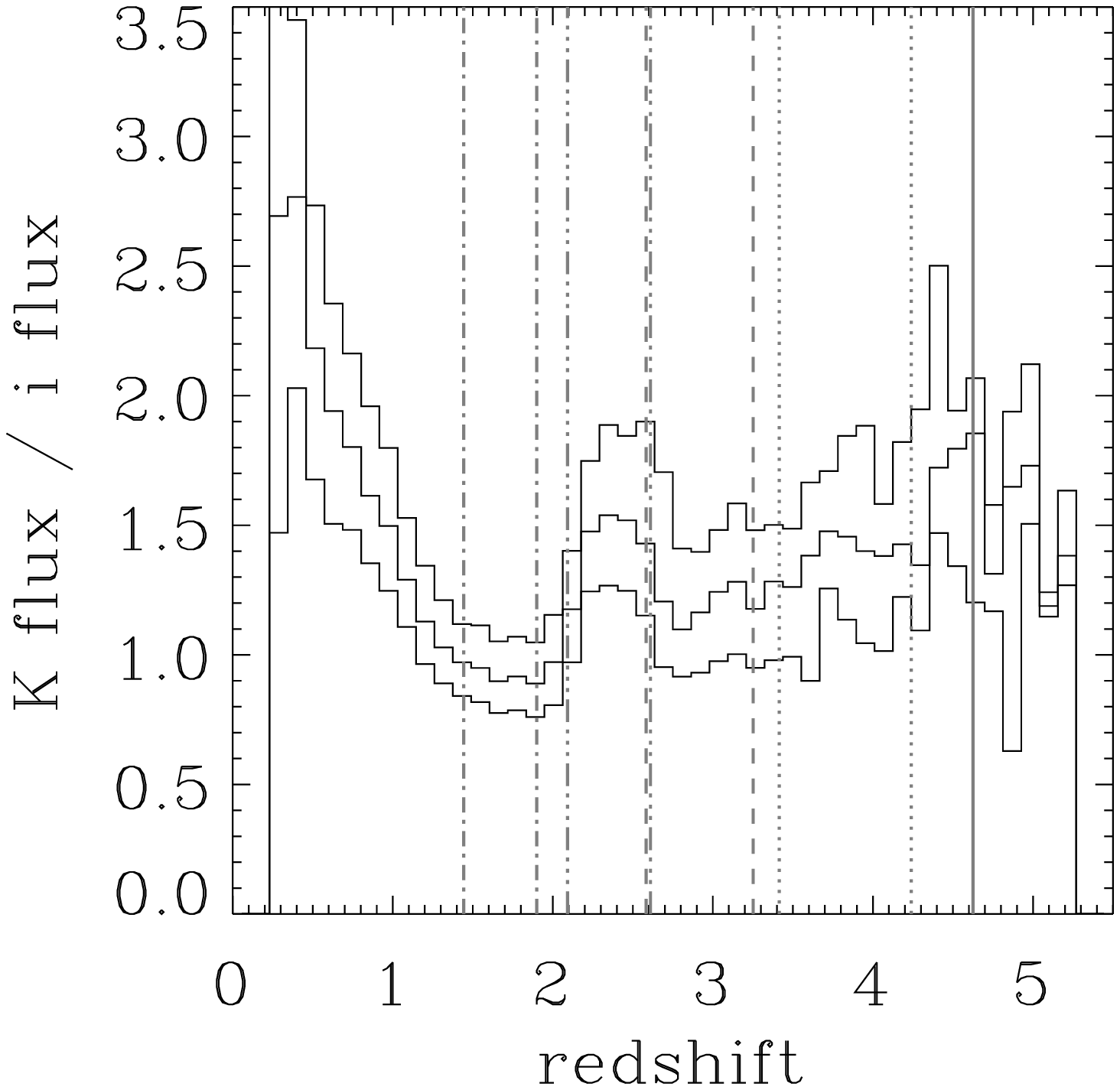}
\includegraphics[width=0.24\textwidth,clip=]{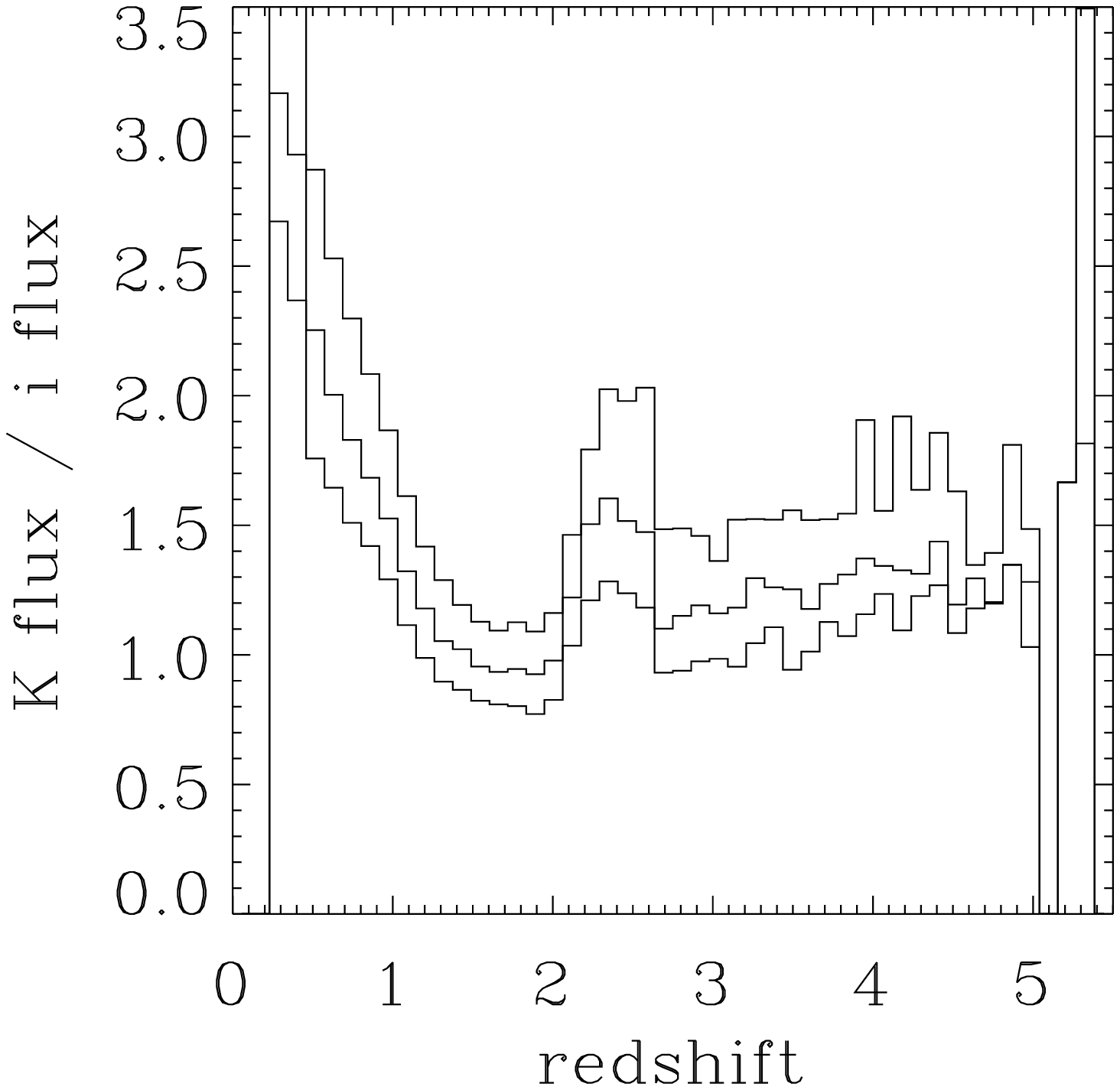}
\caption{Same as \figurename~\ref{fig:allzexfitgalex}, but for
  the 25,510 \sdss\ quasars that have \ukidsslas\ observations in all
  four \ukidss\ bandpasses.}\label{fig:allzexfitukidss}
\end{figure}

\figurename~\ref{fig:allzexfitukidss} presents relative-flux--redshift
diagrams for the four \nir\ fluxes in the \xdqsoz\ model that contains
optical and \nir\ data. The agreement between the \xdqsoz\ model and
the data is again excellent and the \xdqsoz\ model captures all of the
photometric redshift information contained in the \nir. We
see that much of the variation in the color--redshift relation in the
\nir\ is driven by the H$\alpha$ line \citep[see
  also][]{Glikman06a,Assef10a,Peth11a}.

The model--data comparisons given in this section are only a
small fraction of the model-assessment diagnostics that we
performed. For example, we do not show the optical fits here in models
that contain \uv\ or \nir\ data, nor do we show the \uv\ models and
\nir\ models in the full \ugriz-\uv+\nir\ model, as all of these
comparisons are very similar to the ones shown here.

\section{Targeting and photometric quasar classification with \xdqsoz}\label{sec:targeting}

We can use the \xdqsoz\ model to photometrically classify and target
quasars by calculating the probability that an object is a quasar
based on its broadband fluxes. The probability that an object is a
quasar in a redshift range $\Delta z$ is obtained by integrating the
probability that an object is a redshift $z$ quasar over redshift. We
start by using \eqnname~(\ref{eq:quasardeltazprob})
\begin{align}
p(z,\mathrm{quasar}| \mathrm{fluxes}) & \propto p(z,\setofrfj |
\fluxi,\mathrm{quasar})\,p(\fluxi , \mathrm{quasar})\,,
\end{align}
where $\setofrfj$ is the set of fluxes relative to $\fluxi$ and
$\fluxi$ is the \iband\ flux of the object. The normalization factor
is given by
\begin{equation}
p(\mathrm{fluxes}) = p(\mathrm{fluxes},\mathrm{star}) + 
\int_0^\infty \dd z\,p(z,\mathrm{fluxes},\mathrm{quasar})\,.
\end{equation}
Because the apparent magnitude factor $p(\fluxi,\mathrm{quasar})$ does
not depend on redshift, the integral over redshift is only over the
$p(z,\setofrfj|\fluxi,\mathrm{quasar})$ factor, which is modeled as a
simple sum of Gaussian distributions.

For any given object we can simplify the mixture of $n$-dimensional
Gaussian distributions to a mixture of one-dimensional Gaussian
distributions for the redshift of the object. First, we find the bin
in the \iband\ magnitude that best matches the object's \iband\ magnitude
and use the mixture-of-Gaussians representation of the
relative-flux--redshift density in this bin. Assuming that the
$n$-dimensional mixture of Gaussians has amplitudes $\alpha_k$, means
$\mm_k$, and variance matrices $\VV_k$, we can condition each of the
components on the measured relative flux $\rr=\setofrfj$ of the object
and its uncertainty variance matrix $\SSS$ to find (\eg,
\appendixname~B of \citealt{Bovy09a})
\begin{align}\label{eq:means}
m_{z,k} &= m^k_z + \VV^k_{zr}\,\TT^{-1,k}_{rr}\,(\rr - \mm^k_{r})\,\\
\sigma^2_{z,k} &= V^k_{zz} - \VV^k_{zr}\,\TT^{-1,k}_{rr}\,\VV^{T,k}_{zr}\,\label{eq:vars}
\end{align}
while the amplitudes of these one-dimensional Gaussian distributions
are given by the posterior probability that the object was drawn from
component $k$
\begin{equation}\label{eq:alphazk}
\alpha_{z,k} = \frac{\alpha_k\,\mathcal{N}\left(\rr|\mm^k_{r},\TT^k_{rr}\right)}{\sum_l\alpha_l\,\mathcal{N}\left(\rr|\mm^l_{r},\TT^l_{rr}\right)}\,.
\end{equation}
In these expressions $\mm^k_{r}$ and $m^k_z$ are the relative flux and
the redshift parts of $\mm_k$, respectively; $\TT^k_{rr} = \VV^k_{rr}
+ \SSS$; $\VV^k_{rr}$, $\VV^k_{zr}$, and $V^k_z$ are the
relative-flux--relative-flux, redshift--relative-flux, and
redshift--redshift parts of $\VV_k$, respectively; and
$\mathcal{N}(\cdot|\cdot,\cdot)$ is the multivariate Gaussian
distribution. $\TT^k_{rr}$ includes the uncertainty variance matrix
$\SSS$ because the necessary uncertainty convolution simply reduces to
adding the observational uncertainty variance matrix to the intrinsic
variance matrix for each Gaussian component.

Integrating this one-dimensional mixture of Gaussian distributions
over an arbitrary redshift range results in a sum over error
functions. Remembering that our model lives in $\log$ redshift space
\begin{equation}
\begin{split}
\int_{z_{\mathrm{min}}}^{z_{\mathrm{max}}} \dd z \,p(z,\setofrfj|\fluxi,\mathrm{quasar}) & = p(\setofrfj|\fluxi,\mathrm{quasar}) \\
& \quad \times
\sum_k \frac{\alpha_{z,k}}{2}
\left(\mathrm{erf}\left[\frac{\log z_{\mathrm{max}} - m_{z,k}}{\sqrt{2}\,\sigma_{z,k}}\right]-\mathrm{erf}\left[\frac{\log z_{\mathrm{min}} - m_{z,k}}{\sqrt{2}\,\sigma_{z,k}}\right]\right)\,,
\end{split}
\end{equation}
where the error function erf$[x] \equiv 2 \int_0^x e^{-t^2}\dd
t\,/\sqrt{\pi}$. The first factor on the right-hand side of this
\eqnname\ is the integral over the entire redshift range $[0,\infty]$,
which simplifies to
\begin{align}
\begin{split}
p(\setofrfj|\fluxi,\mathrm{quasar}) 
&= \int_0^\infty \dd z\,p(z,\setofrfj|\fluxi,\mathrm{quasar})\\
 & = \sum_k\alpha_k\,\mathcal{N}\left(\rr|\mm^k_{r},\TT^k_{rr}\right)\,\,
\end{split}
\end{align}
\ie, the denominator in \eqnname~(\ref{eq:alphazk}).

\begin{figure}
\includegraphics{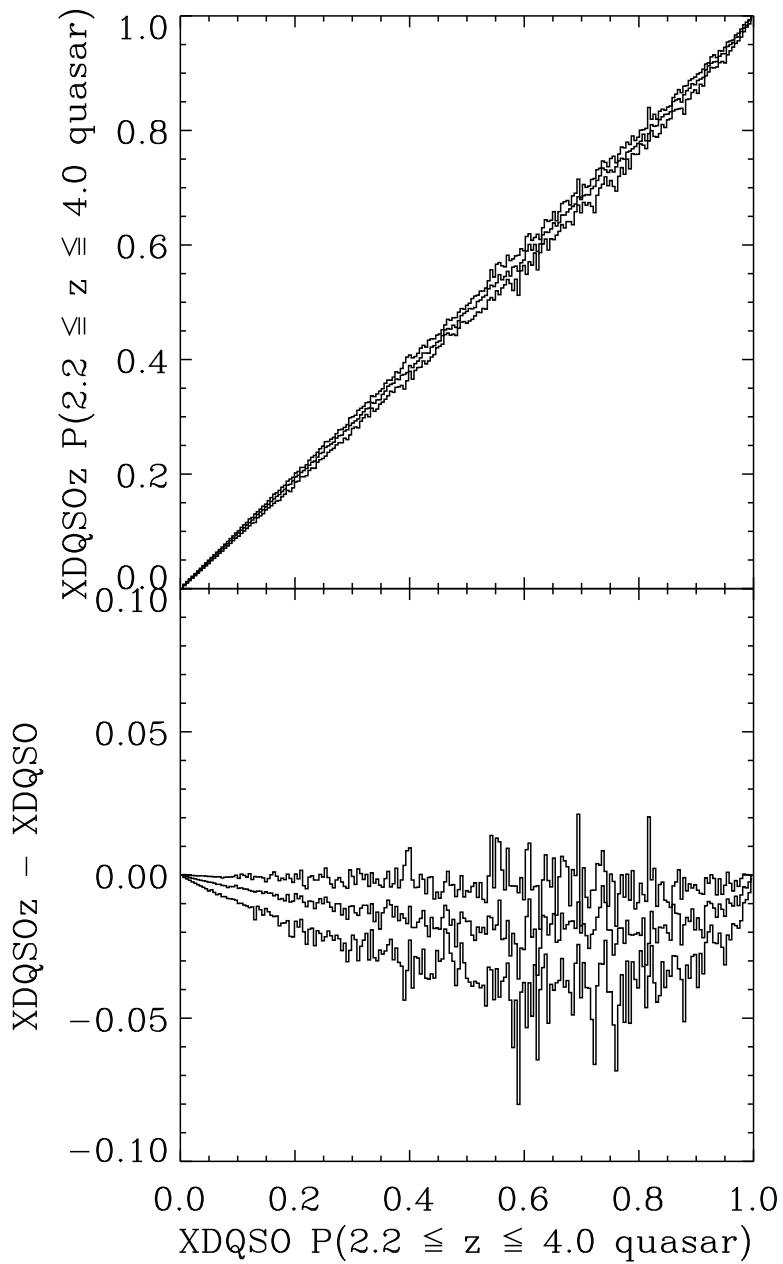}
\caption{Comparison between mid-redshift ($2.2 \leq z \leq 4.0$)
  quasar probabilities computed using \xdqso, that is, based on
  flux-density models in broad redshift ranges, and \xdqsoz, \ie,
  obtained by integrating flux--redshift--density models over the
  relevant redshift range, for 490,793 objects in \sdss\ stripe 82
  based on single-imaging-run flux measurements. Conditional 25, 50,
  and 75\,percent quantiles are shown.}\label{fig:xdqsovsxdqsoz}
\end{figure}

We can compare the quasar probabilities obtained by integrating the
\xdqsoz\ model over redshift to those from the \xdqso\ technique,
which models the distribution of quasar fluxes in three wide redshift
bins. \figurename~\ref{fig:xdqsovsxdqsoz} shows the probabilities that
490,793 objects are medium-redshift ($2.2 \leq z \leq 4.0$)
quasars obtained by the two methods for objects in the \sdss\ imaging
stripe 82. It is clear that most of the objects cluster tightly around
the one-to-one line and that the two models are essentially the same
for this redshift range.

\begin{figure}
\includegraphics[height=0.8\textheight]{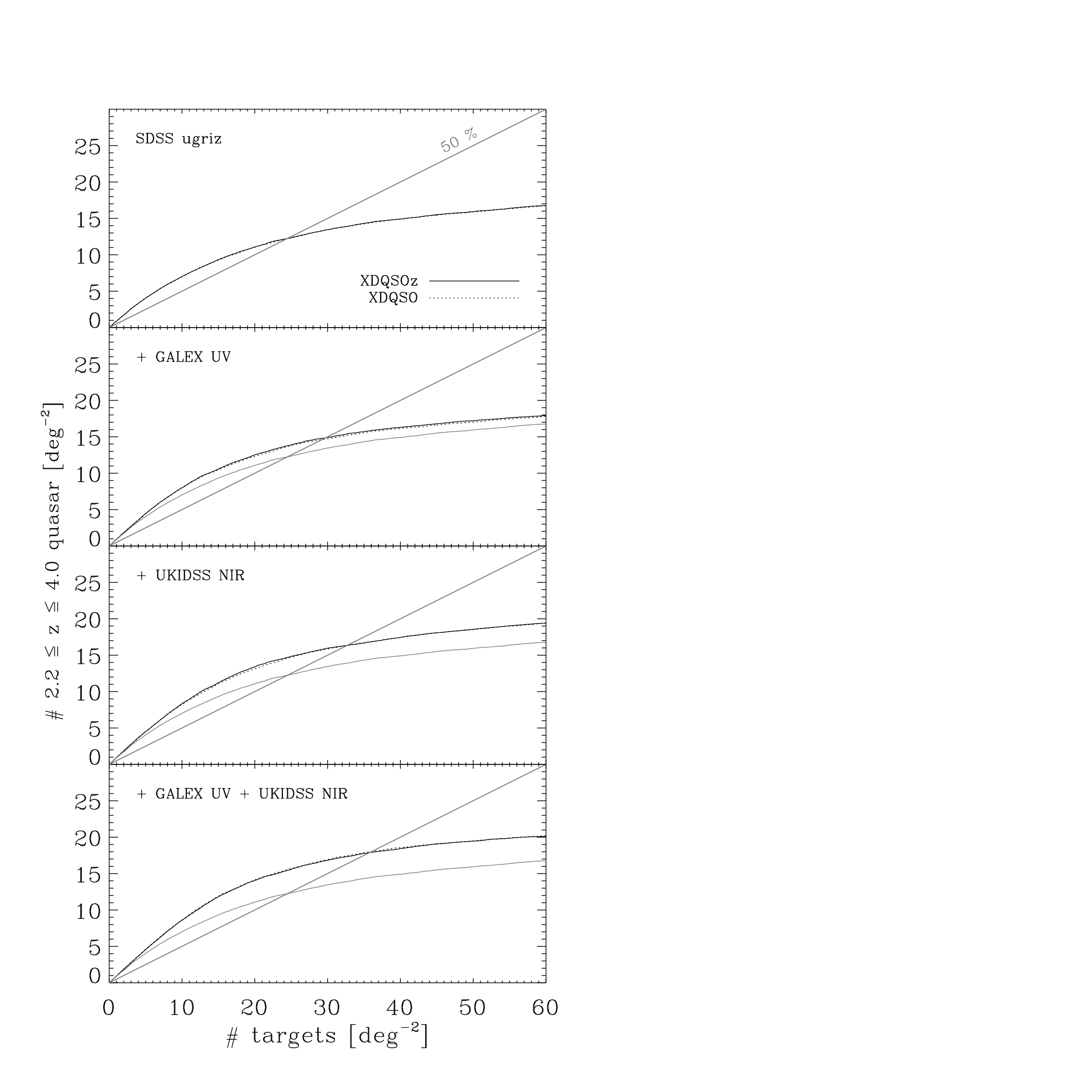}
\caption{Mid-redshift ($2.2 \leq z \leq 4.0$) quasar selection
  efficiency for \xdqso\ and \xdqsoz\ as a function of target density
  for objects in \sdss\ Stripe 82 based on single-imaging-run flux
  measurements. The top panel bases selection solely on
  \sdss\ \ugriz\ fluxes, the lower panels add \galex\ \nuv\ and
  \fuv\ medium-deep measurements as well as
  \ukidss\ \yjhk\ photometry, both of which are available for almost
  all Stripe-82 sources, through force-photometering \galex\ and
  \ukidsslas\ imaging data at \sdss\ positions. The 50\,percent
  selection efficiency is indicated and the \ugriz-only curve for
  \xdqsoz\ is repeated in gray in each
  panel.}\label{fig:xdqsovsxdqsozefficiency}
\end{figure}

\figurename~\ref{fig:xdqsovsxdqsozefficiency} shows the efficiency of
quasar targeting using both the \xdqso\ and the \xdqsoz\ method for
targeting medium-redshift ($2.2 \leq z \leq 4.0$) quasars. This test
uses a sample of medium-redshift quasars spectroscopically confirmed
by \boss\,---which also re-targets quasars previously identified in 
earlier surveys---in stripe 82. This quasar sample 
is expected to be highly complete, because it
was targeted using the superior imaging in stripe 82
where there is variability information \citep{palanque11} and where a 
number of campaigns prior to \boss\ have also obtained extensive spectroscopy. 
The sample has on average of 30 $z \geq 2.2$ quasars
deg$^{-2}$ down to $g \approx 22$ mag, which is close to the number
expected from current quasar luminosity functions
\citep[\eg,][]{Hopkins07a}. We only use regions of stripe 82 that have
more than 15 $z \geq 2.2$ quasars deg$^{-2}$. See \citet{Ross11a} and
\citet{Bovy11a} for a more detailed description of the \boss\ quasar
target selection in general and this test set in particular.

The top panel of \figurename~\ref{fig:xdqsovsxdqsozefficiency} shows
selection based on \sdss\ \ugriz\ fluxes alone. We see that the
performance of the \xdqsoz\ and the \xdqso\ techniques is essentially
identical. The lower panels of this figure show how the
selection improves when we add \galex\ \uv\ and
\ukidsslas\ \nir\ observations, both of which are available for
essentially all objects in \sdss\ stripe 82. The \xdqso\ models with
\uv\ and \nir\ data are models trained with these \uv\ and
\nir\ fluxes in the broad redshift ranges used by \xdqso. For all
intended purposes the \xdqsoz\ technique performs identically to the
\xdqso\ technique for targeting medium-redshift quasars.

We have also checked the performance of the \xdqsoz\ technique as
compared to the kernel-density-estimation based photometric quasar
classification technique of \citet{Richards04a,Richards09a}. We find
results that are similar to those for the \xdqso\ technique as shown
in \tablename~3 of \citet{Bovy11a}: at low and medium redshift the
\xdqsoz\ technique performs slightly better than the \xdqso\ technique
(and thus better than the KDE technique), while at high-redshift ($z >
3.5$) \xdqsoz\ performs slightly worse than \xdqso. This behavior is
expected because the quasar training data do not include much data
at high redshift. We thus do not probe the color--redshift
relation at high redshift as well as the KDE approach, which included
additional high-redshift data. Because we use the same stellar model
as \xdqso, the same problem with sampling regions of low stellar
density that we encountered for \xdqso\ persists for \xdqsoz.

In summary, the \xdqsoz\ technique performs almost identically to the
\xdqso\ method for photometrically classifying objects as
quasars---and thus for quasar targeting. \xdqsoz\ has the advantage
over \xdqso\ and any other photometric quasar classification scheme
that it can classify quasars in arbitrary redshift ranges ``on the
fly'' (\ie, without retraining the model).

\begin{figure}
\includegraphics[height=0.7\textwidth]{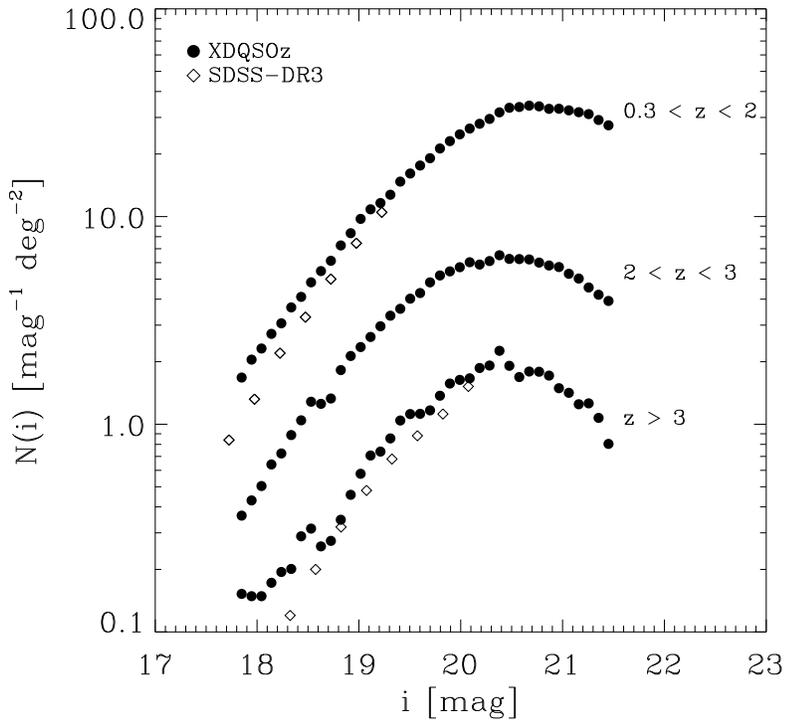}
\caption{Apparent $i$-band magnitude distribution of all point sources
  in the expected \sdssiii\ \boss\ footprint with \xdqsoz\ quasar
  probability larger than 0.5 over the indicated redshift range and
  dereddened $i$ between 17.8 mag and 21.5 mag. Diamonds indicate
  number counts from the \sdss\ spectroscopic survey
  \citep{Richards06a}.}\label{fig:xdqsoz_imag}
\end{figure}

We have computed \xdqsoz\ quasar probabilities for all 160,904,060
point sources with dereddened \iband\ magnitude between 17.75 and 22.45
mag in the 14,555 deg$^2$ of imaging from SDSS Data Release 8
\citep{Aihara11a} in three redshift ranges (0.3 $< z < 2$, $2 < z <
3$, and $z > 3$). \figurename~\ref{fig:xdqsoz_imag} shows the apparent
$i$-band magnitude distribution of all of the objects with $17.8 \leq
i \leq 21.5$ mag in the expected \boss\ spectroscopic footprint
\citep{Eisenstein11a} with \xdqsoz\ probability larger than 0.5 over
the specified redshift range. These apparent-magnitude distributions
are smooth and well-behaved for low and medium redshifts. They also
agree at the bright end with number counts derived from spectroscopic
observations \citep{Richards06a}. At the faint end ($i \gtrsim 21$)
the \iband\ number counts start to decline due to increasing
photometric uncertainties and incompleteness of the \sdss\ imaging
near the faint limit of \sdss.

The \boss\ aims to detect the baryon acoustic feature (\baf) in the
Ly$\alpha$ forest of background redshift $z \geq 2.2$ quasars. Not all
quasars contribute equally to this measurement and, as shown by
\citet{McDonald07a} and \citet{Mcquinn11a}, both brighter quasars and
quasars near redshift $z \approx 2.5$ are the most valuable. Combining
Ly$\alpha$ \baf\ weights with the quasar probabilities as a function of
redshift produced by \xdqsoz, we can calculate the expected value of a
quasar for the Ly$\alpha$ \baf\ measurement. Defining a value function
$w(g,z)$, where $g$ is the dereddened $g$-band magnitude of the
object, the expected value of an object is
\begin{equation}
\langle \mathrm{quasar\ value} \rangle = \int_0^\infty \dd z\,
w(g,z)\,p(z,\mathrm{quasar}|\mathrm{flux})\,. 
\end{equation}
By targeting objects with the highest expected value for a particular
Ly$\alpha$ \baf\ survey---which is dependent on the exact
observational characteristics of that survey---we could optimize the
targeting of quasars for that \baf\ measurement.

\begin{figure}
\includegraphics[height=0.7\textheight]{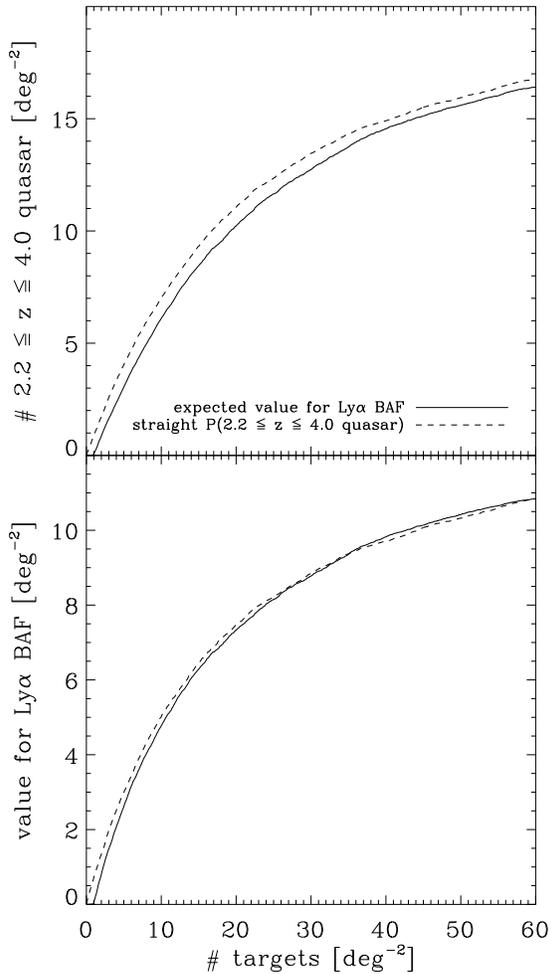}
\caption{Comparison between ``value-based'' and straight
  probability-based quasar selection for \boss. ``Value-based''
  selection ranks targets on the expected signal-to-noise ratio of the
  Lyman-$\alpha$ forest, while probability-based selection ranks on
  $P(2.2 \leq z \leq 4.0$ quasar). The top panel shows the number of
  mid-redshift quasars found by each method as a function of the
  target density; the bottom panel shows the value of the selected
  quasars. Note that some $z < 2.2$ quasars are valuable for the
  Lyman-$\alpha$ forest BAF measurement.}\label{fig:xdqsozvsvalue}
\end{figure}

The top panel of \figurename~\ref{fig:xdqsozvsvalue} shows the number
of medium-redshift quasars found by applying this value-based
targeting for \boss\ using the value function of
\citet{McDonald07a}. Value-based targeting finds about 1 quasar
deg$^{-2}$ less than targeting based on the ranked medium-quasar
probability list. The bottom panel shows that value-based targeting
finds as much value as the straight probability-based targeting---but
not more---such that the \baf\ measurement based on both samples
should be equally precise. Straight probability-based targeting thus
finds the same value while assembling a larger overall quasar
sample. In addition, value-based targeting optimizes one experiment in
a specific survey, whereas straight probability-based targeting
returns information that is broadly applicable to a range of
experiments and a range of surveys. Thus, in general, there is little
to be gained from pursuing value-based targeting for \boss.

\begin{figure}
\includegraphics[height=0.7\textheight]{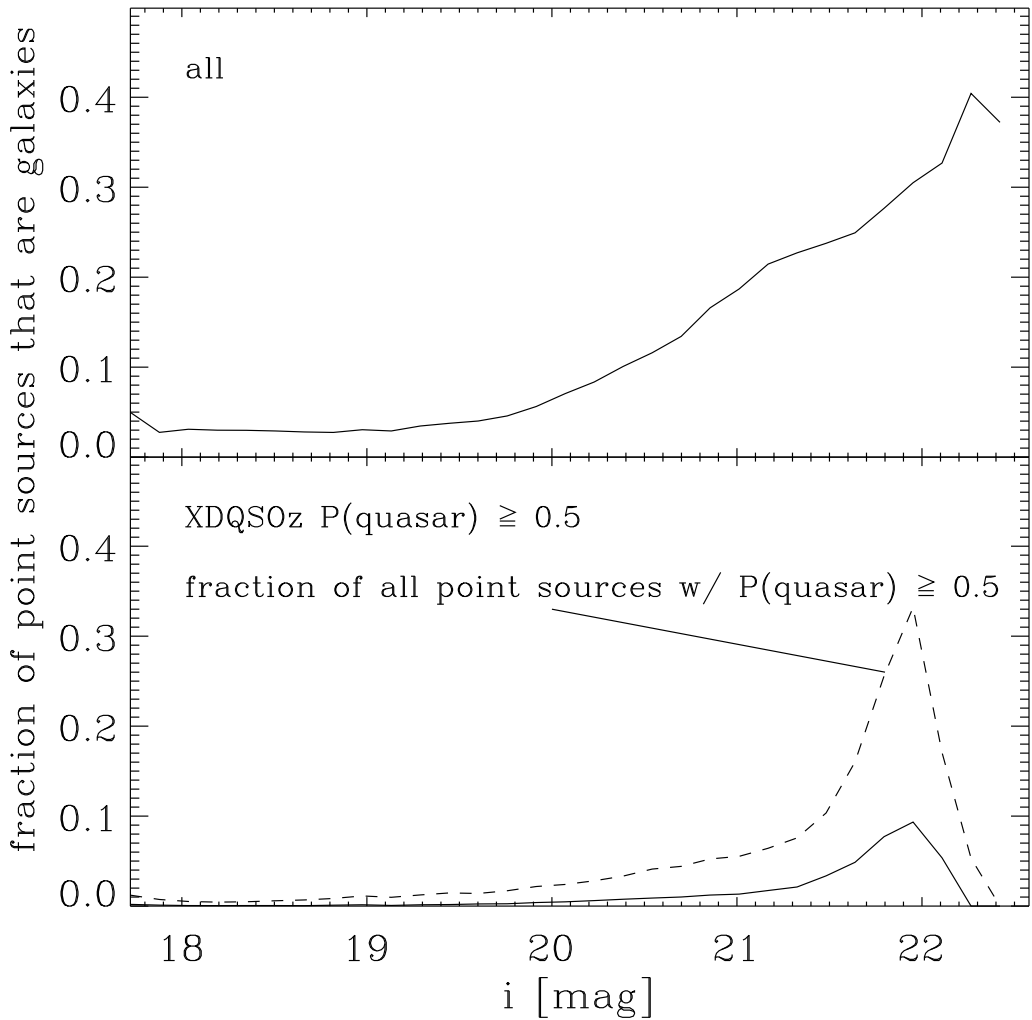}
\caption{Point-like galaxy contamination of color-based quasar
  selection: The top panel shows the fraction of point sources in a
  single imaging-pass of \sdss\ stripe 82 that are extended in the
  co-added stripe-82 imaging \citep{Abazajian09a} as a function of
  the $i$-band magnitude. The bottom panel shows the fraction of such
  point-like galaxies that have an \xdqsoz\ quasar probability (over
  all redshifts) larger than 0.5. The dashed curve in the bottom panel
  shows the fraction of all point-sources that have an \xdqsoz\ quasar
  probability larger than 0.5. Even though point-like galaxies start
  to dominate the number counts around $i$ = 22 mag, they only make up
  a small fraction of photometrically selected quasars at all
  magnitudes.}\label{fig:pointgals}
\end{figure}

To investigate whether the \xdqsoz\ quasar selection technique is
limited by contamination from galaxies that appear point-like at the
faint flux levels to which we push quasar classification, we look at
the fraction of objects that appear point-like in a single \sdss\
imaging-pass but are extended in co-added data on \sdss\ imaging
stripe 82. We match the point sources in a typical \sdss\ imaging run
to the co-added galaxy catalog on stripe 82 \citep{Abazajian09a} and
assess the fraction of point sources that are extended in the co-added
data as a function of the $i$-band magnitude. This is shown in the top
panel of \figurename~\ref{fig:pointgals}. We see that the fraction of
point sources that are extended in the co-added imaging is only a few
percent at relatively bright magnitudes, but almost reaches
50\,percent at $i = 22$ mag. To assess whether these point-like
galaxies are a significant contaminant for the \xdqsoz\ quasar
selection, we calculate their quasar probabilities (over all
redshifts). The fraction of point-like sources that are extended in
the co-added imaging and that have quasar probabilities larger than
0.5 is shown as a function of the $i$-band magnitude in the lower panel of
\figurename~\ref{fig:pointgals}.  For comparison, the fraction of all
point sources with quasar probability larger than 0.5 is shown as the
dashed curve. Point-like galaxies make up only a small ($\lesssim
10$\,percent) fraction of \xdqsoz-selected photometric
quasars. However, because galaxies unlike stars cluster similarly to
quasars, even this small contamination fraction might significantly
degrade precision quasar-clustering measurements without improved
star--galaxy separation or proper modeling. Given the rising fraction
of point-like galaxies with increasing magnitude in the top panel of
\figurename~\ref{fig:pointgals}, point-like galaxies are likely to be
the major contaminant for quasar selection at $i > 23$ mag.

\section{Photometric redshifts with \xdqsoz}\label{sec:photoz}

We can use the \xdqsoz\ flux--redshift density model to derive full
posterior probability distributions for the redshift of photometric
quasars taking the photometric uncertainties of the object fully into
account. Because the main advantages of the \xdqsoz\ technique for
photometric redshift estimation are that it a) returns full PDFs and
b) allows auxiliary data such as that furnished by \uv\ and
\nir\ surveys to be included, we focus on those points
here. \ugriz-only photometric quasar redshifts suffer from various
degeneracies that are inherent to the \ugriz\ filter system. While
including appropriate apparent-magnitude dependent redshift
priors---as we do here---can partially relieve these degeneracies
somewhat, no photometric redshift technique can entirely remove these
degeneracies and \xdqsoz\ is no exception (as we will see
below). Crucially, even low signal-to-noise ratio \uv\ and \nir\ data
\emph{can} cleanly resolve these degeneracies.

For each object the posterior probability distribution for its
redshift, based on its measured broadband fluxes, is calculated by
finding the apparent-magnitude bin that best matches the object's
dereddened \iband\ magnitude. The posterior probability distribution
is given by the mixture of 60 one-dimensional Gaussian distributions,
with means, variances, and amplitudes given in
\eqnname~(\ref{eq:means}), (\ref{eq:vars}), and (\ref{eq:alphazk}),
respectively. This posterior probability distribution can be
calculated based on \ugriz\ fluxes, or with additional \uv\ or
\nir\ information if available.

\begin{figure}
\includegraphics[width=0.4\textwidth,clip=]{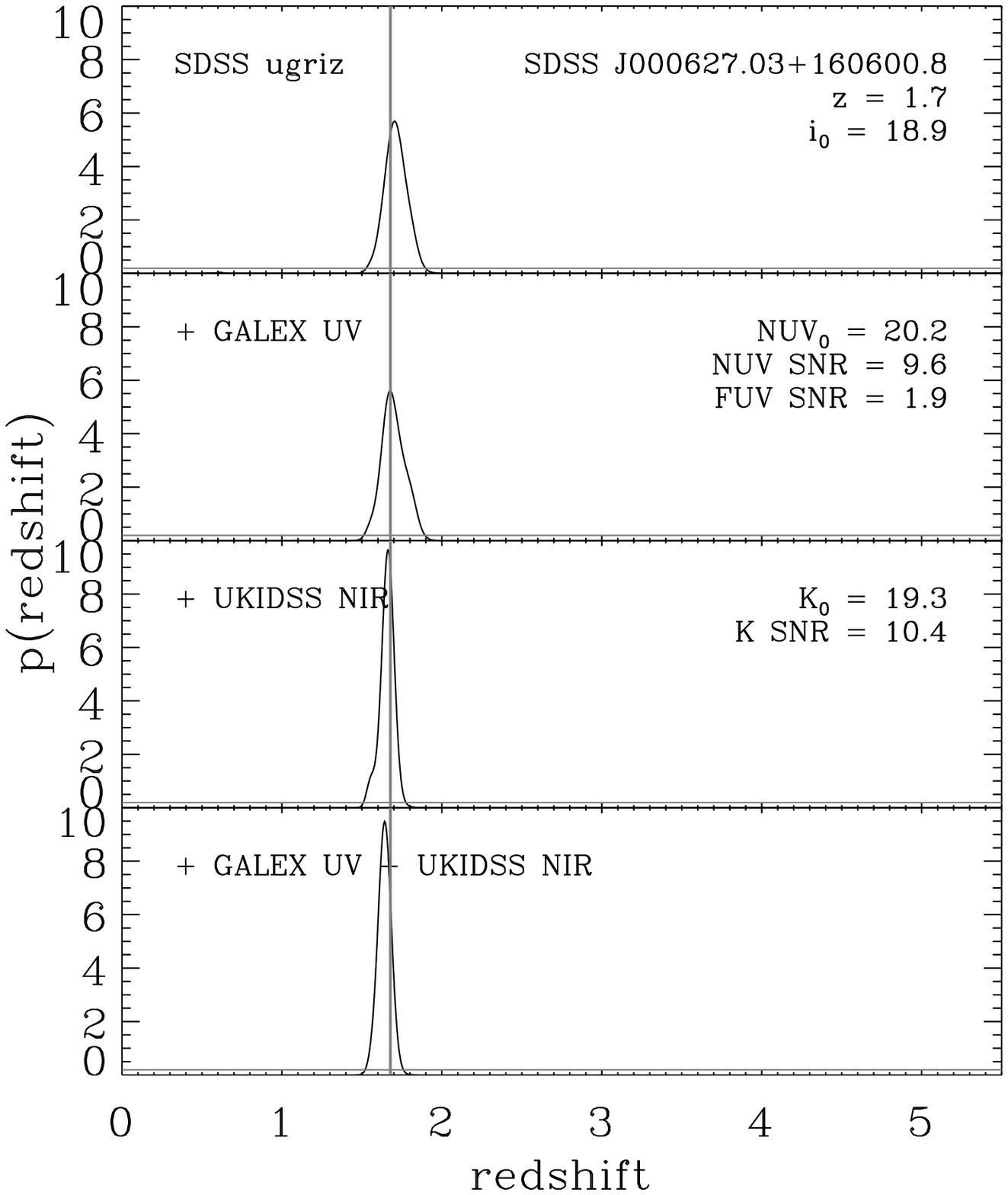}
\includegraphics[width=0.4\textwidth,clip=]{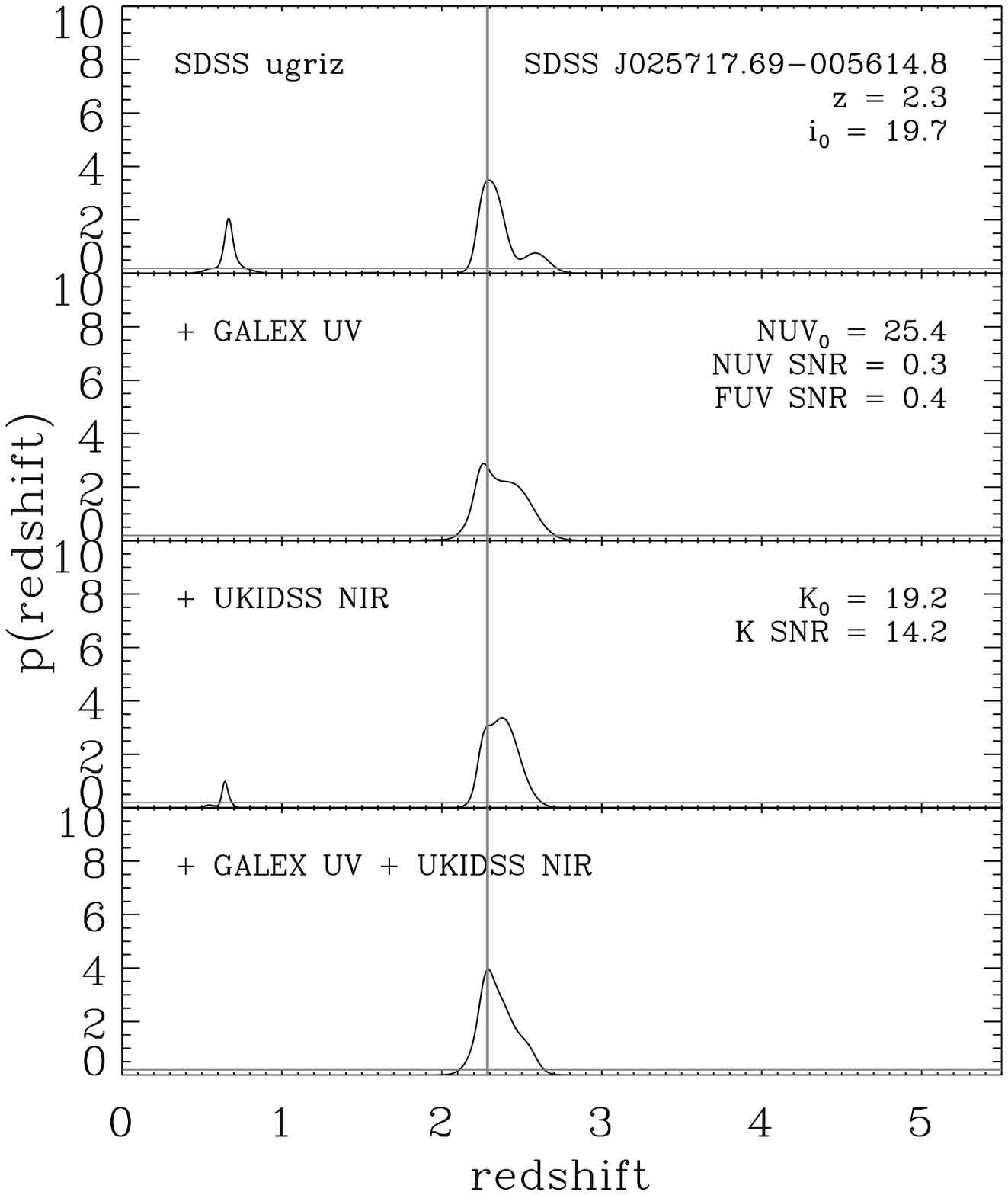}\\
\includegraphics[width=0.4\textwidth,clip=]{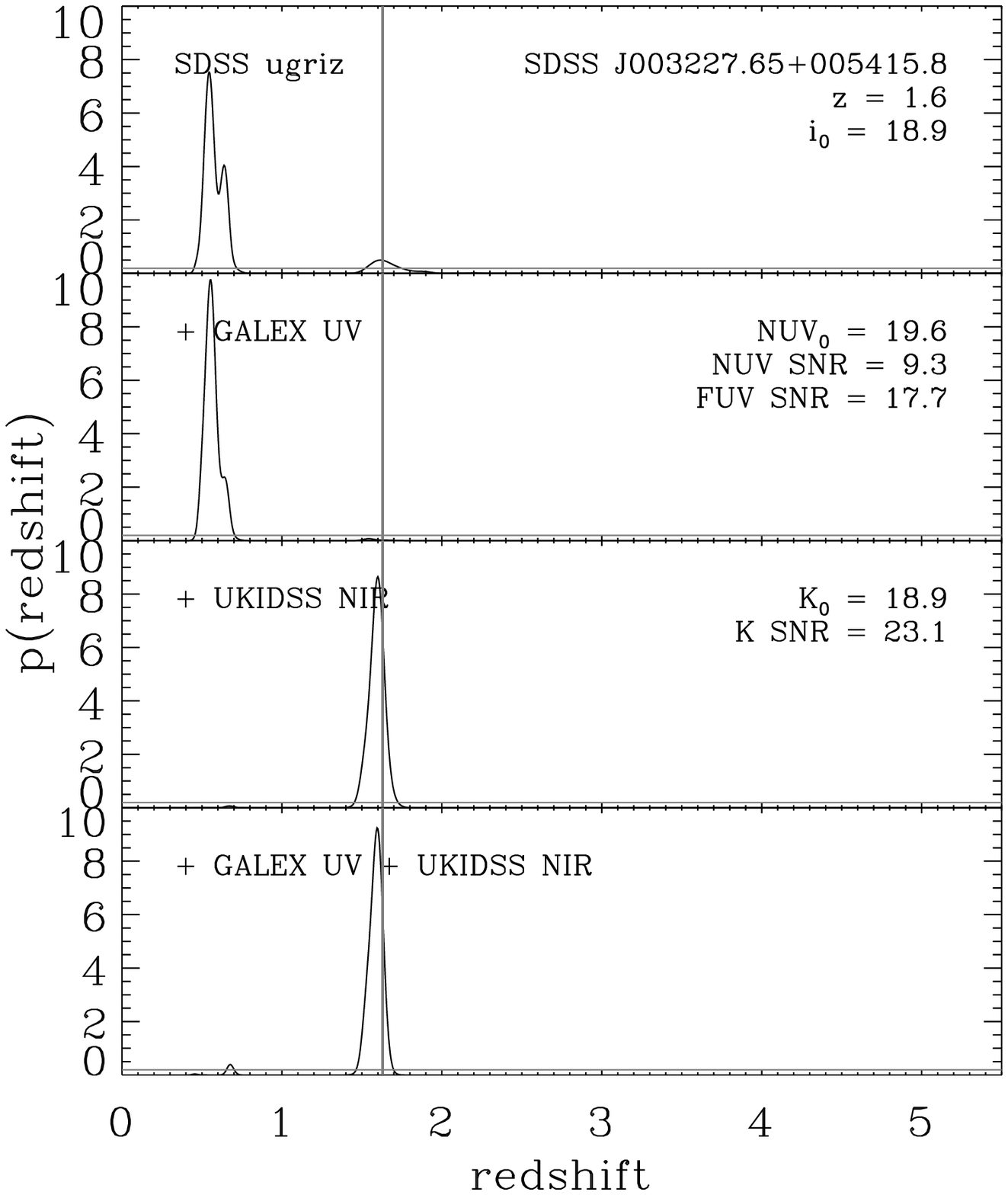}
\includegraphics[width=0.4\textwidth,clip=]{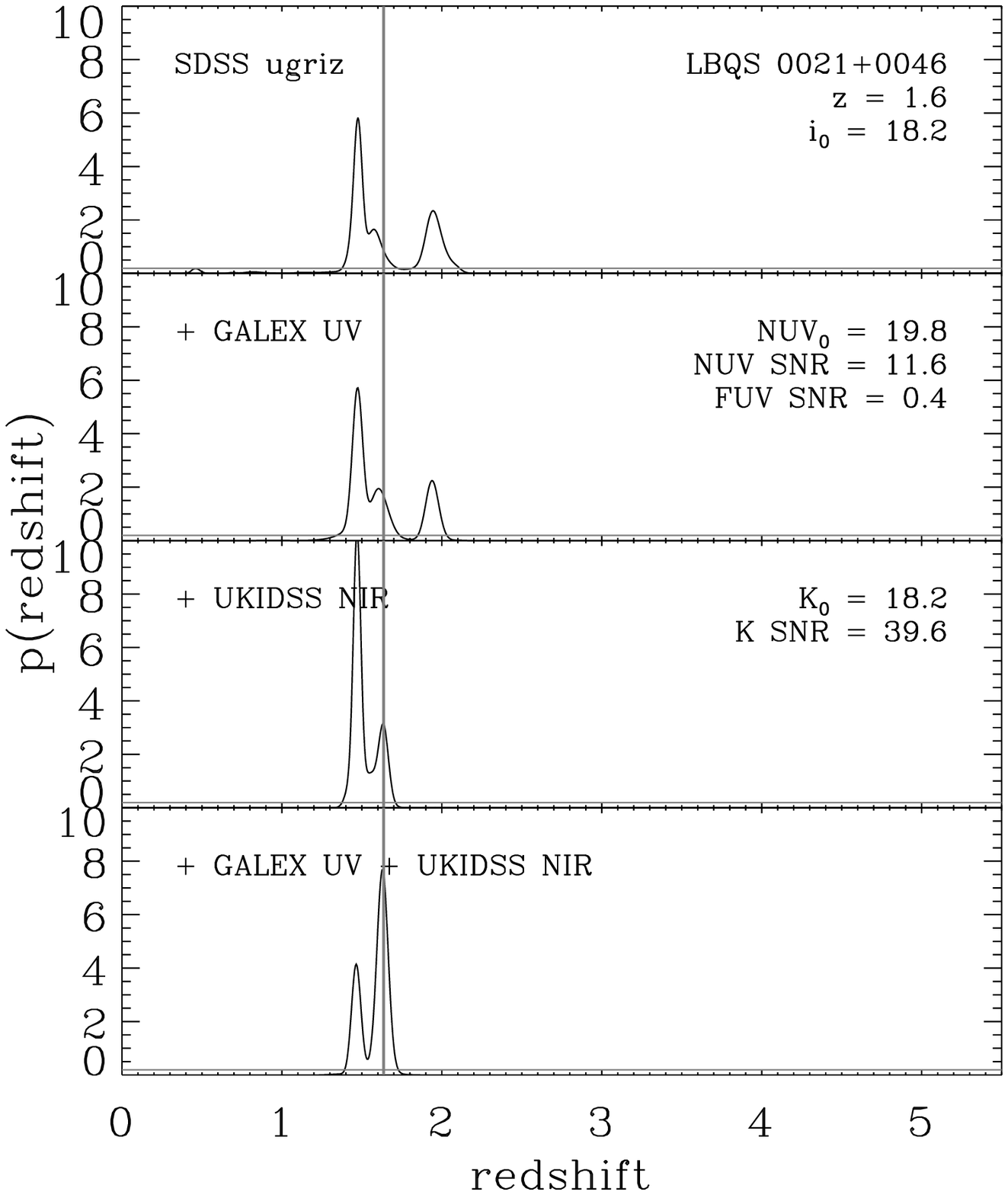}
\caption{Posterior distribution functions for the photometric redshift
  of four quasars from the \sdss\ DR7 quasar catalog. The top panel in each
  plot shows the redshift posterior distribution function based
  only on \ugriz\ fluxes; the lower panels add \uv\ (\nuv\ and \fuv)
  and \nir\ measurements (in \yjhk). The vertical line shows the
  spectroscopic redshift. The horizontal line represents the uniform
  distribution over $0.3 \leq z \leq 5.5$.}\label{fig:predictdr7qso}
\end{figure}

We show four examples of such redshift PDFs in
\figurename~\ref{fig:predictdr7qso} for objects from the \sdss\ DR7
quasar catalog that have measurements in all of the \uv\ and
\nir\ filters. These objects are from the sample used to train the
flux--redshift quasar model, but, as we discuss in more detail below,
we nevertheless believe that they provide an adequate representation
of the performance of the \xdqsoz\ technique. These objects are chosen
to demonstrate the power and weaknesses of the \ugriz, \uv, and
\nir\ data for photometric redshift estimation, and are therefore not
a random subset of the data. We discuss the overall performance below.

The top left panel shows an example where the \ugriz\ fluxes suffice
to accurately and precisely measure the redshift, and how the
(relatively high signal-to-noise ratio) \uv\ and \nir\ measurements
tighten the PDF significantly. The top right panel shows an example
where the extremely low \uv\ flux basically vetoes the low-redshift
peak that is present in the \ugriz-only redshift PDF. If one were to
use a simple non-detection \galex\ catalog at 5$\sigma$ this result
would not have been clear, because this object could have had the mean
$z \approx 0.8$ \uv\ flux and still not be detected by \galex. The
ability of the \xdqsoz\ technique to use and interpret low
signal-to-noise ratio data is therefore crucial in this example.

A weakness of the auxiliary \uv\ data is apparent in the lower left
panel. Here we see a $z = 1.6$ quasar that is much brighter than the
average quasar at this redshift in the \uv, such that the addition of
the \uv\ data mistakenly chooses the low-redshift peak of the
degenerate \ugriz\ redshift PDF. However, the \nir\ data are able to
overcome this error and the addition of all the data confidently
assigns this object a close-to-correct redshift. The lower right panel
of \figurename~\ref{fig:predictdr7qso} shows another amusing example.

In addition to testing the \xdqsoz\ technique using the \sdss\ DR7
quasar sample, we have also drawn a sample of quasars located in the
\sdss\ imaging in stripe 82 discovered as part of the \twoslaq\ survey
\citep{Croom09a} and \boss\ \citep{Ross11a,palanque11}. These quasars
are generally fainter than the \sdss\ quasars and therefore they
represent a stringent, independent test of the \xdqsoz\ technique's
ability to return accurate redshift PDFs at faint magnitudes. We have
specifically selected all $0.3 \leq z \leq 5.5$ quasars with
dereddened $i \geq 19.1$ mag from the \twoslaq\ sample and all quasars
on the \sdss\ imaging stripe 82 newly discovered by \boss\ and use
their single-pass \sdss\ photometry. Because most of these objects lie
in the \sdss\ equatorial stripe, many of them have measurements from
\galex\ and \ukidsslas.

\begin{figure}
\includegraphics[width=0.4\textwidth,clip=]{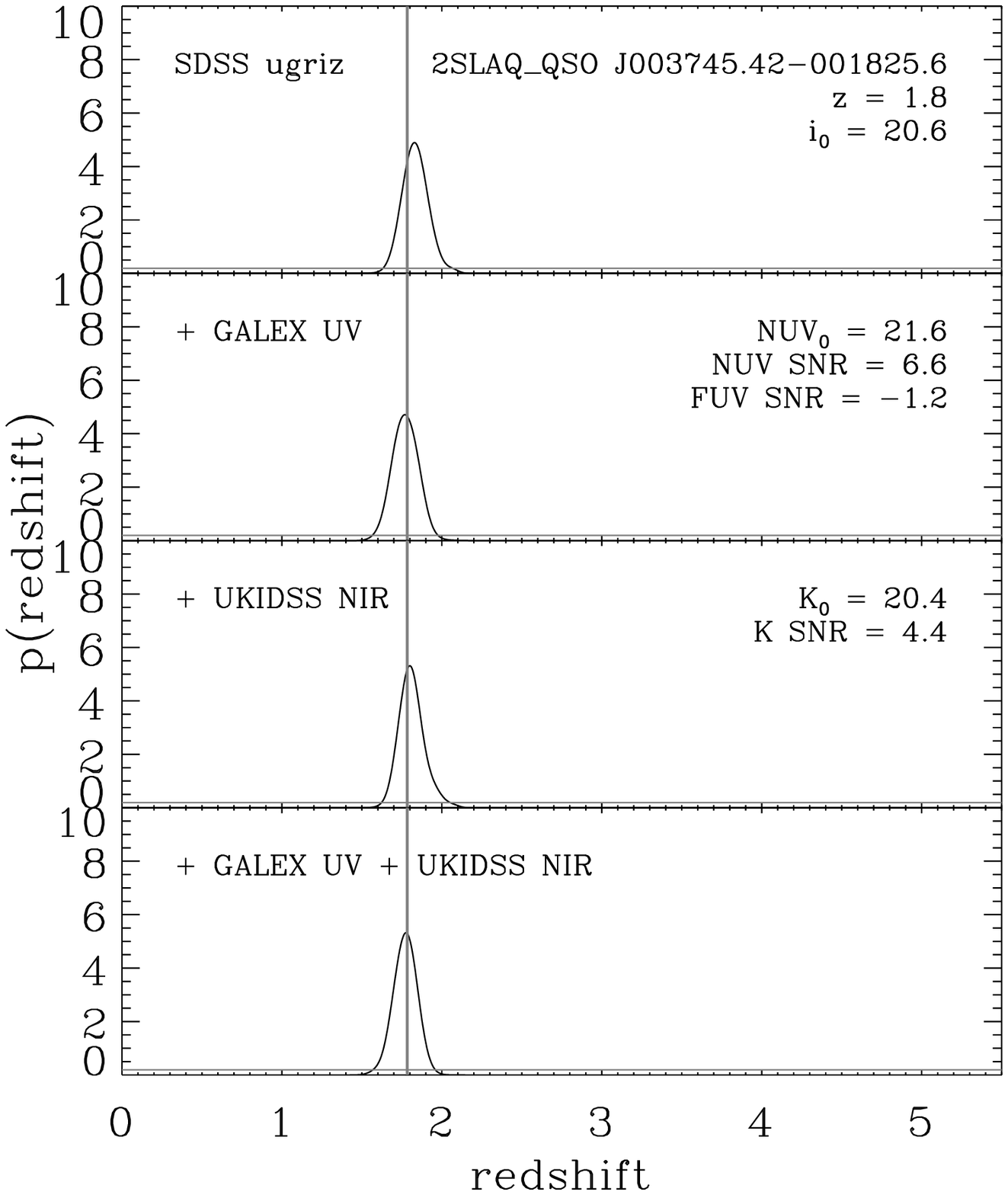}
\includegraphics[width=0.4\textwidth,clip=]{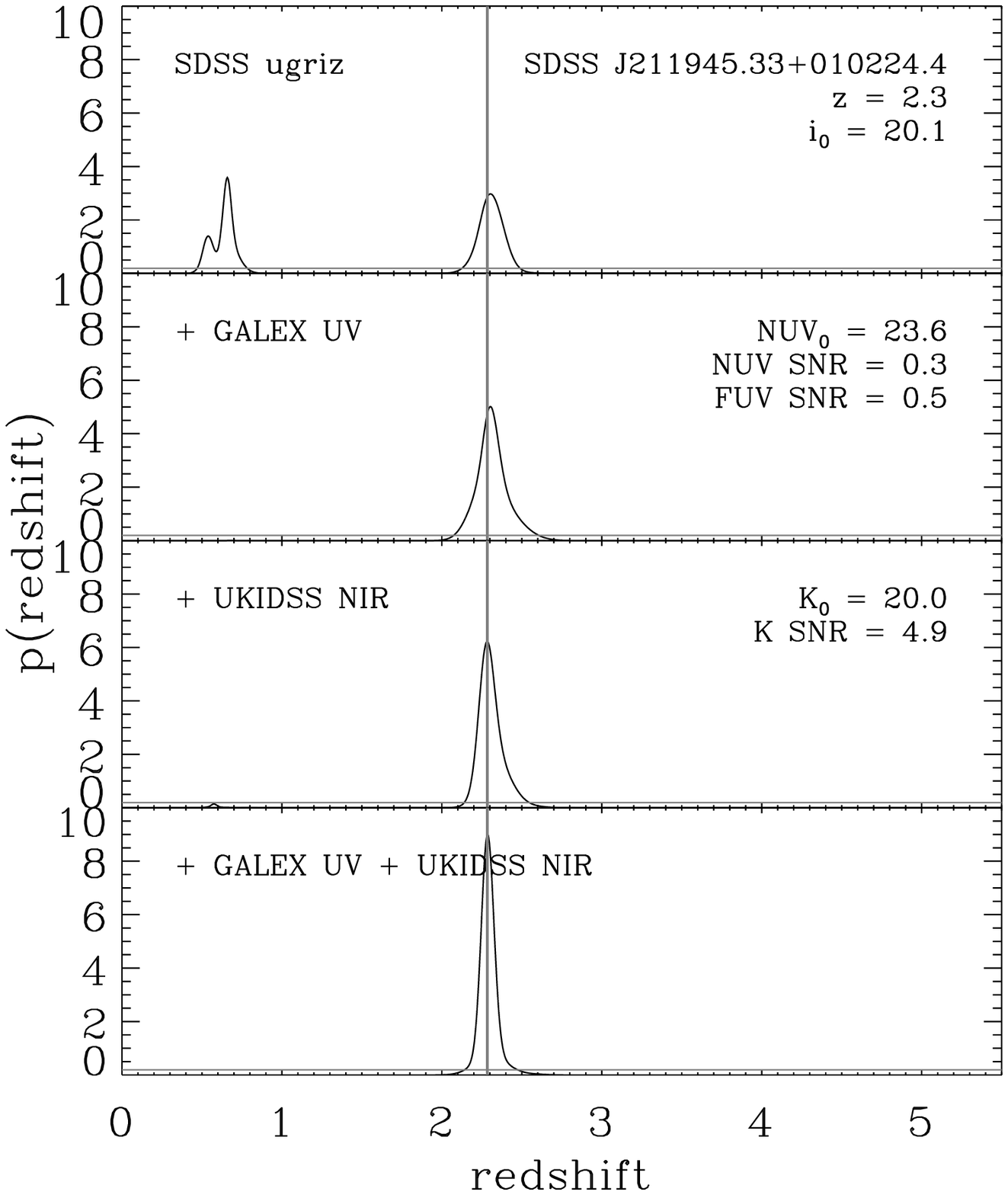}\\
\includegraphics[width=0.4\textwidth,clip=]{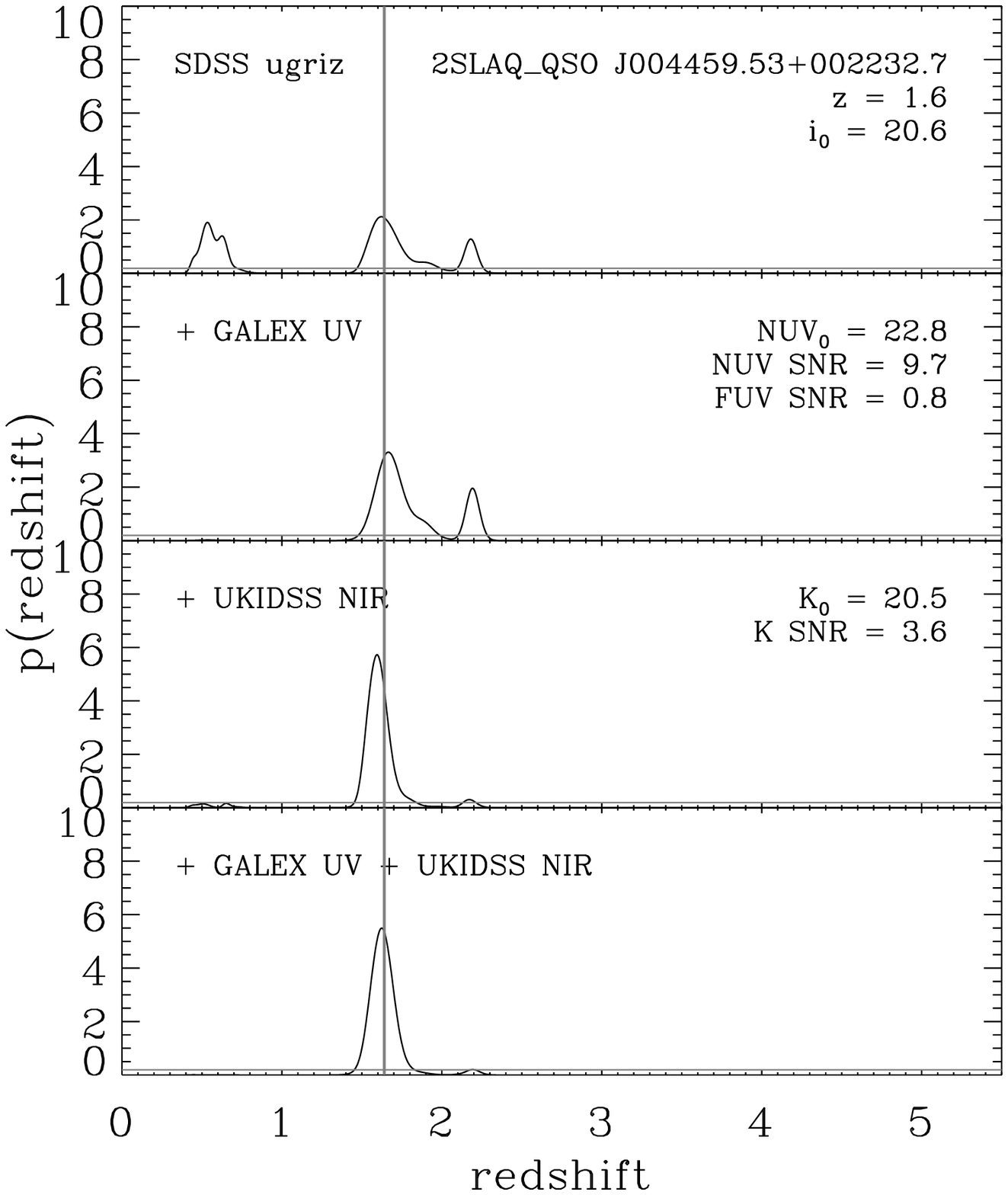}
\includegraphics[width=0.4\textwidth,clip=]{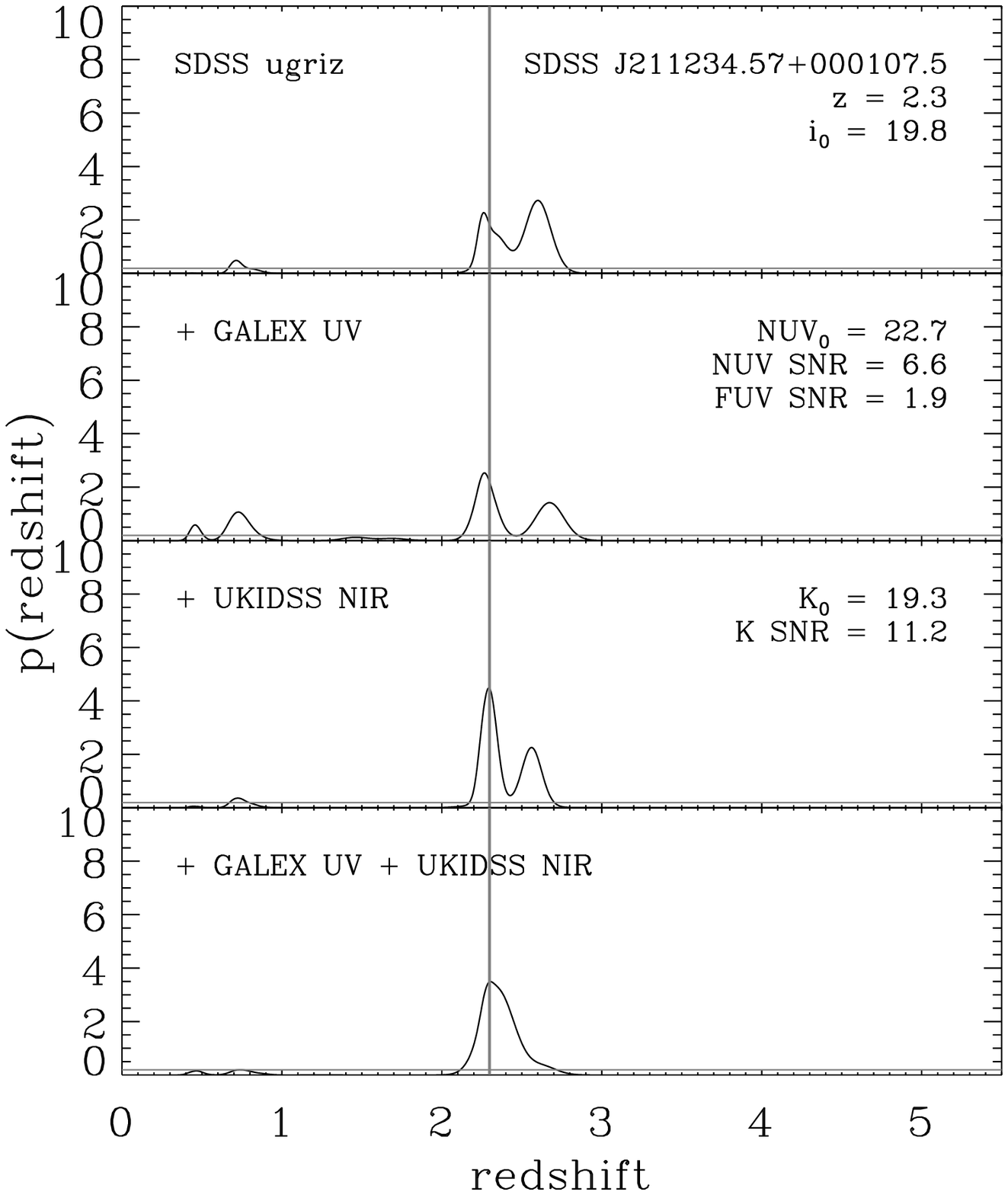}
\caption{Same as \figurename~\ref{fig:predictdr7qso} but for
  objects from the fainter test sample, composed of quasars from the
  \slaq\ survey and from the \boss\ in
  \sdss\ stripe 82.}\label{fig:predict2slaqboss}
\end{figure}

\figurename~\ref{fig:predict2slaqboss} shows posterior probability
distributions for the redshift of two objects from the
\twoslaq\ catalog and two from the \boss\ sample. The trends that were
apparent for the \sdss\ DR7 quasars in
\figurename~\ref{fig:predictdr7qso} are also evident for these fainter
objects. The \uv\ and \nir\ fluxes for these objects have, in general,
been measured much less precisely than those discussed above, but the
auxiliary data still provide valuable extra information about the
redshift.

While most of the examples in \figurename s~\ref{fig:predictdr7qso}
and \ref{fig:predict2slaqboss} have a considerable posterior
probability mass associated with the correct redshift, even when
multiple peaks are present in the redshift PDF, this situation is
generic---in that inspections of many redshift PDFs show that it is rare
to have no posterior probability mass associated with the
spectroscopic redshift.

\begin{figure}
\includegraphics[height=0.8\textheight,clip=]{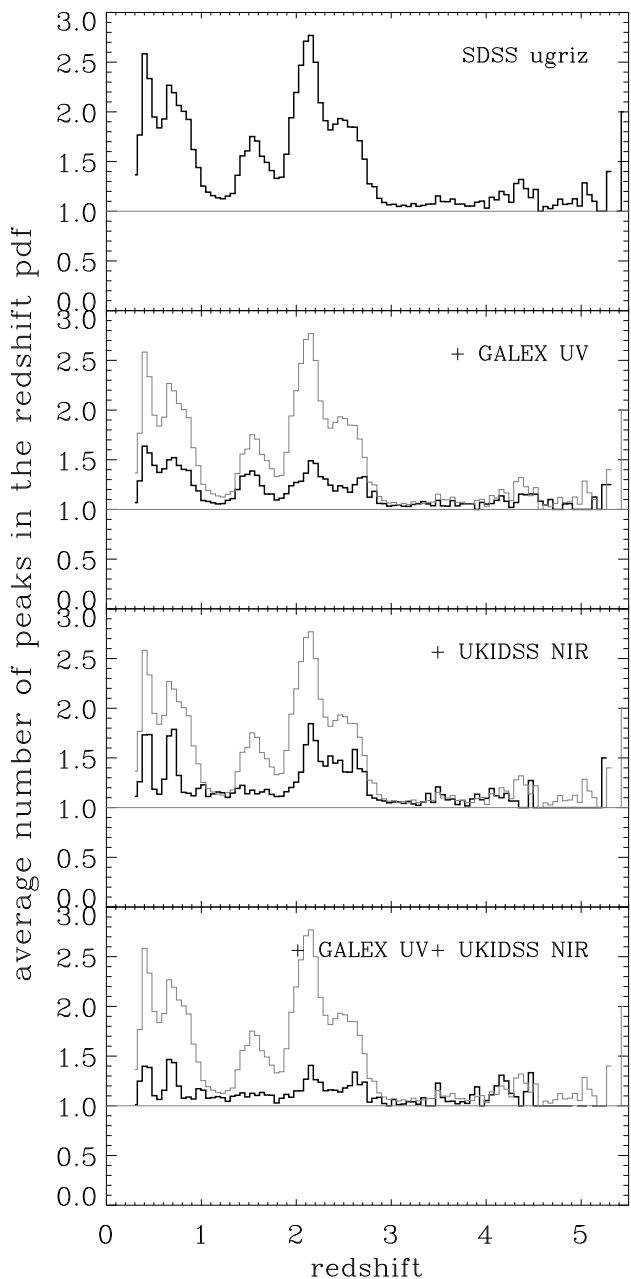}
\caption{Average number of peaks in the posterior distribution
  function for the photometric redshift as a function of spectroscopic
  redshift for the \sdss\ DR7 quasar sample. A peak is defined as a
  contiguous region where the posterior distribution exceeds the
  uniform distribution on 0.3 $\leq z \leq 5.5$.  The top
  panel uses photometric redshift predictions from only \ugriz\ data;
  the lower panels add \uv\ and \nir\ data. The optical-only curve is
  repeated in the lower panels in gray.}\label{fig:peaks}
\end{figure}

As a simple statistic for the degeneracy in the redshift PDF we
examine the number of distinct peaks as a function of redshift. A
single peak in the PDF is defined here as the widest contiguous region
where the PDF is above the uniform distribution between redshift 0.3
and 5.5 (\ie, flat in redshift). The top panel of
\figurename~\ref{fig:peaks} shows the average number of such peaks as
a function of redshift. This statistic clearly shows the main
degeneracies of \ugriz-based photometric quasar redshifts. Basically
the entire $z < 1$ region, a region around $z = 1.5$, and the $ 2.0
\leq z \leq 2.7$ redshift range are degenerate. Higher redshift
quasars are readily identified as such using the \ugriz\ colors
\citep[\eg,][]{Fan99b}. From the lower panels we see that the addition
of \uv\ and \nir\ data softens all of these degeneracies. Essentially
no degeneracies remain using the combination of all the \uv, optical,
and \nir\ data (lower panel of
\figurename~\ref{fig:peaks}). Additionally, requiring that distinct
peaks in the photometric-redshift PDF need to have a minimum
integrated probability (\eg, defining a peak as a contiguous region
above the uniform distribution with $> 0.05$ integrated redshift
probability) gives similar results for the number of peaks and the
improvement when adding \uv\ and \nir\ data.

\begin{figure}
\includegraphics[height=0.8\textheight,clip=]{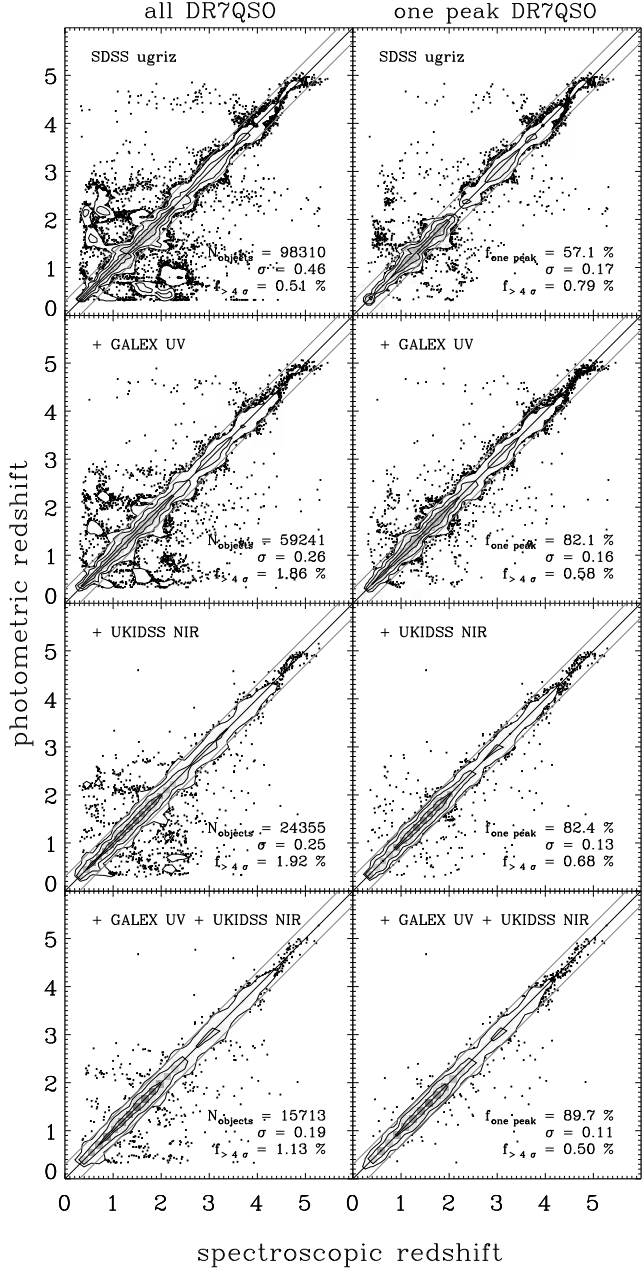}
\caption{Spectroscopic versus photometric redshift for quasars from
  the \sdss\ DR7 quasar catalog. Photometric redshifts are maximum
  {\em a posteriori} redshifts, \ie, they are at the peak of the photometric
  redshift posterior distribution function. The left column shows all
  sources; the right column shows sources that have only a single peak
  in their photometric redshift posterior distribution, that is, they
  have only one contiguous region in their posterior distribution
  where the distribution exceeds the uniform distribution on 0.3
  $\leq z \leq$ 5.5. The top row shows predictions based only
  on \ugriz\ fluxes, the lower panels add \uv\ and \nir\ information,
  restricted to those objects that were observed in both \nuv\ and
  \fuv\ for \galex, and in all four \yjhk\ \ukidss\ filters. The
  one-to-one line is shown in black and the $|\Delta z| = 0.3$ lines
  are shown in gray.}\label{fig:zspeczphotdr7qso}
\end{figure}

\figurename~\ref{fig:zspeczphotdr7qso} shows the traditional
spectroscopic-redshift vs. photometric-redshift diagram for quasars in
the \sdss\ DR7 quasar sample, for various combinations of wavelength
regimes. The right panels restrict the sample to those objects for
which the redshift PDF has only a single peak and as such can be
accurately described by a single photometric redshift (plus
uncertainty). In the top left panel all of the \ugriz-related
degeneracies are clearly present and the right panel shows that by
restricting the sample to single-peaked PDFs most of these
degeneracies vanish, albeit at the cost of entire redshift
ranges---most notably redshift-range $2.0 \lesssim z \lesssim 2.5$
quasars. The addition of \uv\ and especially that of
\nir\ observations a) greatly reduces the degeneracies as witnessed by
the diminishing structure in the spectroscopic vs. photometric
redshift plane and the increasing fraction of objects with a single
peaked redshift PDF, and b) significantly reduces the scatter. In the
individual panels we report the number of 4$\sigma$ outliers rather
than the number of $|\Delta z| > 0.3$ objects; the latter number is
somewhat meaningless without comparing it to the scatter, but to guide
the eye we have included the $|\Delta z|=0.3$ lines. The scatter is
calculated without outlier-rejection. We note that---here and in the
test below---the distribution of the \iband\ magnitude is unchanged when
restricting the sample to objects with measured \galex\ or
\ukidss\ fluxes; restricting to objects with \nir\ fluxes actually
creates a fainter sample, because many faint quasars in the
\sdss\ imaging stripe 82 have been observed by \ukidsslas, while many
brighter quasars are located outside of the \ukidsslas\ footprint.

With the addition of \uv\ and \nir\ data most objects have accurate
and precise single-peaked photometric redshifts over the entire $0.3
\leq z \leq 5.5$ redshift range: 97\,percent of all objects with
\uv\ and \nir\ data and 99\,percent of the subset with single-peaked
redshift PDFS have photometric redshifts within $|\Delta z| < 0.3$;
for $|\Delta z| < 0.1$ these numbers are 84\,percent and 86\,percent
respectively. This is a significant improvement over \ugriz-only
photometric redshifts, where we find 86\,percent of objects within
$|\Delta z| < 0.3$. Similarly, \citet{Weinstein04a} found 83\,percent
of objects in this range.

Photometric and spectroscopic redshifts in
\figurename~\ref{fig:zspeczphotdr7qso} are compared for objects in the
sample used to train the \xdqsoz\ technique. As such, one might object
that this is not a fair representation of the performance of the
\xdqsoz\ technique. But, the relationship between the photometrically
estimated redshift PDF and the spectroscopic redshift of a training
object for \xdqsoz\ is through the many-parameter flux--redshift
density model. This model includes reweighting objects in the training
set according to a redshift prior. There is therefore no direct
connection between output photometric redshifts and input
spectroscopic redshifts--- as there is, for example, in
nearest-neighbor approaches to photometric redshift estimation
(\citealt{Ball07a,Ball08a}). The fact that
\figurename~\ref{fig:zspeczphotdr7qso} contains all of the expected
redshift degeneracies for \ugriz-based photometric redshifts is
further proof of this independence: if there were a dependent
connection we would not suffer from these degeneracies.

\begin{figure}
\includegraphics[height=0.8\textheight,clip=]{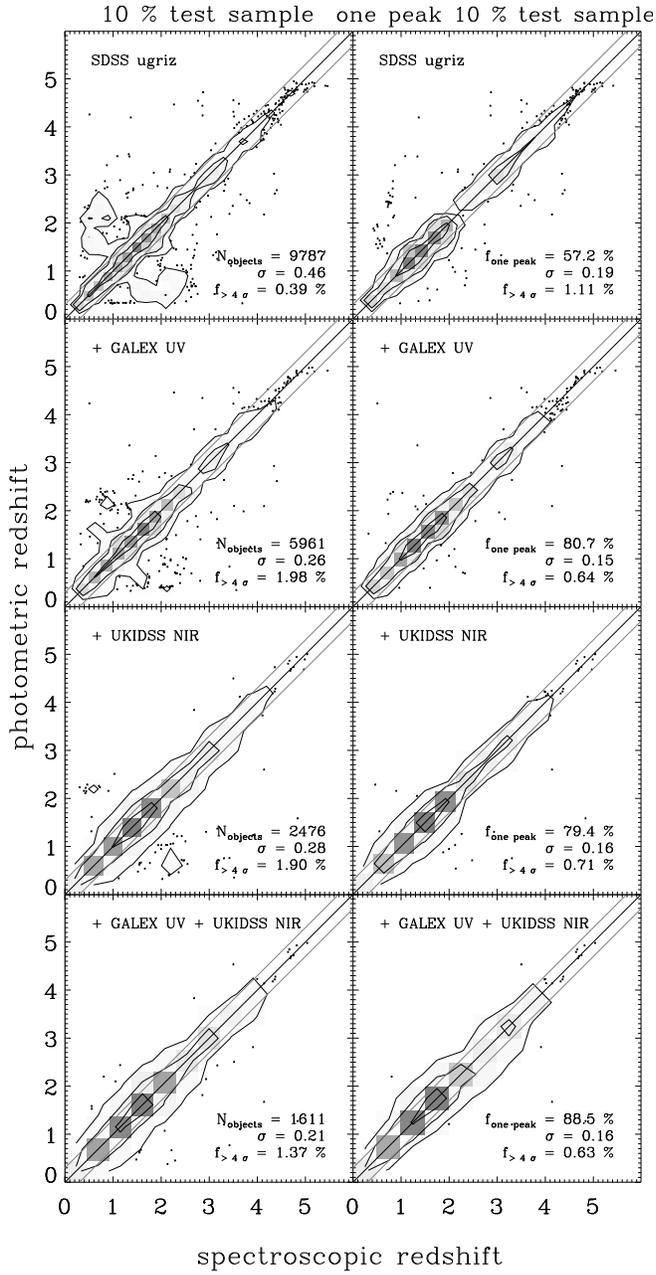}
\caption{Same as \figurename~\ref{fig:zspeczphotdr7qso}, but
  for a random sample of 10\,percent of objects from the \sdss\ DR7
  quasar catalog, with the model trained on the remaining 90\,percent
  of quasars.}\label{fig:zspeczphottestqso}
\end{figure}

To further test this issue we have divided our sample into a
90\,percent training sample and a 10\,percent test sample, as
described above in \sectionname~\ref{sec:modelconstruction}. We redo
the spectroscopic-redshift vs.\ photometric-redshift comparison for the
10\,percent sample using the model trained in the 90\,percent of
remaining data---the results are in
\figurename~\ref{fig:zspeczphottestqso}. Because the 10\,percent sample
is much smaller than the full \sdss\ DR7 quasar sample the statistics
are noisier, but the trends are the same as in
\figurename~\ref{fig:zspeczphotdr7qso}.

\begin{figure}
\begin{center}
\includegraphics[width=1.\textwidth,clip=]{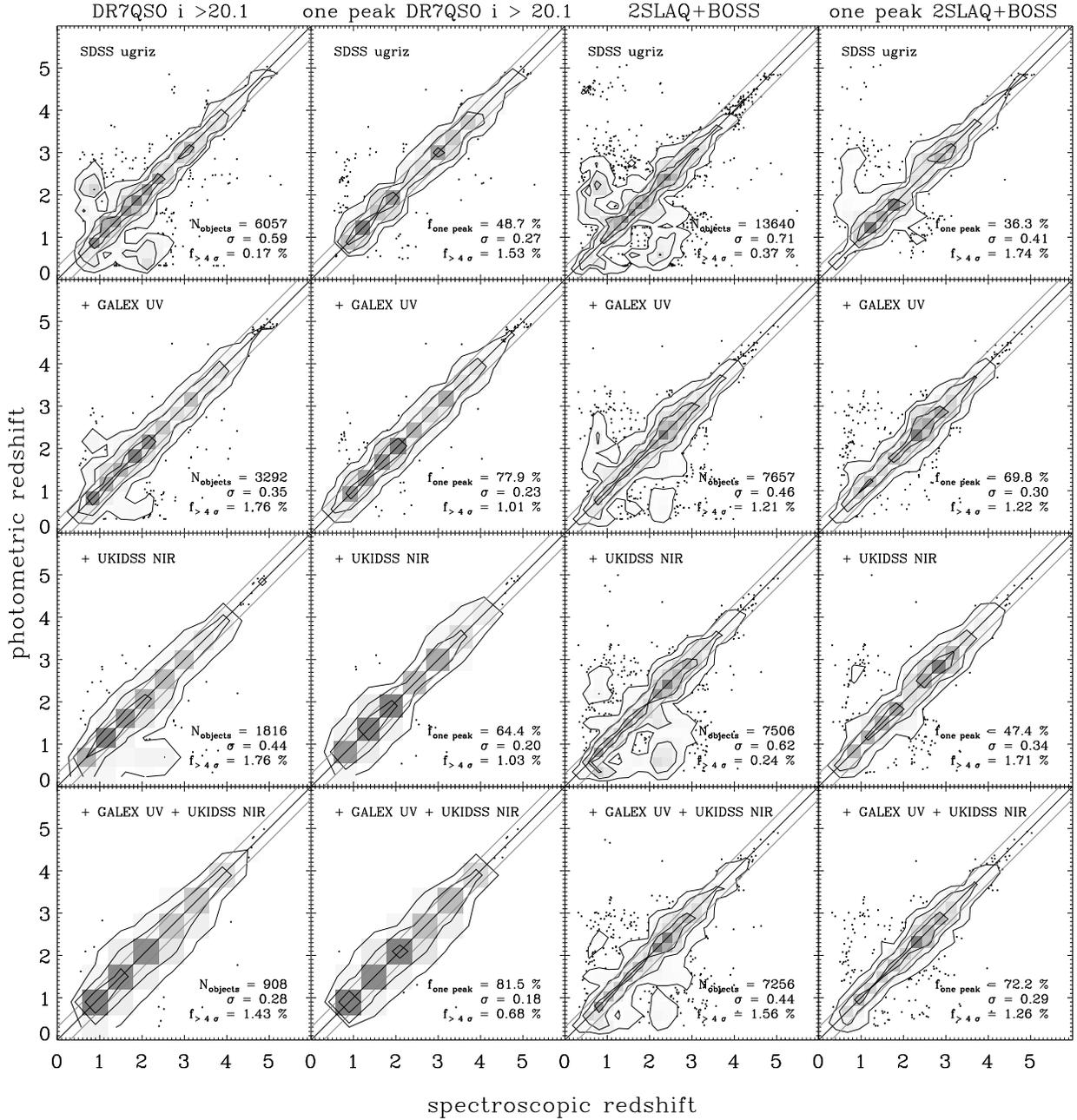}
\caption{Spectroscopic versus photometric redshifts for the $i > 20.1$
  subset of the \sdss\ DR7 quasar catalog as well as for the faint
  test sample composed of quasars from the \slaq\ survey and the
  \boss. The two columns on the left are as for
  \figurename~\ref{fig:zspeczphotdr7qso}, but restricted to those
  objects with $i > 20.1$ mag.  The two columns on the right are as for
  the leftmost columns, but for the \slaq\ + \boss\ test
  sample.}\label{fig:zspeczphot2slaqboss}
\end{center}
\end{figure}

To test the \xdqsoz\ technique at fainter magnitudes, in
\figurename~\ref{fig:zspeczphot2slaqboss} we compare spectroscopic
redshifts to photometric redshifts for $i > 20.1$ objects in the
\sdss\ DR7 quasar catalog and for objects in the combined
\twoslaq\ and \boss\ sample. The trends in this figure are the
same as those for the brighter \sdss\ quasar sample and the scatter is
somewhat larger; however, the photometric redshifts remain clustered
around the spectroscopic redshifts with no discernible bias. Even for
the faint \twoslaq\ and \boss\ sample, the addition of low
signal-to-noise ratio \uv\ and \nir\ data leads to a significant
increase in accuracy and precision.

\begin{figure}
\includegraphics[width=0.7\textwidth,clip=]{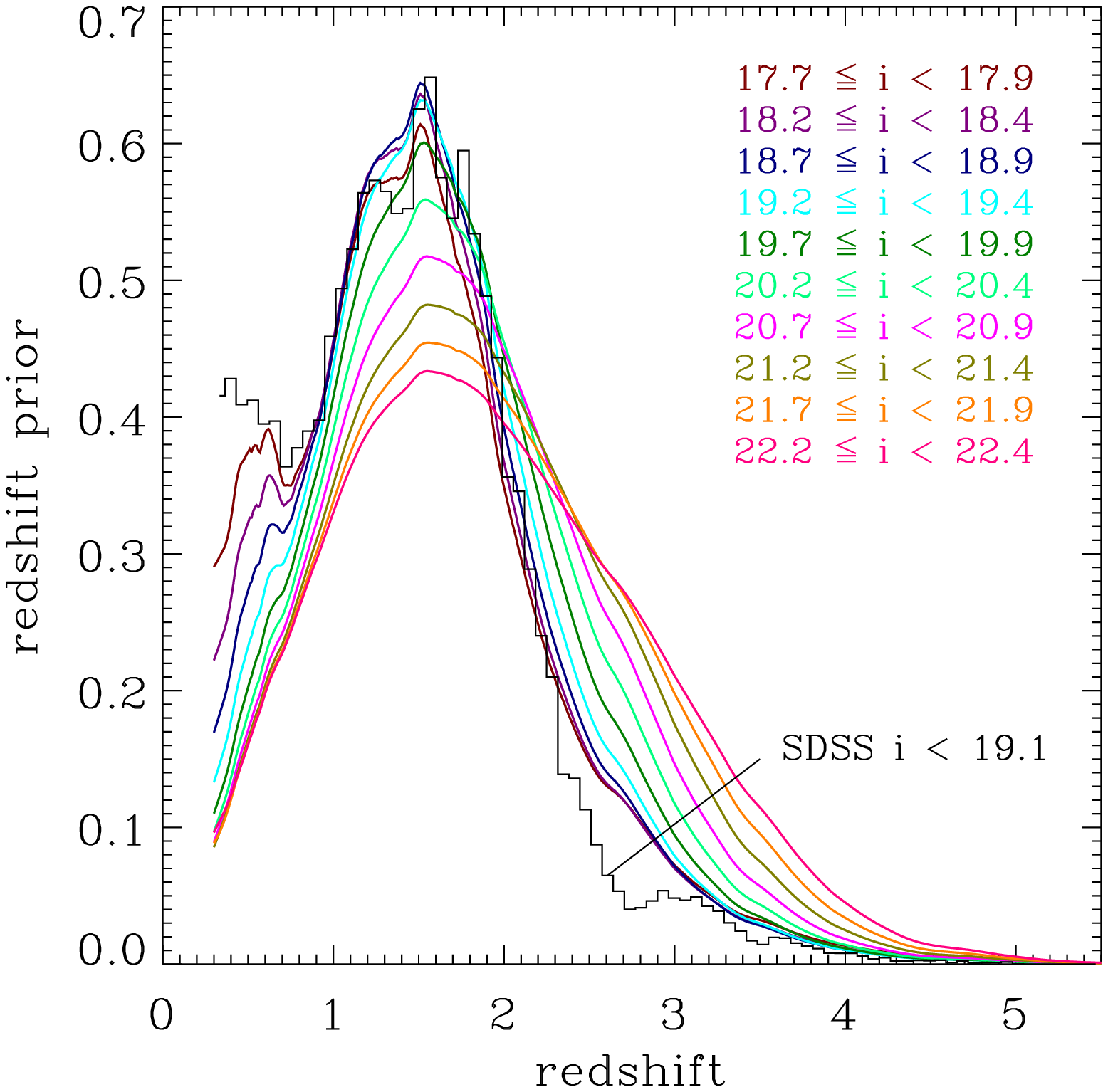}\\
\includegraphics[width=0.7\textwidth,clip=]{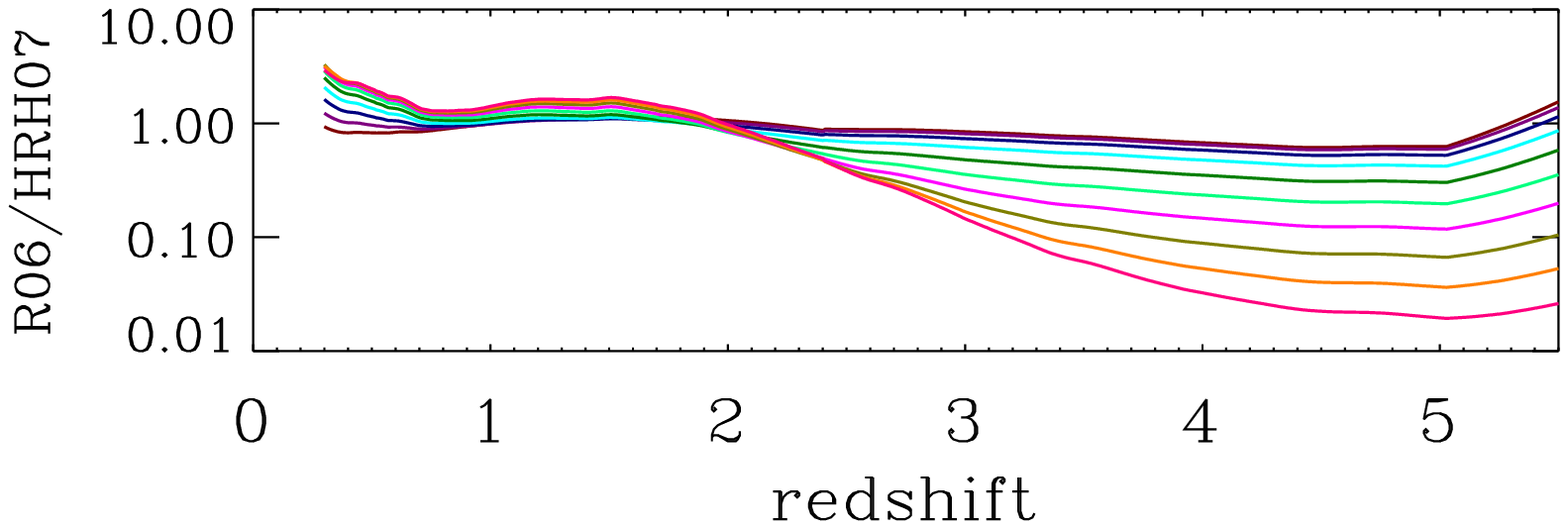}
\caption{Prior distribution for the redshift in a few \iband\ bins
  (\emph{top panel}). The histogram shows the redshift distribution of
  69,994 quasars from the \sdss\ DR7 quasar catalog with dereddened
  \iband\ magnitude $< 19.1$, where the quasar catalog is highly
  complete (except for the redshift range 2.5 $\leq z \leq 3.2$). The
  bottom panel shows the difference in prior when using the
  \citeauthor{Richards06a} (\citeyear{Richards06a}; hereafter R06)
  quasar luminosity function rather than the fiducial
  \citeauthor{Hopkins07a} (\citeyear{Hopkins07a}; hereafter HRH07)
  model.}\label{fig:redshiftprior}
\end{figure}

\begin{figure}
\includegraphics[height=0.7\textwidth]{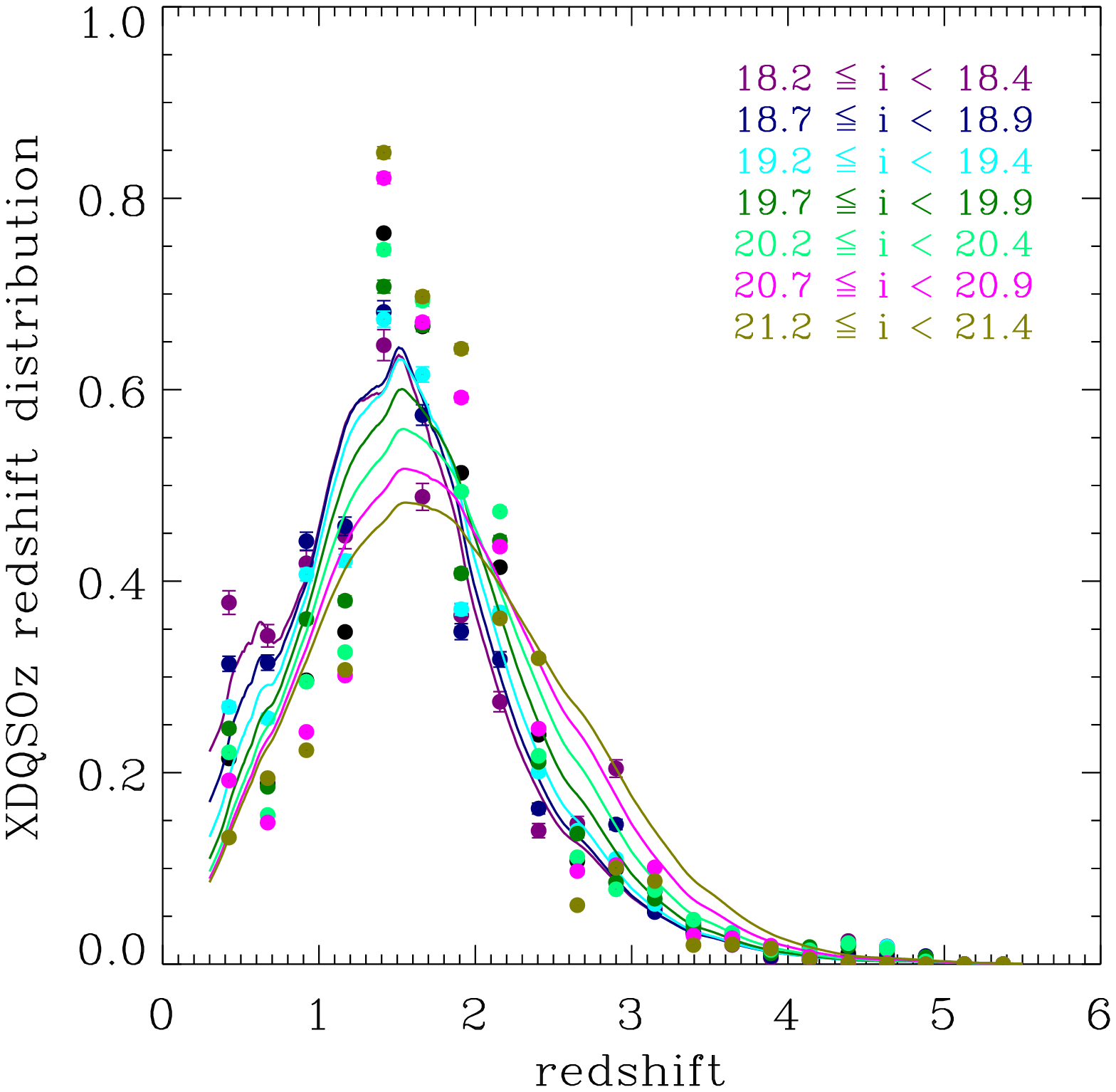}
\caption{Distribution of the peak of the photometric redshift
  distribution for all objects in the expected \sdssiii\ \boss\ survey
  in a few apparent-magnitude ranges. The curves are the redshift
  priors calculated from the \citet{Hopkins07a} luminosity-function
  model. The color coding is the same as in
  \figurename~\ref{fig:redshiftprior}. The overall shape of the
  redshift distribution is similar to the prior distribution, except
  for the drop in $2.5 \leq z \leq 3.5$ due to the decreased
  efficiency of photometric quasar classification based on \sdss\
  photometry.}\label{fig:xdqsoz_redshift}
\end{figure}

Using the technique described in this section we have computed
photometric redshifts for all point sources in the expected \boss\
spectroscopic footprint with \xdqsoz\ quasar probabilities larger than
0.5 and $17.8 \leq i \leq 21.5$ mag. The distribution of peaks of the
photometric-redshift distribution for these objects is shown in
\figurename~\ref{fig:xdqsoz_redshift} in a few apparent-magnitude bins
(those bins from \figurename~\ref{fig:redshiftprior} that lie within
the $17.8 \leq i \leq 21.5$ apparent-magnitude range. The overall
shape of the redshift distribution in each $i$-band bin is similar to
the redshift prior calculated from the \citet{Hopkins07a}
luminosity-function model. However, the redshift-dependent efficiency
of photometric quasar classification and redshift estimation is
apparent in this comparison and the low classification efficiency at
$2.5 \lesssim z \lesssim 3.5$ depresses the distribution in that range
while increasing the significance of the $z \approx 1.5$ peak.

All of the results in this section have assumed the \citet{Hopkins07a}
redshift prior, shown in \figurename~\ref{fig:redshiftprior}. Using
the difference between the Hopkins, Richards, \& Hernquist (2007) and
\citet{Richards06a} redshift prior, given in the bottom panel of
\figurename~\ref{fig:redshiftprior}, we can assess the difference in
photometric redshift distribution when using these two alternatives to
the quasar luminosity function. The bottom panel of
\figurename~\ref{fig:redshiftprior} shows that the only significant
difference between these two models is at relatively high redshift ($z
\gtrsim 2.5$) and near the \sdss\ detection limit ($i \gtrsim 21$
mag). Strongly single-peaked photometric redshift distribution
functions, such as many of those shown in \figurename
s~\ref{fig:predictdr7qso} and \ref{fig:predict2slaqboss}, are not
affected by even order-of-magnitude changes in the redshift prior,
especially when \uv\ or \nir\ data are available. It is clear from
\figurename s~\ref{fig:zspeczphotdr7qso} and
\ref{fig:zspeczphot2slaqboss} that, on average, the influence of a
different redshift prior will be limited, as the main differences lie
at higher redshift, where the \sdss\ colors provide relatively
unambiguous photometric redshifts (as shown by the lack of
degeneracies at higher redshift in the photometric versus
spectroscopic redshift plane). As the photometric redshift
distributions are only marginally affected by the use of a different
prior, classification based on integration over these redshift PDFs
also does not depend strongly on the details of the redshift prior.

\section{Discussion}\label{sec:discussion}

\subsection{Comparison with other methods}\label{sec:compare}

We have previously discussed other photometric redshift estimation
techniques for quasars in \sectionname~\ref{sec:general}. Comparing
\figurename~\ref{fig:zspeczphotdr7qso} to similar diagrams in
\citet{Budavari01a,Richards01a,Ball07a,Ball08a} we see, at least
qualitatively, that the \xdqsoz\ technique performs similarly when
applied to the \ugriz\ fluxes of bright, high signal-to-noise ratio
objects. We did not expect to perform better as the near-degeneracies
in the \ugriz\ color--redshift plane are real and the quasar locus is
broad. The advantage of the \xdqsoz\ technique over these other
techniques is that it can be applied to faint objects and that it can
incorporate \uv\ and \nir\ observations, even at low signal-to-noise
ratio, to improve photometric redshift estimation and quasar
classification.

No other method exists to calculate photometric quasar probabilities
over arbitrary redshift ranges. By comparing with state-of-the-art
photometric quasar classification using kernel-density estimation or
Gaussian mixture density deconvolution \citep{Richards04a,Bovy11a}, we
have shown that the photometric quasar probabilities obtained by
integrating the photometric redshift PDF over redshift are as good as
those trained on the redshift range in question.

\subsection{Including additional information}

Two additional sources of information relevant to photometric quasar
classification and redshift estimation stand out as the next steps
toward a full quasar model, although neither of these is currently
available over the large areas of the sky surveyed by projects such as
the \sdss: photometric variability and
differential-chromatic-refraction-induced astrometric offsets for
quasars. Of these, photometric variability is the easiest to include
in the quasar classification technique discussed here, as we can
ignore the redshift information contained in the variability, because
this information seems to be limited \citep{Macleod11b}. As such, a
photometric variability likelihood for quasars and stars could be
multiplied with the flux--redshift likelihood employed and modeled
here to perform simultaneous color and variability selection. The
combination of photometric variability and color information will lead
to accurate photometric quasar classification and redshift estimation
in the \lsst\ era.

The strong spectral features of quasars induce positional offsets to
standard differential-chromatic-refraction corrections
\citep{Kaczmarczik09a}. These positional offsets are
redshift dependent much as quasar colors are redshift dependent
because of spectral features moving through individual filters (see,
\eg, \figurename~\ref{fig:allzexfit}). Thus, these offsets could be
used to break redshift degeneracies. Accomplishing this in the
\xdqsoz\ flux--redshift density context necessitates adding the
astrometric offsets into the density model. Because the astrometric
offsets are zenith angle dependent, these models would have to be
constructed for a range of airmasses, or airmass could be added as an
additional dimension. As astrometric redshifts are a subtle and
difficult-to-measure effect, the deconvolution aspect of the
\exd\ density-estimation technique could be useful.

\subsection{Generalized photometric object classification and characterization}

The technique described in this \documentname\ is a step toward a
generalized method for object classification and characterization from
broadband photometric data, which will become increasingly relevant in
this era of major wide-field imaging surveys. While our quasar model
includes redshift in addition to the broadband fluxes of an object,
our star model does not because stars do not possess a cosmological
redshift. Stars are characterized by other properties---\eg, distance
and metallicity---that are often estimated photometrically
\citep[\eg,][]{Juric08a,Ivezic08b}. As we are interested here in
quasar classification and characterization, our model implicitly
marginalized over stellar properties. However, as part of a general
object classification pipeline, these properties should be 
included---and the technique developed in this paper could be applied.

More importantly, the general framework outlined in
\sectionname~\ref{sec:general} and the specific implementation in
\sectionname s~\ref{sec:xdqsoz} and \ref{sec:targeting} show that we
can perform classification when different models are characterized by
different parameters---even different \emph{numbers} of
parameters. This aspect is especially relevant in the context of
quasar selection based on variability. Quasar variability is commonly
modeled as a stochastic Gaussian Process \citep{Kelly09a,Kozlowski10a}
characterized by a small number of parameters. Recently it has been
shown that this framework allows for a clean selection of quasars
because most stars---the main contaminants for quasar targeting
currently---in general do not vary over long time baselines. In this type of selection,
however, quasars and stars are often modeled (or fit) using the same
stochastic model, which is inappropriate for the stars
(\citealt{Schmidt10a,Macleod11a}; however, see
\citealt{Butler10a}). 

The use of a stochastic model for variability-based star--quasar separation
is particularly problematic for RR Lyrae stars---a common contaminant in
color-based classification of quasars in some redshift ranges. RR Lyraes are known
to vary periodically rather than stochastically. In the framework we use in this
\documentname\ all classes of objects can be described using models
appropriate for the class---\eg, stochastically varying objects with a
cosmological redshift for the quasars and non-variable sources for
most stars---because object classification only uses
\emph{marginalized} probabilities, that is, probabilities marginalized
over the internal properties of each class
(cf.~\eqnname~[\ref{eq:quasardeltazprob}]). Describing each class with
a model appropriate for that class should lead to better
classification and simultaneous object classification of sources into
\emph{all} classes.

As photometric quasar classification moves to ever fainter flux
levels, contamination from point-like galaxies becomes increasingly
important. As discussed by \citet{Bovy11a}, unresolved galaxies are
implicitly taken into account in our model because our training set
of ``stars'' is actually a set of non-variable point-like objects that
therefore includes faint galaxies. This model of galaxies again
implicitly marginalizes over galaxy properties, most notably the
redshift of the galaxy. Photometric redshift estimation for galaxies
is closely related to obtaining photometric redshifts for quasars, but
the galaxy photometric redshift techniques tend to rely more on
templates in their model building, while quasars are modeled in a more
empirical manner \citep[\eg,][]{Benitez00a}. However, this distinction
is not fundamental and the general framework discussed here still
applies. Template-based models are just another way of obtaining the
probability $p(\mathrm{fluxes}|\mathrm{galaxy})$ or
$p(\mathrm{fluxes},z|\mathrm{galaxy})$.

\subsection{Quasar tracks}\label{sec:track}

The \xdqsoz\ model of \sectionname~\ref{sec:xdqsoz} also contains the
distribution of broadband fluxes as a function of redshift
$p(\mathrm{fluxes}|z,\mathrm{quasar})$ such as is used to compute mean
quasar color tracks. This probability density is obtained from the
full \xdqsoz\ model flux--redshift density by conditioning on
redshift. For the relative flux this leads to a mixture of Gaussians
with means, variances, and amplitudes given by expressions similar to
those in \eqnname s~(\ref{eq:means}) and
(\ref{eq:alphazk})---essentially, relative flux and redshift need to
be interchanged in those equations. Properties of this distribution
can be calculated from the mixture of Gaussians. For example, the mean
quasar relative flux (or color) as a function of redshift is obtained
by weighting the means of the Gaussian components using the
redshift-dependent amplitudes. Code to calculate the mean quasar color
track is included in the package described in the \appendixname.

\section{Conclusion}\label{sec:conclusion}

In this \documentname\ we have introduced a new approach to
photometric quasar classification that can simultaneously classify
quasars and characterize their redshifts based on broadband
photometry. This technique, \xdqsoz, is an extension of the
\xdqso\ technique of \citet{Bovy11a} that adds the unknown redshift as
an extra parameter to the quasar model to obtain the likelihood
$p(z,\mathrm{fluxes}|\mathrm{quasar})$ that is central to both quasar
classification and photometric redshift estimation. We have shown that
this combined approach is both the best current quasar classification
technique---it has similar performance as the \xdqso\ method---and a
competitive photometric redshift method. Compared to other approaches
to photometric redshift estimation for quasars it has the advantage
that it can incorporate additional \uv\ and \nir\ data, even at low
signal-to-noise ratio, and can be extended to fainter flux levels
where photometric uncertainties are significant. Using samples of
quasars drawn from the \sdss, \twoslaq, and \boss\ spectroscopic
catalogs we have demonstrated this increased performance down to $g
\approx 22$ mag. The addition of \uv\ and \nir\ data to the
photometric redshift estimation problem essentially breaks all of the
redshift degeneracies inherent to the \ugriz\ filter set.

Code to use the \xdqsoz\ technique for classification and redshift
estimation, including the ability to calculate full posterior
probability distributions for the redshift, are made publicly
available. This code is briefly described in
the \appendixname.


\acknowledgements It is a pleasure to thank the anonymous referee and
Paul Martini, Gordon Richards, and David Weinberg for helpful comments
and discussions. J.B. and D.W.H. were partially supported by NASA
(grant NNX08AJ48G) and the NSF (grant AST-0908357). A.D.M.
acknowledges support under the NASA ADAP program (grant NNX08AJ28G).
J.F.H acknowledges support provided by the Alexander von Humboldt
Foundation in the framework of the Sofja Kovalevskaja Award endowed by
the German Federal Ministry of Education and Research. D.W.H. and
A.D.M. are research fellows of the Alexander von Humboldt Foundation
of Germany.

We gratefully acknowledge NASA's support for construction, operation,
and science analysis for the \galex\ mission, developed in cooperation
with the Centre National d'Etudes Spatiale of France and the Korean
Ministry of Science and Technology.

This work is based in part on data obtained as part of the UKIRT
Infrared Deep Sky Survey (\ukidss).

Funding for the SDSS and SDSS-II has been provided by the Alfred
P. Sloan Foundation, the Participating Institutions, the National
Science Foundation, the U.S. Department of Energy, the National
Aeronautics and Space Administration, the Japanese Monbukagakusho, the
Max Planck Society, and the Higher Education Funding Council for
England. The SDSS Web Site is http://www.sdss.org/.

The SDSS is managed by the Astrophysical Research Consortium for the
Participating Institutions. The Participating Institutions are the
American Museum of Natural History, Astrophysical Institute Potsdam,
University of Basel, University of Cambridge, Case Western Reserve
University, University of Chicago, Drexel University, Fermilab, the
Institute for Advanced Study, the Japan Participation Group, Johns
Hopkins University, the Joint Institute for Nuclear Astrophysics, the
Kavli Institute for Particle Astrophysics and Cosmology, the Korean
Scientist Group, the Chinese Academy of Sciences (LAMOST), Los Alamos
National Laboratory, the Max-Planck-Institute for Astronomy (MPIA),
the Max-Planck-Institute for Astrophysics (MPA), New Mexico State
University, Ohio State University, University of Pittsburgh,
University of Portsmouth, Princeton University, the United States
Naval Observatory, and the University of Washington.

SDSS-III is managed by the Astrophysical Research Consortium for the
Participating Institutions of the SDSS-III Collaboration including the
University of Arizona, the Brazilian Participation Group, Brookhaven
National Laboratory, University of Cambridge, University of Florida,
the French Participation Group, the German Participation Group, the
Instituto de Astrofisica de Canarias, the Michigan State/Notre
Dame/JINA Participation Group, Johns Hopkins University, Lawrence
Berkeley National Laboratory, Max Planck Institute for Astrophysics,
New Mexico State University, New York University, the Ohio State
University, the Penn State University, University of Portsmouth,
Princeton University, University of Tokyo, the University of Utah,
Vanderbilt University, University of Virginia, University of
Washington, and Yale University.

\appendix

\section{Code}\label{sec:code}

The \xdqsoz\ code for target selection, classification, and
photometric redshift estimation is publicly available at
\begin{quote}
\url{http://www.sdss3.org/svn/repo/xdqso/tags/}~.
\end{quote}
The code can be downloaded by \texttt{svn export} of the most recent
tag. The documentation of the most recent version of the code can be
found at
\begin{quote}
\url{http://www.sdss3.org/svn/repo/xdqso/tags/v0\_6/doc/build/html/index.html}~.
\end{quote}
Future updates will have documentation available at a similar URL.

The \xdqso/\xdqsoz\ package contains routines for quasar
classification using the \xdqso\ and \xdqsoz\ techniques. It also
contains code to calculate posterior probability distributions for the
quasar redshift of objects based on input \texttt{psfflux} and
\texttt{psfflux\_ivar}; these can be found in standard \sdss\ data
files such as the `sweeps' files\footnote{See
  \url{http://data.sdss3.org/datamodel/files/PHOTO\_SWEEP/RERUN/calibObj.html}~.}

The \xdqsoz\ models for the quasar color--redshift density are
contained in the \texttt{data/} directory. They are in the form of
FITS files containing the \exd\ models for all of the bins in apparent
magnitude, with one file for each combination of \sdss\ with
\galex\ and \ukidss. Each FITS file contains \nbins\ extensions, where
extension $k$ contains a structure with the amplitudes (tag
\flag{xamp}), means (tag \flag{xmean}), and covariance matrices (tag
\flag{xcovar}) for bin $k$ in \iband\ magnitude. The zeroth dimension
of the Gaussian represents the natural logarithm of the redshift,
followed by \sdss, \galex, and \ukidss\ fluxes (where relevant) in
this order, and ordered as NUV/FUV for the \galex, and YJHK for the
\ukidss.

\clearpage

\clearpage

\clearpage

\clearpage

\clearpage

\clearpage

\clearpage

\clearpage

\clearpage

\clearpage

\clearpage

\clearpage

\clearpage

\clearpage

\clearpage

\clearpage

\clearpage

\clearpage

\end{document}